\documentclass[aps,prb,reprint,showpacs]{revtex4-1}
\usepackage{amsmath,amssymb,bbm,hyperref,graphicx,color,comment,txfonts}
\usepackage[english]{babel}

\newcommand{\mathematica}{\textsc{Mathematica}}

\newcommand{\R}{\mathbb{R}}

\newcommand{\id}{\mathbbm{1}}
\newcommand{\dtt}{\tilde{\partial}_t}
\newcommand{\oq}{(\omega,\mathbf{q})}
\newcommand{\s}{\mathcal{S}}

\newcommand{\cg}{\mathrm{cg}}
\newcommand{\kcg}{k_{\mathrm{cg}}}
\newcommand{\ma}{\mathrm{MA}}
\newcommand{\mar}{\mathrm{MAR}}

\newcommand{\scgmar}{\mathcal{S}_{\cg}^{\mar}}
\newcommand{\zrcg}{Z_{R,\mathrm{cg}}}
\newcommand{\zicg}{Z_{I,\mathrm{cg}}}
\newcommand{\teff}{T_{\mathrm{eff}}}
\newcommand{\dip}{\mathrm{det}}
\newcommand{\qnb}{\bar{\phi}_q = \bar{\phi}_q^{*} = 0}
\newcommand{\qn}{\phi_q = \phi_q^{*} = 0}
\newcommand{\rqn}{\rho_{cq} = \rho_{qc} = 0}
\newcommand{\ezr}{\eta_{\mathit{ZR}}}
\newcommand{\ezi}{\eta_{\mathit{ZI}}}
\newcommand{\dqn}{\partial_q^2 \bigr\rvert_{q = 0}}

\newcommand{\abs}[1]{\left\lvert #1 \right\rvert}
\renewcommand{\Re}{\mathop{\mathrm{Re}}}
\renewcommand{\Im}{\mathop{\mathrm{Im}}}
\DeclareMathOperator{\Tr}{Tr}
\DeclareMathOperator{\tr}{tr}

\newcommand{\unitvec}[1]{\hat{\mathbf{#1}}}

\begin{document}

\date{\today}

\title{Non-equilibrium Functional Renormalization for
  Driven-Dissipative Bose-Einstein Condensation}

\author{L.\,M. Sieberer$^{1,2}$}
\author{S.\,D. Huber$^{3,4}$}
\author{E. Altman$^{4,5}$}
\author{S. Diehl$^{1,2}$}

\affiliation{$^1$Institute for Theoretical Physics, University of Innsbruck,
  A-6020 Innsbruck, Austria}
\affiliation{$^2$Institute for Quantum Optics and
  Quantum Information of the Austrian Academy of Sciences, A-6020 Innsbruck,
  Austria}
\affiliation{$^3$Theoretische Physik, Wolfgang-Pauli-Stra\ss e 27, ETH Zurich,
  CH-8093 Zurich, Switzerland}
\affiliation{$^4$Department of Condensed Matter Physics, Weizmann Institute of
  Science, Rehovot 76100, Israel}
\affiliation{$^5$Department of Physics, University of California, Berkeley, CA
  94720, USA}

\begin{abstract}
  We present a comprehensive analysis of critical behavior in the
  driven-dissipative Bose condensation transition in three spatial dimensions.
  Starting point is a microscopic description of the system in terms of a
  many-body quantum master equation, where coherent and driven-dissipative
  dynamics occur on an equal footing. An equivalent Keldysh real time functional
  integral reformulation opens up the problem to a practical evaluation using
  the tools of quantum field theory. In particular, we develop a functional
  renormalization group approach to quantitatively explore the universality
  class of this stationary non-equilibrium system. Key results comprise the
  emergence of an asymptotic thermalization of the distribution function, while
  manifest non-equilibrium properties are witnessed in the response properties
  in terms of a new, independent critical exponent. Thus the driven-dissipative
  microscopic nature is seen to bear observable consequences on the largest
  length scales. The absence of two symmetries present in closed equilibrium
  systems -- underlying particle number conservation and detailed balance,
  respectively -- is identified as the root of this new non-equilibrium critical
  behavior. Our results are relevant for broad ranges of open quantum systems on
  the interface of quantum optics and many-body physics, from exciton-polariton
  condensates to cold atomic gases.
\end{abstract}

\pacs{67.25.dj,64.60.Ht,64.70.qj,67.85.Jk}
\maketitle

\section{Introduction}
\label{sec:introduction}

In recent years, there has been tremendous progress in realizing systems with
many degrees of freedom, in which matter is strongly coupled to
light.\cite{carusotto13} This concerns vastly different experimental platforms:
In ensembles of ultracold atoms, the immersion of a Bose-Einstein condensate
(BEC) into an optical cavity has allowed to achieve strong matter-light
coupling, and lead to the realization of open Dicke
models;\cite{esslingerdicke,esslingerdicke3,esslingerQEDreview} In the context
of semiconductor quantum wells in optical cavities, non-equilibrium Bose
condensation has been achieved\cite{Kasp2006,lagoudakis08,Roumpos24042012} --
here the effective degrees of freedom, the exciton-polaritons, result from a
strong hybridization of cavity light and excitonic matter degrees of
freedom.\cite{SnokeBook,keeling10,carusotto13} Further promising platforms,
which are at the verge of the transition to true many-body systems, are arrays
of microcavities\cite{clarke-nature-453-1031,hartmann08,houck12,koch13} or
trapped ions,\cite{blatt12,britton12} as well as optomechanical
setups.\cite{marquardt09,chang11,ludwig13}

Those systems have three key properties in common. First, they are strongly
driven by external fields, such as lasers, placing them far away from
thermodynamic equilibrium even under stationary conditions. Equilibrium detailed
balance relations therefore are not generically present. Second, they exhibit
the characteristics of quantum optical setups, in that coherent and dissipative
dynamics occur on an equal footing, but at the same time are also genuine
many-body systems. Finally, a third characteristic is the absence of the
conservation of particle number. In particular, the admixture of light opens up
strong loss channels for the effective hybrid light-matter degrees of freedom,
and it becomes necessary to counterpoise these losses by continuous pumping
mechanisms in order to achieve stable stationary flux equilibrium states. The
pumping mechanisms can be either coherent or incoherent. In the latter case,
e.g., single particle pumping directly counteracts the incoherent single
particle loss; once it starts to dominate over the losses, a second order phase
transition results on the mean-field level, in close analogy to a laser
threshold.

At this point a clear difference between the quantum optical single mode problem
of a laser and a driven-dissipative many-body problem becomes apparent: While
the inclusion of fluctuations in the treatment of a laser smears out the mean-field
transition, in a system with a continuum of spatial degrees of freedom a genuine
out-of-equilibrium second order phase transition with true universal critical
behavior can be expected. The theoretical challenge is then to understand the
universal phenomena that can emerge due to the many-body complexity in a driven
non-equilibrium setting.

In this work we address this challenge, focusing on a key representative that
shows all the above characteristics: The driven-dissipative Bose condensation
transition, relevant to experiments with exciton-polariton condensates, or more
generally to any driven-dissipative system equipped with a $U(1)$ symmetry of
global phase rotations tuned to its critical point. We provide a comprehensive
characterization of the resulting non-equilibrium critical behavior in three
dimensions, extending and corroborating results presented
recently.\cite{sieberer13:_dynam_critic_phenom_driven_dissip_system} A key
finding concerns the existence of an additional, independent critical exponent
associated with the non-equilibrium drive. It describes universal decoherence at
long distances, and is observable, e.g., in the single particle response, as
probed in homodyne detection of exciton-polariton systems.\cite{utsu08} This
entails evidence that the microscopic non-equilibrium character bears observable
consequences up to the largest distances in driven Bose
condensation. Furthermore an asymptotic thermalization mechanism for the low
frequency distribution function is found. Such a phenomenon has been observed
previously in other
contexts.\cite{mitra06,diehl08:_quant,diehl10:_dynam_phase_trans_instab_open,dalla10:_quant,torre13:_keldy,oztop12:_excit,wouters06:_absen,mitra11:_mode_coupl_induc_dissip_therm,mitra12:_therm}
Here it is reflected in a symmetry that is emergent in the critical system on
the longest scales.

By contrast, in systems at true thermal equilibrium, this symmetry is present at
all scales as a microscopic symmetry. It then places severe restrictions on the
relations between the noise and the coherent and dissipative dynamics in the
system,\cite{graham73,deker75} leading to fluctuation dissipation relations
valid at all frequencies and wavelengths. This is the case, e.g., in the models
for dynamical universality classes established by Hohenberg and Halperin
(HH).\cite{hohenberg77:_theor} Non-equilibrium perturbations to these models
that have been discussed in the literature concern, e.g., modifications of the
noise term by spatial anisotropies, violating fluctuation-dissipation relations
on a microscopic scale. For models without conserved order parameter, such as
model A (MA) of HH, it has been shown that this does not lead to the existence
of new universal critical behavior, but rather to a modification of
non-universal amplitude
ratios.\cite{tauber02,henkel07:_field_theor_approac_noneq_dynam} Genuinely
non-equilibrium universal critical behavior has been found in several classical,
driven systems with different microscopic origins. Examples include models with
conserved order parameter with spatially anisotropic temperature,\cite{tauber02}
the driven-diffusive lattice gas, \cite{schmittmann95} reaction-diffusion
systems,\cite{doi76:_secon,peliti85:_path,cardytauber96,Canet2006} the problems
of directed percolation\cite{obukhov80,hinrichsen00} and self-organized
criticality,\cite{jensen98} or kinetic roughening phenomena such as described by
the Kardar-Parisi-Zhang
equation.\cite{kardar86,halpinhealy95,Canet2010,Kloss2012}

At the technical level, the purpose of this paper is to lay out a general
framework for addressing universal critical phenomena in open markovian
many-body quantum systems. This framework may be further generalized and applied
to a large variety of non-equilibrium situations, such as driven or
driven-dissipative systems with different
symmetries,\cite{ates12,olmos13,lee13,sarkar13} driven-dissipative systems with
disorder,\cite{janot13} and even superfluid
turbulence.\cite{berges08,bergesgasenzer10,gasenzer12,gasenzer13} We start from
a microscopic, second quantized description of the system in terms of quantum
master equations, and show how to translate the master equation into a Keldysh
real-time functional integral, which opens up the toolbox of well-established
techniques of quantum field theory. Next, we develop a functional
renormalization group (FRG) approach based on the Wetterich
equation,\cite{wetterich93:_exact} which allows us to compute both the dynamical
critical behavior as well as certain non-universal aspects of the problem. For
example, in addition to determining critical exponents we can also extract a
Ginzburg scale which marks the extent of the critical fluctuation regime.

The paper is organized as follows. In the next section we present our key
results and sketch the resulting physical picture. Section~\ref{sec:model}
introduces to our model and provides the mapping of the master equation to an
equivalent Keldysh functional integral. Using this framework, in
Sec.~\ref{sec:preparatory-analysis} we reproduce the results from mean-field and
Bogoliubov theory, and show how the physics of a semi-classical
driven-dissipative Gross-Pitaevskii equation emerges naturally as a low
frequency limit of the full quantum master equation. We highlight the additional
challenges which arise from the need to treat a continuum of spatial degrees of
freedom in order to capture critical behavior, and show in
Sec.~\ref{sec:funct-renorm-group} how they are properly addressed by means of
the FRG approach. The precise manifestation of the non-equilibrium character of
the problem is worked out in Sec.~\ref{sec:relat-equil-dynam}. A detailed
comparison of our non-equilibrium versus more conventional equilibrium models
highlights a symmetry which is only present in thermal equilibrium and expresses
detailed balance. We summarize the computation of the flow equations in
Sec.~\ref{sec:non-eq-frg-flow} and explain the hierarchical structure of the
universal critical behavior implied by the flow. In
Sec.~\ref{sec:numer-integr-flow} we discuss the numerical analysis of the flow
equations. We conclude in Sec.~\ref{sec:conclusions}.

At this point, we remark that the physical picture described in this work, and summarized in the following section, has been fully confirmed and further developed in a recent complementary perturbative field theoretical study presented in Ref.~~\onlinecite{Tauber2013}. There,  in particular, analytical estimates for the critical exponents are provided.

\section{Key Results and physical picture}
\label{sec:key-results-physical}

\paragraph*{Driven-dissipative Bose condensation transition --}

Driven open quantum systems are commonly modeled microscopically by means of
quantum master equations or in terms of Keldysh functional integrals as shown
below.  Starting from such a microscopic model of a driven Bose condensate we
derive in Sec.~\ref{sec:canon-power-count} an effective long-wavelength
description of the critical dynamics.  The result, after dropping all irrelevant
terms in the sense of renormalization group (RG), is a stochastic equation of
motion for the order parameter, which may be cast in Langevin form,
\begin{equation}
  \label{eq:1}
  i \partial_t \psi = \left[ - \left( A - i D \right) \Delta - \mu - i \kappa_1
    + 2 \left( \lambda - i \kappa \right)
    \abs{\psi}^2 \right] \psi + \xi.
\end{equation}
Such a dissipative stochastic Gross-Pitaevskii equation has been used as a model
for exciton-polariton
condensates.\cite{wouters07:_excit_noneq_bose_einst_conden_excit_polar,keeling04:_polar_conden_local_excit_propag_photon,szymaifmmode06:_noneq_quant_conden_incoh_pumped_dissip_system,keeling08:_spont_rotat_vortex_lattic_pumped_decay_conden,wouters10:_super_critic_veloc_noneq_bose_einst_conden,wouters10:_energ}
This equation includes terms describing coherent dynamics, as well as ones
capturing the dissipative processes and the drive. The coherent terms are the
inverse mass $A = 1/\left( 2 m \right)$, the chemical potential $\mu$ and the
elastic two-body interaction $\lambda$, whereas dissipative contributions
include a kinetic coefficient $D$, the effective single-particle loss rate
$\kappa_1$ as the difference between single-particle loss and pump rates, as
well as two-body loss $\kappa$.  The loss and gain processes induce noise, which
is taken into account by the Gaussian white noise source $\xi$ of strength
$\gamma$ with zero mean, $\langle \xi(t,\mathbf{x}) \rangle = 0$, and
correlations
\begin{equation}
  \label{eq:2}
  \langle \xi(t,\mathbf{x}) \xi^{*}(t',\mathbf{x}') \rangle = \frac{\gamma}{2} \delta(t - t') \delta(\mathbf{x} - \mathbf{x}').
\end{equation}
Unlike the models of critical dynamics classified in
Ref.~\onlinecite{hohenberg77:_theor}, the coherent and dissipative terms in a
driven condensate stem from completely independent physical processes. In
particular, this implies that the steady state of the Langevin
equation~\eqref{eq:1} is not characterized by a thermal (Gibbs) distribution of
the fields, and this leads to the distinct critical behavior analyzed in this
paper.

Equation~\eqref{eq:1} admits a time-independent homogeneous mean-field solution
$\abs{\psi_0}^2 = - \kappa_1/\left( 2 \kappa\right)$ if the single-particle pump
rate exceeds the corresponding loss rate, i.e., the effective single-particle
loss rate $\kappa_1$ becomes negative, and the chemical potential is set to be
$\mu = 2 \lambda \abs{\psi_0}^2$. Thus at the mean-field level a continuous
transition is tuned by varying the single particle pump rate: $\psi_0$ vanishes
as $\kappa_1$ goes from negative values to zero. Mean-field theory, however,
breaks down in the vicinity of the phase transition as the inclusion of
fluctuations may induce non-trivial scaling behavior or even render the
transition first order.\cite{colemanweinberg,halperinlubenskyma} In this paper
we verify that the system described by~\eqref{eq:1} in three spatial dimensions
indeed has a critical point characterized by universal dynamics. We argue that
this dynamics is governed by a ``Wilson-Fisher'' like fixed point, but with
another layer of dynamical critical behavior that is not found in non-driven
systems.

\paragraph*{Universality and extent of the critical domain --}

\begin{figure}
  \centering 
  \includegraphics[width=8.7
  cm]{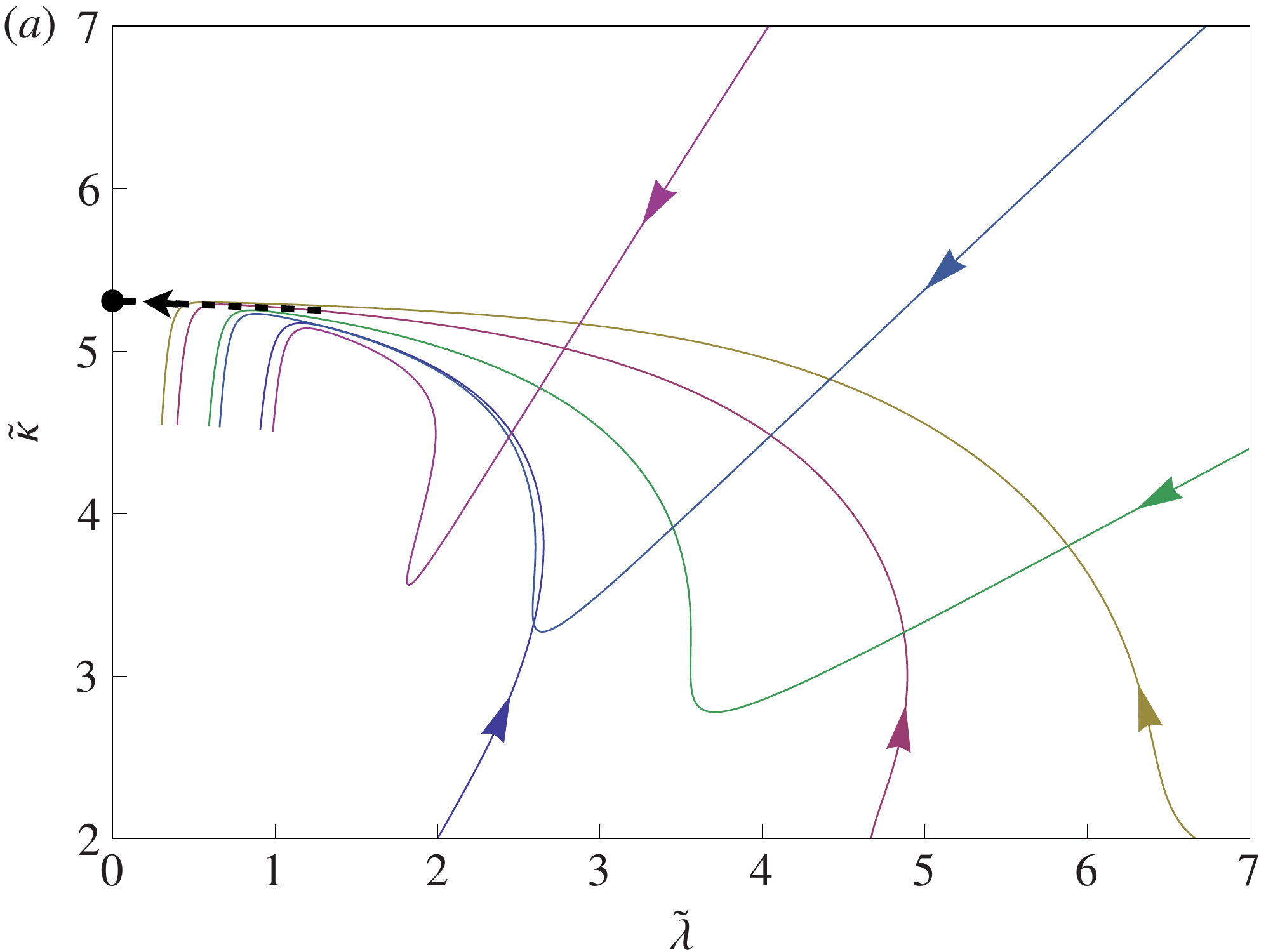}
  \includegraphics{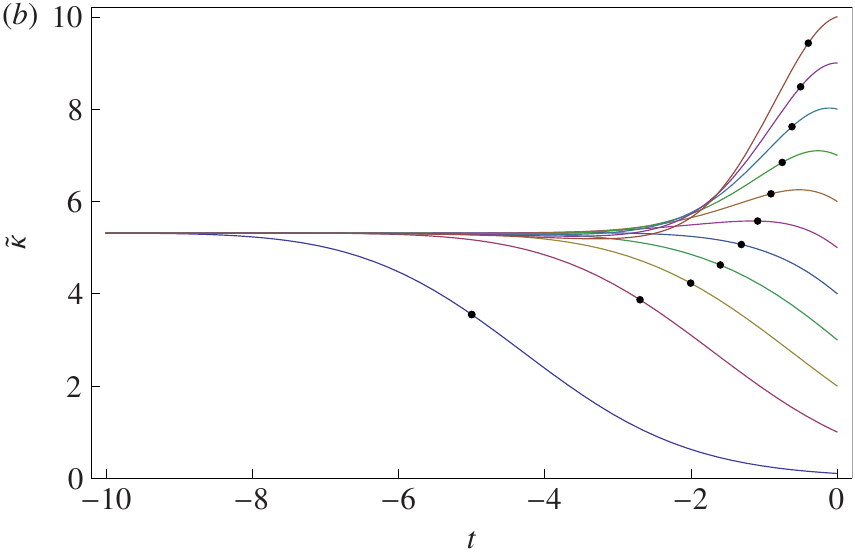}
  \caption{(Color online) Emergence of universality: (a) We show the flow of the
    complex renormalized two-body coupling $\tilde{u}_2 = \tilde{\lambda} + i
    \tilde{\kappa}$ (see Sec.~\ref{sec:resc-flow-equat}) for various initial
    values $\tilde{u}_{2 \Lambda}$. As a result of fine-tuning the initial
    values $w_{\Lambda}$ of the dimensionless mass parameter close to
    criticality, all flow trajectories approach the Wilson-Fisher fixed point
    $\tilde{u}_{2*} = i 5.308$ (indicated by the black dot) before eventually
    bending away. Shown are numerical solutions to the flow equations for $r_{K
      \Lambda} = 10, r_{u_3 \Lambda} = 1, \tilde{\kappa}_{3 \Lambda} = 0.01$,
    and values of $\tilde{u}_{2 \Lambda}$ lying on a rectangle with sides
    $\tilde{\lambda} \in [0,10], \tilde{\kappa} = 2, 10$ and $\tilde{\lambda} =
    10, \tilde{\kappa} \in [2,10]$. (See Sec.~\ref{sec:resc-flow-equat} for
    definitions of the parameters.) (b) Flow of $\tilde{\kappa}$ as a function
    of the dimensionless infrared cutoff $t = \ln \left( k/\Lambda \right)$ for
    various starting values $\tilde{\kappa}_{\Lambda}$. Dots on the lines
    indicate the extent of the critical domain, which is set by the Ginzburg
    scale Eq.~\eqref{eq:3}. Initial values are the same as in (a), apart from
    $\tilde{\kappa}_{\Lambda} = 0.1, 1, 2, \dotsc, 10$ and $r_{u_2 \Lambda} =
    10$.}
  \label{fig:universality}
\end{figure}

Our main technical tool for the analysis is a functional RG carried out for the
dynamical problem. Emergence of a universal critical point is evident from the
flow of the coupling constants to a fixed point independent of the initial
conditions, as long as the system is tuned to the phase transition (cf.\ also
Sec.~\ref{sec:numer-integr-flow}). This is demonstrated in
Fig.~\ref{fig:universality} (a), showing the flow of the real and imaginary
parts of the complex interaction parameter ${\tilde
  u}_2={\tilde\lambda}+i{\tilde\kappa}$ (see Sec.~\ref{sec:resc-flow-equat}). At
the fixed point the real parts of all couplings vanish, which implies that the
effective long-wavelength dynamics is purely dissipative. Integrating out fast
fluctuations in the course of the RG flow, therefore, leads to a loss of
coherence.

An important aspect of the phase transition which we analyze in detail here
concerns the extent of the critical domain, which is delimited by the Ginzburg
momentum scale $k_G$. Knowledge of this non-universal scale is important for
assessing the requirements from experiments aimed at measuring the critical
phenomena. We find it to be given by (cf.\ Sec.~\ref{sec:numer-integr-flow})
\begin{equation}
  \label{eq:3}
  k_G = \gamma_{\Lambda} \kappa_{\Lambda}/\left( 2 C D_{\Lambda}^2 \right),
\end{equation} 
where $C \approx 14.8$ and $\gamma_{\Lambda}$, $\kappa_{\Lambda}$, and
$D_{\Lambda}$ are, respectively, the noise-strength, two-body loss rate and
dissipative kinetic coefficient appearing in the description of the system at a
mesoscopic scale $\Lambda$ (see Sec.~\ref{sec:truncation}). Here we confirm this
behavior quantitatively by a full numerical solution of the flow equations
outside the critical domain, highlighting the capability of the FRG approach to
compute universal and non-universal physics in a single framework.

\paragraph*{Asymptotic thermalization of the distribution function --}

An interesting result of the RG analysis is that the distribution function of
the order parameter field at the critical point effectively thermalizes at long
wavelengths and low frequencies.  The effective thermalization is manifest as an
emergent symmetry of the equations of motion at the fixed point that is not
present at the mesoscopic level, cf.\ Sec.~\ref{sec:relat-equil-dynam}.  For
this reason the dynamical critical exponent $z$ is the same as that of MA of the
equilibrium classification. The presence of this symmetry implies a
fluctuation-dissipation theorem (FDT), or, more physically speaking, a detailed
balance condition valid at asymptotically long wavelengths.

In order to better understand this aspect, consider an equilibrium problem with
detailed balance. This means that all subparts of the system are in equilibrium
with each other.  In other words, temperature is invariant under the system's
partition in such a state. This statement is easily translated into a RG
language: Natural system partitions are the momentum shells. Partition
invariance of the temperature thus becomes a scale invariance of temperature
under renormalization, which successively integrates out high momentum
shells. The ``equilibrium symmetry'' expresses precisely this physical
intuition.

\begin{figure}
  \centering
  \includegraphics{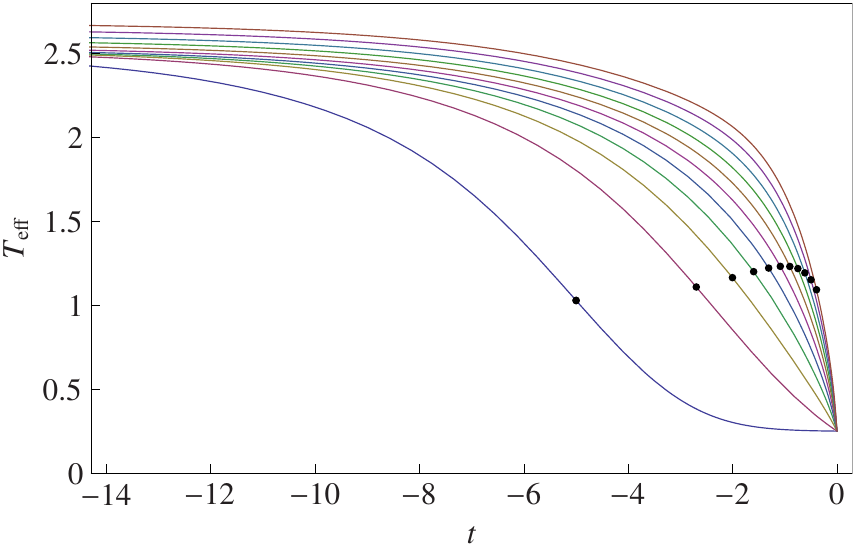}  
  \caption{(Color online) Scale dependence at criticality of the effective
    temperature $\teff = \bar{\gamma}/\left( 4 \abs{Z} \right)$, where
    $\bar{\gamma}$ denotes the Keldysh mass and $Z$ is the wave-function
    renormalization (see Secs.~\ref{sec:truncation}
    and~\ref{sec:resc-flow-equat}). For $t \to - \infty$ the effective
    temperature saturates to a constant value. Initial values are the same as in
    Fig.~\ref{fig:universality} (b).}
  \label{fig:teff}
\end{figure}

In a non-equilibrium problem such as the driven condensate we discuss, this
symmetry is in general absent at arbitrary momentum scales. In order to
demonstrate how it emerges at long scales, we compute the scale dependence of an
effective temperature, entering the (non-equilibrium) FDT, cf.\
Sec.~\ref{sec:relat-equil-dynam}. Indeed, we find scale dependent behavior at
high momenta, which becomes universal and scale independent within the critical
region delimited by the Ginzburg scale, cf.\ Fig.~\ref{fig:teff}.

We note that, in principle, it is conceivable that the system might allow for
different stationary scaling solutions far from equilibrium with different
universal scaling behavior, not captured in the present approach. Indeed, in two
dimensions, such a scenario is realized.\cite{Altman2013} In three dimensions,
however, such a behavior could be present only beyond a threshold value for the
microscopic strength of violation of detailed balance.

\paragraph*{Hierarchical shell structure of non-equilibrium criticality --}

A key result of the RG analysis is the hierarchical organization of the
non-equilibrium criticality. This structure consists of three shells of critical
exponents.  The innermost shell in this hierarchy contains the two independent
exponents $\nu, \eta$ describing the static (spatial) critical behavior of the
classical $O(2)$ model.\footnote{The continuous planar rotations of $O(2)$
  reflect the continuous phase rotation symmetry $U(1) \cong \mathit{SO}(2)$ of
  the driven open Bose system.} We find that the static exponents coincide with
those of an \textit{ab initio} computation of the classical $O(2)$ exponents at
the same level of approximation. Thus the non-equilibrium conditions do not
modify the static critical behavior.

The intermediate shell contains the so-called dynamical exponent $z$ which
describes the dynamical (temporal) critical behavior. This intermediate shell is
already present in models for equilibrium dynamical criticality. Crucially, it
extends the static critical behavior but does not modify it. In fact there is a
certain dynamical fine structure: The same static universality class splits up
into various dynamical universality classes, classified in models A to J by
HH.\cite{hohenberg77:_theor} Again, we find the dynamic exponents to coincide
with the one of an \emph{ab initio} computation for one of HH's models (MA) --
the non-equilibrium conditions do not modify the dynamical critical behavior
either. A stronger physical consequence of this finding is discussed in the next
subsection.

The unique element found only in the driven system is the outer shell of the
aforementioned hierarchy. The related exponent $\eta_r$ identified in
Ref.~\onlinecite{sieberer13:_dynam_critic_phenom_driven_dissip_system}, which we
refer to as the ``drive exponent'', physically describes universal decoherence
of the long-wavelength dynamics as explained above. Crucially, $\eta_r$ relates
to the dynamical MA in the same way as MA relates to the classical $O(2)$ model:
It adds a new shell, but does not ``feed back'' or modify the inner shells of
the hierarchy. In Sec.~\ref{sec:relat-equil-dynam} we argue that this exponent
manifestly witnesses non-equilibrium conditions.

\paragraph*{Independence of the drive exponent and maximality of the extension --}

It is important to demonstrate the independence of the drive exponent: At a second
order phase transition, many critical exponents can be defined, each
characterizing a different observable. However, only few of them are
independent, i.e., cannot be expressed in terms of a smaller set by means of
scaling relations.

{The independence of the four critical exponents identified with our FRG approach is
  manifest in the block diagonal structure of the linearized RG flow 
  in the vicinity of the Wilson-Fisher fixed point, cf.\
  Sec.~\ref{sec:scaling-solutions}: }There are two blocks, and the lowest
eigenvalue of each of them determines an independent critical exponent. In
addition we have the independent anomalous dimension $\eta$ and the dynamical
exponent $z$. 

A general way to determine the number of independent exponents and thereby see
the need for one and only one additional exponent in this system (as compared to
equilibrium MA dynamics) comes from the UV limit of the problem. Any
independent critical exponent must be related to a short-distance mass scale in
the problem.\cite{goldenfeldbook} For example, this can be seen in the case of
the anomalous dimension associated with the spatial two-point correlation
function. An anomalous dimension $\eta$ implies decay of the correlation
function as $\langle \phi^*(\mathbf{x}) \phi(\mathbf{0})\rangle \sim
\abs{\mathbf{x}}^{2-d + \eta}$. Since the physical units of this correlation are
$[L]^{2-d}$ we require a microscopic scale $a$, to fix the units so that
$\langle \phi^*(\mathbf{x}) \phi(\mathbf{0})\rangle \sim a^{-\eta}
\abs{\mathbf{x}}^{2-d + \eta} \sim [L]^{2-d}$. In the same way any non-trivial
independent exponent requires such a microscopic scale.

To determine the number of independent critical exponents in our problem we
therefore need to count the microscopic mass scales in the bare action. The
corresponding quadratic part of the action reads
\begin{multline}
  \label{eq:4}
  \s_{m} = \int d t d^d \mathbf{x} \left[ \left( \phi_c^{*},\phi_q^{*} \right)
    \begin{pmatrix}
      0 & \mu - i \kappa_1   \\
      \mu + i \kappa_1 & i \gamma
    \end{pmatrix}
    \begin{pmatrix}
      \phi_c \\ \phi_q
    \end{pmatrix}
  \right. \\ \left.
    \vphantom{\begin{pmatrix} \phi_c \\ \phi_q
      \end{pmatrix}} + f \left( j_c^* \phi_q + j_q^* \phi_c + \mathrm{c.c.}
    \right) \right],
\end{multline}
with real parameters $\mu ,\kappa_1, \gamma,f$. $\kappa_1$ and $f$, which
describe the tuning parameter of the phase transition and an external ordering
field respectively, have direct counterparts in the equilibrium $O(2)$
model. They give rise to the two critical exponents $\nu$, which characterizes
the divergence of the correlation length, and $\eta$, the anomalous dimension of
the static two-point function. $\gamma$ is introduced in the theory of dynamical
critical phenomena and is associated to the dynamical exponent in the purely
relaxational MA of HH.\cite{hohenberg77:_theor} In the full non-equilibrium
problem however, there is yet another mass scale $\mu$. This scale is at the
origin of the additional independent exponent identified in
Ref.~\onlinecite{sieberer13:_dynam_critic_phenom_driven_dissip_system}.

From this discussion we conclude that the extension of the critical behavior at
the condensation transition is maximal, i.e., no more independent exponents can
exist. This is due to general requirements on the mass matrix above: the
off-diagonal elements must be hermitian conjugates; the lower diagonal must be
anti-hermitian; and the upper diagonal must be zero due to the conservation of
probability.

It is worth noting how this analysis would change if the critical point in
question involved breaking of a $Z_2$ symmetry rather than a continuous $O(N)$
symmetry as we discuss here. Such an Ising transition in a driven system is
relevant for the formation of a super-solid due to interaction of a BEC with the
modes of an optical cavity.\cite{esslingerdicke,esslingerdicke3} In this case
the reality of the Ising fields rules out an imaginary mass term ($\kappa_1 =
0$). Hence the maximal number of independent critical exponents is 3, which
implies that there can be no modification of MA dynamics.

\paragraph*{Interpretation and observability of the drive  exponent --}

The drive critical exponent describes the universal flow behavior of all
possible ratios of coherent vs.\ dissipative couplings (real vs.\ imaginary
parts, see Sec.~\ref{sec:resc-flow-equat}) to zero upon moving to larger and
larger distances. In the competition of coherent and dissipative dynamics,
loosely speaking dissipation always wins. Physically, this should be interpreted
as a universal mechanism of decoherence. The drive exponent therefore is
subleading and not observable in the correlation functions of the
system. However, it is directly observable in the single particle dynamical
response (single particle retarded Green's function).

\begin{figure}
  \includegraphics{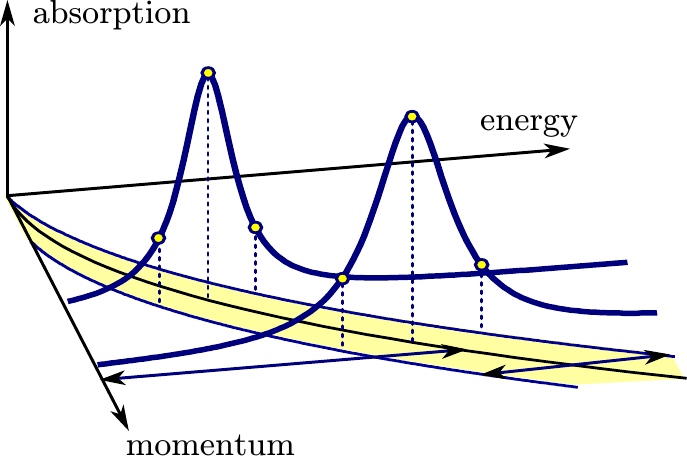}
  \caption{(Color online) Illustration of the observability of the drive
    exponent: The absorption peak for a measurement that observes the single
    particle dynamical response. The drive exponent $\eta_r$ reveals itself in a
    different scaling of the peak location and peak-width as a function of the
    momentum.}
  \label{fig:sketch}
\end{figure}

The dynamical response can be measured with any probe that couples directly to
the field operator $\hat{\psi}(\mathbf{x})$, i.e., any probe that out-couples
single particles from the system. This is the case, e.g., in the angle-resolved
detection of leakage photons in exciton-polariton systems,\cite{utsu08} or in
angle-resolved radio-frequency spectroscopy in ultra-cold
atoms.\cite{Stewart2008} As we argue in Sec.~\ref{sec:wilson-fisher-fixed}, the
excitation spectrum close to the critical point is given by ($q =
\abs{\mathbf{q}}$)
\begin{equation}
  \label{eq:5}
  \omega(q) \sim A_0 q^{z-\eta_r} - i D_0 q^z \sim  A_0 q^{2.223}-i D_0 q^{2.121},
\end{equation}
where $A_0$ and $D_0$ are non-universal constants. This excitation spectrum
leads to a broadened signal in the experiment, cf.\ Fig.~\ref{fig:sketch}. The
drive exponent $\eta_r$ can be observed due to the different scaling with
momentum of the location ($\sim q^{z-\eta_r}$) and the width ($\sim q^z$) of the
measured peak. We note here, however, that technical noise and other
uncertainties in the measurement setup will unavoidably also lead to a
broadening of the spectrum. The small value of $\eta_r = - 0.101$ thus
challenges experiments to verify this prediction.

\section{The Model}
\label{sec:model}

In this section we introduce a generic microscopic description of
driven-dissipative Bose systems, written in terms of a second quantized master
equation. We then show how to translate this model into the Keldysh functional
integral framework, which provides a convenient starting point for obtaining the
long wavelength universal properties of the system. Moreover we introduce the
concept of the effective action, which generalizes the action principle to
include all quantum and statistical fluctuations and is the key object for the
formulation of the FRG.

\subsection{Quantum Master equation}
\label{sec:master-equation}

Our model with particle loss and pumping is described microscopically by a
many-body master equation that determines the time evolution of the system
density operator (units are chosen such that $\hbar = 1$),
\begin{equation}
  \label{eq:6}
  \partial_t \hat{\rho} = - i \left[ \hat{H},\hat{\rho} \right] + \mathcal{L} [\hat{\rho}].
\end{equation}
This equation incorporates both coherent dynamics generated by the Hamiltonian
$\hat{H}$ and dissipation that is subsumed in the action of the Liouville
operator $\mathcal{L}$. The Hamiltonian $\hat{H}$ describes interacting bosonic
degrees of freedom of mass $m$ and is given by (we use the shorthand
$\int_{\mathbf{x}} = \int d^d \mathbf{x}$)
\begin{equation}
  \label{eq:7}
  \hat{H} = \int_{\mathbf{x}} \hat{\psi}^{\dagger}(\mathbf{x}) \left( -
    \frac{\Delta}{2 m} \right) \hat{\psi}(\mathbf{x}) + \frac{g}{2}
  \int_{\mathbf{x}} \hat{\psi}^{\dagger}(\mathbf{x})^2 \hat{\psi}(\mathbf{x})^2,
\end{equation}
where $\hat{\psi}(\mathbf{x})$ are bosonic field operators. Note that we do not
explicitly introduce any system chemical potential, as the density of the system
will be fixed by the balance of pumping and losses. Two-body interactions are
described by a density-density interaction with coupling constant $g$. In the
following we shall be interested in dynamically stable systems which are
characterized by a positive coupling constant $g > 0$. This modeling of
interactions is valid on length scales which are not sufficient to resolve
details of the microscopic interaction potential.

In our model, dissipative dynamics comes in the form of one-body pumping ($p$)
and losses ($l$) as well as two-body losses ($t$). Accordingly, the Liouville
operator can be decomposed into the sum of three terms $\mathcal{L} =
\sum_{\alpha} \mathcal{L}_{\alpha}$ with $\alpha = p, l, t$ which have the
common Lindblad structure
\begin{equation}
  \label{eq:8}
  \mathcal{L}_{\alpha}[\hat{\rho}] = \gamma_{\alpha} \int_{\mathbf{x}} \left(
    \hat{L}_{\alpha}(\mathbf{x}) \hat{\rho} \hat{L}_{\alpha}^{\dagger}(\mathbf{x}) -
    \frac{1}{2} \left\{ \hat{L}_{\alpha}^{\dagger}(\mathbf{x})
      \hat{L}_{\alpha}(\mathbf{x}), \hat{\rho} \right\} \right),
\end{equation}
with local Lindblad or quantum jump operators $\hat{L}_{\alpha}(\mathbf{x})$
that create ($p$) and destroy ($l$) single particles; for $\alpha = t$ two
particles are destroyed at the same instant in time, i.e., the quantum jump
operators are given by
\begin{equation}
  \label{eq:9}
  \hat{L}_p(\mathbf{x}) = \hat{\psi}^{\dagger}(\mathbf{x}), \quad
  \hat{L}_l(\mathbf{x}) = \hat{\psi}(\mathbf{x}), \quad
  \hat{L}_t(\mathbf{x}) = \hat{\psi}(\mathbf{x})^2.
\end{equation}
These processes occur at rates $\gamma_p, \gamma_l$, and $\gamma_t$,
respectively.

The net effect of single-particle pumping and losses is determined by the
relative size of the respective rates: For $\gamma_p > \gamma_l$, there is an
effective gain of single particles. Nevertheless, Eq.~\eqref{eq:6} leads (in a
suitably chosen rotating frame, as we will show below) to a stationary state
$\hat{\rho}_{\mathrm{ss}}$ in which the gain of single particles is balanced by
two-body losses. In this situation, a finite condensate amplitude builds up,
\begin{equation}
  \label{eq:10}
  \langle \hat{\psi}(\mathbf{x})
  \rangle_{\mathrm{ss}} = \tr \left( \hat{\psi}(\mathbf{x})
    \hat{\rho}_{\mathrm{ss}} \right) = \psi_0 \neq 0, \quad \gamma_p > \gamma_l.
\end{equation}
That is, in stationary state the system is in a condensed phase in which the
symmetry of the dynamics described by Eq.~\eqref{eq:6} under global $U(1)$
transformations of the field operators $\hat{\psi}(\mathbf{x}) \mapsto
\hat{\psi}(\mathbf{x}) e^{i \phi}$ is broken. When the loss rate $\gamma_l$
exceeds the pumping rate $\gamma_p$, on the other hand, no condensate emerges in
stationary state, and the expectation value of the bosonic field operator is
zero,
\begin{equation}
  \label{eq:11}
  \langle \hat{\psi}(\mathbf{x}) \rangle_{\mathrm{ss}} = 0, \quad \gamma_p \leq \gamma_l.
\end{equation}
Equations~\eqref{eq:10} and~\eqref{eq:11} can be derived from the master
equation~\eqref{eq:6} in mean-field approximation by making the ansatz of a
coherent stationary state $\hat{\rho}_{\psi} = \lvert \psi \rangle \langle \psi
\rvert$, where we assume that the amplitude in $\lvert \psi \rangle =
\frac{1}{\mathcal{N}} \exp \left( \psi \int_{\mathbf{x}}
  \hat{\psi}^{\dagger}(\mathbf{x}) \right) \lvert 0 \rangle$ is spatially
homogeneous but possibly time-dependent. Proper normalization of the coherent
state is ensured by the choice $\mathcal{N} = e^{V \abs{\psi}^2}$ with the
system volume $V$. The time-dependence of the condensate amplitude is determined
by taking the time derivative on both sides of the equality $\psi = \tr \left(
  \hat{\psi}(\mathbf{x}) \hat{\rho}_{\psi} \right)$ and using the master
equation~\eqref{eq:6}, which results in
\begin{equation}
  i \partial_t \psi = \left[ g \abs{\psi}^2 + \frac{i}{2} \left( \gamma_p -
      \gamma_l - 2 \gamma_t \abs{\psi}^2 \right) \right] \psi.
\end{equation}
For $\gamma_p > \gamma_l$ this equation allows for a solution of the form $\psi
= \psi_0 e^{- i \mu t}$, where the condensate density is determined by the
imaginary part of the term in brackets on the right-hand side (RHS) as
\begin{equation}
  \abs{\psi_0}^2
  = \frac{\gamma_p - \gamma_l}{2 \gamma_t}.
\end{equation}
The parameter $\mu$ is then given by $\mu = g \abs{\psi_0}^2$. We obtain the
steady state density matrix of Eq.~\eqref{eq:10} by means of a transformation to
a rotating frame with the unitary operator $\hat{U} = \exp \left( i \mu \hat{N}
  t \right)$, where the particle number operator is $\hat{N} = \int_{\mathbf{x}}
\hat{\psi}^{\dagger}(\mathbf{x}) \hat{\psi}(\mathbf{x})$: We have
$\hat{\rho}_{\mathrm{ss}} = \hat{U} \hat{\rho}_{\psi} \hat{U}^{\dagger}$, which
is indeed time-independent, and recover Eqs.~\eqref{eq:10}
and~\eqref{eq:11}. Under the transformation to this rotating frame, the
Hamiltonian acquires a contribution $- \mu \hat{N}$, whereas the Liouvillian
$\mathcal{L}$ remains invariant. In the following we will always be working in
the rotating frame.

In summary, the steady state phase diagram of our model exhibits two phases: A
symmetric one characterized by Eq.~\eqref{eq:11} and an ordered one where the
global $U(1)$ symmetry is broken by a finite condensate amplitude Eq.~\eqref{eq:10}
with definite phase. These two phases are separated by a continuous phase
transition with order parameter $\psi_0$. The transition is crossed by tuning
the single-particle pumping rate from $\gamma_p < \gamma_l$ in the ``symmetric''
to $\gamma_p > \gamma_l$ in the ``symmetry-broken'' or ``ordered'' phase.

In the following we shall be interested in the critical behavior that is induced
by tuning $\gamma_p - \gamma_l$ to zero. Powerful tools for investigating
critical phenomena at a second order phase transition are provided by a
multitude of variants of the RG. The particular flavor we employ here is the FRG
in the formulation of Wetterich\cite{wetterich93:_exact} (for reviews see
Refs.~\onlinecite{berges02:_nonper,salmhofer01:_fermion_renor_group_flows,pawlowski07:_aspec,delamotte08:_introd_nonper_renor_group,rosten12:_fundam,boettcher12:_ultrac_funct_renor_group}),
which builds upon the use of functional integrals. Therefore, as a first step
towards implementing a FRG investigation of our model, we will reformulate the
physics that is encoded in the quantum master equation~\eqref{eq:6} in terms of
Keldysh functional integrals.\cite{kamenev09:_keldy,kamenevbook}

\subsection{Keldysh functional integral}
\label{sec:keldysh-path-integr}

The Keldysh approach provides a means to tackle general non-equilibrium problems
in the language of functional integrals. For the model at hand, the dynamics
described by the master equation~\eqref{eq:6} can be represented equivalently as
a Keldysh partition function (see App.~\ref{sec:mark-diss-acti}): By
$\Psi_{\sigma} = \left( \psi_{\sigma},\psi_{\sigma}^{*} \right)^T$ for $\sigma =
+,-$ we denote Nambu spinors of fields on the forward- and backward-branch of
the closed time contour, respectively. Then, collecting time and space in a
single variable $X = (t,\mathbf{x})$ and using the abbreviation $\int_X = \int d
t \int d^d \mathbf{x}$, the Keldysh partition function reads
\begin{equation}
  \label{eq:12}
  \mathcal{Z}[J_{+},J_{-}] = \int \mathcal{D}[\Psi_{+},\Psi_{-}] \, e^{i
    \s[\Psi_{+},\Psi_{-}] + i \int_X \left( J_{+}^{\dagger} \Psi_{+}
      - J_{-}^{\dagger} \Psi_{-} \right)}.  
\end{equation}
The fields $J_{\sigma} = \left( j_{\sigma},j_{\sigma}^{*} \right)^T$ are external
sources inserted here for the purpose of calculating correlation functions of
the bosonic fields in the usual manner by means of functional
differentiation. When they are set to zero, $J_{+} = J_{-} = 0$, the partition
function reduces to unity,\cite{kamenev09:_keldy,kamenevbook} i.e., we have the
normalization $\mathcal{Z}[0,0] = 1$. While the Keldysh approach can in
principle be utilized to study time evolution, here we are assuming
translational invariance in time, as appropriate for the investigation of steady
state properties.

In complete analogy to the separation of coherent and dissipative contributions
to the time evolution of the density operator in Eq.~\eqref{eq:6}, the action
$\s$ in the functional integral Eq.~\eqref{eq:12} can be decomposed as $\s = \s_H +
\s_D$ into a Hamiltonian part $\s_H$ and a part $\s_D$ corresponding to the
dissipative Liouvillian $\mathcal{L}$ in the master equation. The former is
given by (from now on we will use units such that $2 m = 1$)
\begin{equation}
  \label{eq:13}
  \s_H = \sum_{\sigma = \pm} \sigma \int_X \left[ \psi_{\sigma}^{*} \left(
      i \partial_t + \Delta + \mu \right) \psi_{\sigma} -
    \frac{g}{2} \left( \psi_{\sigma}^{*} \psi_{\sigma} \right)^2 \right].
\end{equation}
As a general rule (see App.~\ref{sec:mark-diss-acti}), normally ordered
operators in Eq.~\eqref{eq:6} acting on the density matrix $\hat{\rho}$ from the
left (right) result in corresponding fields on the $\sigma = +$ ($\sigma = -$)
contour. Consequently, the commutator with the Hamiltonian in Eq.~\eqref{eq:6}
is transferred into the two contributions to Eq.~\eqref{eq:13} with a relative
minus sign.

The same rule applies to the dissipative part in the master
equation~\eqref{eq:6}. Passing from the Liouvillian $\mathcal{L}$ on to a
dissipative action $\s_D$, quantum jump operators $\hat{L}_{\alpha}$ are
replaced by corresponding jump fields $L_{\alpha,\sigma}$ on the $\sigma = +$
($\sigma = -$) contour. (In App.~\ref{sec:mark-diss-acti} we will discuss
regularization issues related to normal ordering of Lindblad operators.)  As
above we have the three contributions $\s_D = \sum_{\alpha} \s_{\alpha}$ that
are due to single-particle pumping ($p$) and losses ($l$) as well as two-body
losses ($t$). The form of the jump fields can directly be inferred from
Eq.~\eqref{eq:9} as
\begin{equation}
  L_{p,\sigma} = \psi_{\sigma}^{*}, \quad L_{l,\sigma} = \psi_{\sigma}, \quad
  L_{t,\sigma} = \psi_{\sigma}^2.
\end{equation}
Then, for the dissipative parts of the action we find the expression
\begin{equation}
  \label{eq:14}
  \s_{\alpha} = - i \gamma_{\alpha} \int_X \left[ L_{\alpha,+} L_{\alpha,-}^{*} -
    \frac{1}{2} \left( L_{\alpha,+}^{*} L_{\alpha,+} +
      L_{\alpha,-}^{*} L_{\alpha,-} \right) \right].
\end{equation}
As we can see, the transition from a description of a specific problem in terms
of a master equation to one in terms of Keldysh functional integrals reduces to
the application of simple rules. For our model,
Eqs.~\eqref{eq:12},~\eqref{eq:13} and~\eqref{eq:14} provide us with a convenient
starting point for the investigation of the steady state phase transition
described in the previous section.

While the translation rules from the master equation to the Keldysh functional
integral are most simply applied in a basis of fields $\psi_{\pm}$ that can be
ascribed to the forward and backward branches of the Keldysh contour,
subsequently we will find it advantageous to introduce so-called classical and
quantum fields, given by the symmetric and anti-symmetric combinations
\begin{equation}
  \label{eq:15}
  \begin{pmatrix}
    \phi_c \\ \phi_q
  \end{pmatrix}
  = M
  \begin{pmatrix}
    \psi_{+} \\ \psi_{-}
  \end{pmatrix},
  \quad M =
  \frac{1}{\sqrt{2}}
  \begin{pmatrix}
    1 & 1 \\
    1 & -1
  \end{pmatrix}.
\end{equation}
Condensation is described by a time-independent, homogeneous expectation value
of the fields on the $\sigma = \pm$ contours, $\langle \psi_{+}(X) \rangle =
\langle \psi_{-}(X) \rangle = \psi_0$, cf.~Eq.~\eqref{eq:10}. In the basis of
classical and quantum fields, this is expressed as $\langle \phi_c(X) \rangle =
\phi_0 = \sqrt{2} \psi_0, \langle \phi_q(X) \rangle = 0$, i.e., only $\phi_c$
can condense (and, therefore, become a ``classical'' variable), whereas $\phi_q$
is a purely fluctuating field with zero expectation value by construction.

By means of the transformation Eq.~\eqref{eq:15}, the inverse propagator, determined
by the quadratic part of the action, is cast in the characteristic causality
structure\cite{kamenev09:_keldy,kamenevbook} with retarded, advanced, and Keldysh
components $P^R$, $P^A$, and $P^K$, respectively (in the following we will
denote the two-body coupling constant and loss rate by, respectively, $\lambda =
g/2$ and $\kappa = \gamma_t/2$),
\begin{multline}
  \label{eq:16}
  \s = \int_X \biggl\{ \left( \phi_c^{*},\phi_q^{*} \right)
  \begin{pmatrix}
    0 & P^A\\
    P^R & P^K
  \end{pmatrix}
  \begin{pmatrix}
    \phi_c \\ \phi_q
  \end{pmatrix}  + i 4 \kappa \phi_c^{*} \phi_c \phi_q^{*} \phi_q \\
  - \left[ \left( \lambda + i \kappa \right) \left( \phi_c^{*2}
      \phi_c \phi_q + \phi_q^{*2} \phi_c \phi_q \right) + \mathrm{c.c.}  \right]
  \biggr\}.
\end{multline}
The inverse retarded and advanced single-particle Green's functions are given by
$P^R = P^{A \dagger} = i \partial_t + \Delta + \mu + i \kappa_1$ where $\kappa_1
= \left( \gamma_l - \gamma_p \right)/2$. For the Keldysh component of the
inverse propagator we have $P^K = i \gamma$, where $\gamma = \gamma_l +
\gamma_p$ is the sum of single-particle pumping and loss rates -- both of them
increase the noise level in the system.

The spectrum of single-particle excitations is encoded in the poles of the
retarded propagator in frequency-momentum space or, equivalently, in the zeros
of the inverse propagator. Solving $P^R(Q) = 0$ for $\omega$, where $Q = \left(
  \omega,\mathbf{q} \right)$ collects the frequency and spatial momentum, we
obtain the dispersion relation
\begin{equation}
  \label{eq:17}
  \omega = q^2 - \mu - i \kappa_1.
\end{equation}
For $\kappa_1 > 0$ (i.e., $\gamma_p < \gamma_l$) the pole is located in the
lower complex half-plane, and the effective loss rate $\kappa_1$ takes the role
of an inverse lifetime. One has single-particle excitations that decay
exponentially in time, a situation that is well-known from the general theory of
the analytic structure of correlation functions.\cite{Altland/Simons} As
$\kappa_1$ is tuned to negative values (i.e., as we cross the phase transition),
however, the pole Eq.~\eqref{eq:17} is shifted into the upper complex half-plane,
signaling an instability. After crossing this threshold, the system develops a
condensate, and the proper analytical structure of the retarded propagator is
restored only by taking the tree-level shifts due to the condensate into
account. We will discuss the corresponding modifications of the dispersion
relation Eq.~\eqref{eq:17} below in Sec.~\ref{sec:mean-field-theory}.

Inversion of the $2 \times 2$ matrix in Eq.~\eqref{eq:16} yields the propagator
with retarded, advanced, and Keldysh components,
\begin{equation}
  G =
  \begin{pmatrix}
    G^K & G^R \\
    G^A & 0
  \end{pmatrix}.
\end{equation}
The components along with their respective usual diagrammatic
representation\cite{kamenev09:_keldy,kamenevbook} are given by
\begin{equation}
  \label{eq:18}
  \parbox{2.38in}{\includegraphics{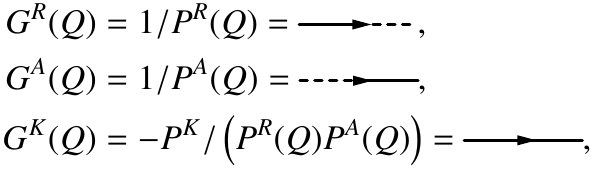}}
\end{equation}
which shows that the poles of $G(Q)$ are determined solely by the zeros of
$P^R(Q)$ and $P^A(Q)$. The Keldysh component $P^K$ of the inverse propagator
enters the expression for $G^K(Q)$ multiplicatively. Therefore, even in a
situation where $P^K$ is a polynomial in frequency and/or momentum, it can not
give rise to further poles in the propagator $G(Q)$.

In the Keldysh formalism elastic two-body collisions and two-body losses are
treated on an equal footing: Both appear in the action Eq.~\eqref{eq:16} as
quartic vertices, however, with a real coupling constant $\lambda$ in the case
of elastic collisions and a purely imaginary coupling constant $i \kappa$ for
two-body losses. The vertices in Eq.~\eqref{eq:16} can further be distinguished
by the number of quantum fields they contain: We have the so-called classical
vertex $- \int_X \left[ \left( \lambda + i \kappa \right) \phi_c^{* 2} \phi_c
  \phi_q + \mathrm{c.c.}  \right]$ which contains only one quantum field and
three classical fields, and two quantum vertices: The first one $i 4 \kappa
\int_X \phi_c^{*} \phi_c \phi_q^{*} \phi_q$ containing two and the second one $-
\int_X \left[ \left( \lambda + i \kappa \right) \phi_q^{* 2} \phi_c \phi_q +
  \mathrm{c.c.} \right]$ containing three quantum fields. Diagrammatically,
these vertices are depicted as
\begin{equation}
  \label{eq:19}
  \parbox{.78in}{\includegraphics{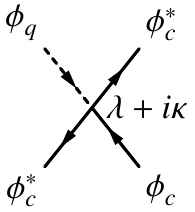}},
  \qquad \parbox{.78in}{\includegraphics{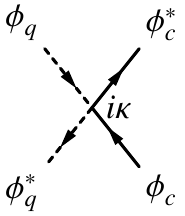}},
  \quad \parbox{.78in}{\includegraphics{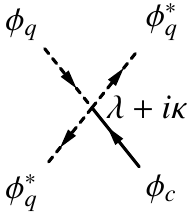}}.
\end{equation}
The fact that there are no vertices consisting only of classical fields is a
manifestation of causality or conservation of probability in the Keldysh
framework.\cite{kamenev09:_keldy,kamenevbook} Below we will find that only the
classical vertex is relevant (in the sense of the RG) once the system is tuned
close to the phase transition.

\subsection{Effective action}
\label{sec:effective-action}

Having established a description of our model in terms of a Keldysh functional
integral, we proceed by introducing the concept of the effective
action,\cite{Amit/Martin-Mayor,Kleinert/Schulte-Frohlinde,Zinn-Justin} which is
central to the FRG. It is also a convenient starting point for a discussion of
the phase transition on the mean-field level (see
Sec.~\ref{sec:mean-field-theory}).

In equilibrium statistical physics the effective action $\Gamma$ is related to
the free energy as a functional of a space-dependent order parameter, and the
equilibrium state is determined as the order parameter configuration that
minimizes $\Gamma$. However, in the present context of non-equilibrium
statistical physics we do not have a sensible notion of a free energy. In fact,
already the Keldysh partition function Eq.~\eqref{eq:12} reduces for vanishing
external sources to a representation of unity $\mathcal{Z}[0,0] = 1$,
independently of the parameters that characterize the
action.\cite{kamenev09:_keldy,kamenevbook} Still, the Keldysh effective action,
defined analogously to its equilibrium counterpart as the Legendre transform of
the generating functional for connected correlation functions, is a very useful
object. From $\Gamma$ we can derive, e.g., field equations that determine the
stationary configurations of classical and quantum fields $\Phi_{\nu} = \left(
  \phi_{\nu},\phi_{\nu}^{*} \right)^T, \nu = c, q$. On a more formal level,
$\Gamma$ is the generating functional of one-particle irreducible
vertices.\cite{Amit/Martin-Mayor,Kleinert/Schulte-Frohlinde,Zinn-Justin} Most
importantly for our model, however, the FRG provides us with a means of
calculating critical exponents for the phase transition by studying the RG flow
of $\Gamma$ as a function of an infrared cutoff $k$.

Our starting point for introducing the effective action is the generating
functional Eq.~\eqref{eq:12} for correlation functions, expressed in the basis of
classical and quantum fields $\Phi_{\nu}$, i.e., the action is given by
Eq.~\eqref{eq:16}, and we introduce classical and quantum sources $J_{\nu} =
\left( j_{\nu},j_{\nu}^{*} \right)^T$ with $\nu = c, q$ according to the Keldysh
rotation
\begin{equation}
  \begin{pmatrix}
    j_c \\ j_q
  \end{pmatrix} =
  M
  \begin{pmatrix}
    j_{+} \\ j_{-}
  \end{pmatrix},
\end{equation}
where the matrix $M$ is defined in Eq.~\eqref{eq:15}. For the generating
functional $\mathcal{W}$ for connected correlation functions and $\mathcal{Z}$
we have the relation
\begin{equation}
  \label{eq:20}
  \mathcal{W}[J_c,J_q] = - i \ln \mathcal{Z}[J_c,J_q].
\end{equation}
The idea is now to express $\mathcal{W}$, which is a functional of the external
sources $J_{\nu}$, in terms of the corresponding field expectation values
$\bar{\Phi}_{\nu} = \langle \Phi_{\nu} \rangle \rvert_{J_c,J_q} = \delta
\mathcal{W}/\delta J_{\nu'}$ where $\nu' = q$ for $\nu = c$ and \textit{vice
  versa}. Introducing these as new variables is accomplished by means of a
Legendre transform:
\begin{equation}
  \label{eq:21}
  \Gamma[\bar{\Phi}_c,\bar{\Phi}_q] = \mathcal{W}[J_c,J_q] + \int_X \left( J_c^{\dagger}
    \bar{\Phi}_q + J_q^{\dagger} \bar{\Phi}_c \right).
\end{equation}
The difference between the in this way defined effective action $\Gamma$ and the
action $\s$ consists in the inclusion of both statistical and quantum
fluctuations in the former. This becomes apparent in the representation of
$\Gamma$ as a functional integral,\cite{berges02:_nonper}
\begin{equation}
  \label{eq:22}
  e^{i \Gamma[\bar{\Phi}_c,\bar{\Phi}_q]} = \int \mathcal{D}[\delta
  \bar{\Phi}_c,\delta \bar{\Phi}_q] \, e^{i \s[\bar{\Phi}_c + \delta
    \bar{\Phi}_c,\bar{\Phi}_q + \delta \bar{\Phi}_q]},
\end{equation}
which holds for the equilibrium states that obey $\delta \Gamma/\delta
\bar{\Phi}_c = \delta \Gamma/\delta \bar{\Phi}_q = 0$ at vanishing external
sources $J_c = J_q = 0$. The most straightforward way of evaluating the
functional integral Eq.~\eqref{eq:22} approximately is by performing a perturbative
expansion around the configuration that minimizes the action
$\s$. To zeroth order this corresponds to mean-field theory, an
approach we will discuss in the following section. In the FRG, the fluctuations
$\delta \bar{\Phi}_{\nu}$ are included stepwise by introducing an infrared
regulator which suppresses fluctuations with momenta less than an infrared
cutoff scale $k$. A short review of this method, adapted to the Keldysh
framework, is provided in Sec.~\ref{sec:frg-approach-keldysh}. We will apply it to
our model in Sec.~\ref{sec:non-eq-frg-flow}.

\section{Preparatory Analysis}
\label{sec:preparatory-analysis}

Here we carry out a basic analysis of the model in preparation for setting up a
full functional RG calculation used to obtain the critical properties at the
phase transition. We summarize the mean-field theory for the effective action
and discuss the generic emergence of infrared divergences near a critical
point. Furthermore, using dimensional analysis we identify the important terms
in the action which are potentially relevant at the critical point. These terms
are then included in the ansatz of the effective action used to carry out the
FRG calculation. Finally we contrast this ansatz with the equilibrium dynamical
models of HH.\cite{hohenberg77:_theor}

\subsection{Mean-field theory}
\label{sec:mean-field-theory}

In Sec.~\ref{sec:master-equation} we identified the precise balance between
single-particle losses and pumping as the transition point,
cf.~Eqs.~\eqref{eq:10} and~\eqref{eq:11}. Here we will derive this result from
the Keldysh functional integral Eq.~\eqref{eq:22}, again employing a mean-field
approximation. We will then proceed by calculating the excitation spectrum above
the stationary mean-field by treating quadratic fluctuations in a Bogoliubov
(tree-level) expansion. While this issue, as well as going beyond the mean-field
approximation by perturbative methods, can equally well be addressed in the
master equation formalism of Sec.~\ref{sec:master-equation},\cite{Li2013} in
performing a perturbative expansion at and below the critical point we encounter
infrared divergences. Proper treatment of these requires RG methods, which are
well-established and elegantly formulated in terms of functional integrals.

Mean-field theory corresponds to a saddle-point approximation of the functional
integral in Eq.~\eqref{eq:22} in which fluctuations around the classical fields
are completely neglected. In the present context, by classical fields we mean
spatially homogeneous solutions to the classical field equations
\begin{equation}
  \label{eq:23}
  \frac{\delta \s}{\delta \phi_c^{*}} = 0, \quad \frac{\delta \s}{\delta
    \phi_q^{*}} = 0.
\end{equation}
As already mentioned above, there are no terms in the action Eq.~\eqref{eq:16}
that have zero power of both $\phi_q^{*}$ and $\phi_q$, and the same is
obviously true for $\delta \s/\delta \phi_c^{*}$. Therefore, the first
equation~\eqref{eq:23} is solved by $\phi_q = 0$. Inserting this condition in
the second equation~\eqref{eq:23}, we have
\begin{equation}
  \label{eq:24}
  \left[ \mu + i \kappa_1 - \left( \lambda - i \kappa \right) \abs{\phi_0}^2 \right] \phi_0 = 0.
\end{equation}
The solution $\phi_c = \phi_0$ is determined by the imaginary part of
Eq.~\eqref{eq:24}: For $\kappa_1 \geq 0$, in the symmetric phase, the classical
field is zero, $\rho_0 = \abs{\phi_0}^2 = 0$, whereas for $\kappa_1 < 0$ we have
a finite condensate density $\rho_0 = - \kappa_1/\kappa$. In a second step, the
parameter $\mu$ is determined by the real part of Eq.~\eqref{eq:24} as $\mu = -
\lambda \kappa_1/\kappa$.

Quadratic fluctuations around the mean-field order parameter can be investigated
in a Bogoliubov or tree-level expansion: We set $\phi_c = \phi_0 + \delta
\phi_c, \phi_q = \delta \phi_q$ in the action Eq.~\eqref{eq:16} and expand the
resulting expression to second order in the fluctuations $\delta
\phi_{\nu}$. The poles of the retarded propagator (which is now a $2 \times 2$
matrix in the space of Nambu spinors $\delta \Phi_{\nu} = \left( \delta
  \phi_{\nu},\delta \phi_{\nu}^{*} \right)^T$) are
then\cite{wouters07:_excit_noneq_bose_einst_conden_excit_polar}
\begin{equation}
  \label{eq:25}
  \omega_{1,2}^R = - i \kappa \rho_0 \pm \sqrt{q^2 \left( q^2 + 2 \lambda \rho_0 \right) - \left(
      \kappa \rho_0 \right)^2}.
\end{equation}
Real and imaginary parts of both branches are shown in
Fig.~\ref{fig:dispersion_relation} in panels (a) and (b), respectively. Due to
the tree-level shifts $\propto \rho_0$ the instability of Eq.~\eqref{eq:17} for
$\kappa_1 < 0$ is lifted: Both poles are located in the lower complex
half-plane, indicating a physically stable situation with decaying
single-particle excitations. For $\kappa = 0$, Eq.~\eqref{eq:25} reduces to the
usual Bogoliubov result,\cite{LL:IX} where for $q \to 0$ the dispersion is
phononic, $\omega_{1,2}^R = \pm c q$, with speed of sound $c = \sqrt{2 \lambda
  \rho_0}$ whereas particle-like behavior $\omega_{1,2}^R \sim q^2$ is recovered
at high momenta. Here, due to the presence of two-body loss $\kappa \neq 0$, the
dispersion is strongly modified: While at high momenta the dominant behavior is
still given by $\omega_{1,2}^R \sim q^2$, at low momenta we obtain purely
dissipative non-propagating modes $\omega_1^R \sim - i \frac{\lambda}{\kappa}
q^2$ and $\omega_2^R \sim - i 2 \kappa \rho_0$. In particular, for $q = 0$ we
have $\omega_1^R = 0$: This is a dissipative Goldstone
mode,\cite{wouters06:_absen,wouters07:_excit_noneq_bose_einst_conden_excit_polar,szymaifmmode06:_noneq_quant_conden_incoh_pumped_dissip_system}
associated with the spontaneous breaking of the global $U(1)$ symmetry in the
ordered phase. The existence of such a mode is not bound to the mean-field
approximation but rather an exact property of the theory guaranteed by the
$U(1)$ invariance of the effective action, even in the present case of a
driven-dissipative condensate.

\subsection{Infrared divergences near criticality}

The discussion of our model on the mean-field level has illustrated some of the
benefits of the Keldysh approach: Not only have we gained a simple physical
picture of the phase transition as a condensation instability in the retarded
and advanced propagators, but we were able to investigate excitations in both
the symmetric and ordered phases quite straightforwardly. Mean-field theory,
however, while providing us with a good qualitative understanding of the
stationary state physics of our model far away from the phase transition, has
major shortcomings when it comes to the discussion of critical phenomena. In
particular, the critical exponents that can be extracted from an analysis of
quadratic fluctuations around the mean-field configuration are not indicative of
the universality class of the phase transition, as they correspond to the RG
flow in the vicinity of a non-interacting (or Gaussian) fixed point. Critical
behavior at the phase transition, however, is encoded in the RG flow in the
vicinity of an interacting (or Wilson-Fisher) fixed point.

In a many-body system, excitations and their interactions get dressed due to
scattering from other particles. The mean-field results of this section can be
taken as the starting point for a calculation of the effective dressed
parameters in a perturbative expansion. In the functional integral
Eq.~\eqref{eq:22}, diagrammatically this amounts to an expansion in the number
of loops around the mean-field configuration. To lowest (one-loop) order, the
correction $\Delta \lambda$ to the real part of the bare classical vertex (the
first diagram in Eq.~\eqref{eq:19}) reads ($v_d = \left( 2^{d + 1} \pi^{d/2}
  \Gamma(d/2) \right)^{-1}$)
\begin{equation}
  \label{eq:26}
  \parbox{1.83in}{\includegraphics{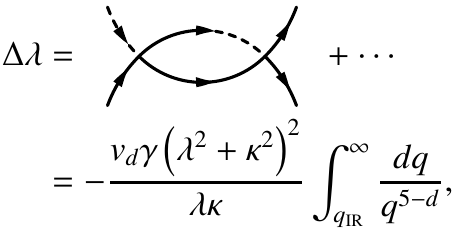}}
\end{equation}
where the elements appearing in the diagram are defined in Eqs.~\eqref{eq:26}
and~\eqref{eq:18} (here, however, lines correspond to propagators of
fluctuations $\delta \Phi_{\nu}$ and acquire an additional $2 \times 2$ matrix
structure in Nambu space), and the ellipsis indicates that all diagrams with
four external legs and one closed loop corresponding to a single internal
momentum integration have to be included. In the integrand we have only kept the
dominant contribution for $q \to 0$, and we have introduced an infrared cutoff
$q_{\mathrm{IR}}$ in order to regularize the divergence at low momenta. Such
infrared divergences, however, appear not only in our specific example of the
loop correction to $\lambda$, but rather are characteristic of perturbative
expansions in symmetry broken phases. They are due to the presence of a massless
Goldstone mode, which results in a pole of the retarded and advanced propagators
at $\omega = q = 0$. This problem is even enhanced as we approach the phase
transition: Then both modes become degenerate, with also the second mode
$\omega_2^R \sim - i 2 \kappa \rho_0$ for $q \to 0$ becoming massless. A method
that allows us to go beyond mean-field theory, therefore, has to provide for a
proper treatment of infrared divergences. In the FRG, this is achieved by
effectively introducing a mass term $\propto k^2$ in the inverse propagators by
hand. In consequence, the integrand in Eq.~\eqref{eq:26} is replaced by
$\int_{q_{\mathrm{IR}}}^{\infty} d q q^{d - 1}/\left( q^2 + k^2 \right)^2$ and
we may safely set $q_{\mathrm{IR}}$ to zero since the effective mass $k^2$ acts
as an infrared cutoff. The resulting loop-corrected coupling is a function of
this cutoff, $\lambda = \lambda(k)$, and we obtain the fully dressed or
renormalized coupling by following the RG flow of the running coupling
$\lambda(k)$ for $k \to 0$. This procedure can be implemented efficiently by
introducing the cutoff in the functional integral Eq.~\eqref{eq:22}. We will
discuss how this is done in practice for the present non-equilibrium
problem\cite{gasenzer08:_towar,berges09:_nonth,Berges201237,berges13:_C,Gezzi2007,Jakobs2007,Karrasch2010,canet11:_gener}
in the following section. Critical exponents can then be extracted from the flow
of the critical system, i.e., when $\kappa_1$ is fine-tuned to zero.

So far we have discussed only corrections to the bare interaction vertices due
to the inclusion of loop diagrams. However, also the propagators appearing in
these diagrams are themselves renormalized. In particular, the inverse
propagator can be written as $P(Q) - \Sigma(Q)$, i.e., as the sum of the bare
inverse propagator $P(Q)$ and a self-energy correction
$\Sigma(Q)$.\cite{Amit/Martin-Mayor,Kleinert/Schulte-Frohlinde,Zinn-Justin} The
self-energy contribution at one-loop order to the retarded propagator is
represented diagrammatically as
\begin{equation}
  \label{eq:27}
  \parbox{2.22in}{\includegraphics{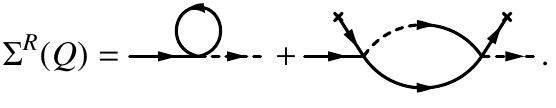}}
\end{equation}
where effective cubic couplings, which are obtained upon expanding the
interaction vertex around the field expectation value, appear in the second
diagram. Lines beginning and terminating in crosses indicate that particles are
scattered out of and into the condensate, respectively. Due to momentum
conservation, the first diagram does not depend on the external momentum $Q =
\left( \omega,\mathbf{q} \right)$ and gives a correction to the constant part of
the inverse propagator, i.e., the so-called mass terms. Since the coupling
$\lambda + i \kappa$ associated with the vertex appearing in this diagram is
complex, both the real and imaginary masses, $\mu$ and $\kappa_1$, are affected
by the loop correction. The second diagram in Eq.~\eqref{eq:27} gives a
frequency- and momentum-dependent contribution to the self-energy. Symmetry
under spatial rotations implies that it depends only on the modulus of the
momentum and we may write $\Sigma^R(Q) = \Sigma^R(\omega,q^2)$. For small
$\omega$ and $q^2$ we can expand $\Sigma^R(\omega,q^2) \approx \Sigma^R(0,0) +
\omega \partial_{\omega} \Sigma^R(0,0) + q^2 \partial_{q^2}
\Sigma^R(0,0)$. Transforming back to the time domain and real space, the
derivatives of the self-energy with respect to frequency and momentum give
corrections to the coefficients of $\partial_t$ and $\Delta$ in the inverse
propagators, which are again complex valued. An imaginary part of the
coefficient of the Laplacian corresponds to an effective dissipative kinetic
coefficient due to the interaction with other particles; A complex prefactor of
the time derivative, on the other hand, has significant consequences for the
physical interpretation of all other couplings, as we will discuss in detail in
later sections.

\subsection{Canonical power counting}
\label{sec:canon-power-count}

While the proper theoretical approach to critical phenomena has to cope
efficiently with the infrared divergences discussed above, such systems also
exhibit an important ordering principle, which is provided by the classification
of couplings according to their canonical scaling dimension. In the following we
will briefly review this procedure, often referred to as canonical power
counting or dimensional analysis. It lays the basis for a suitable choice of
ansatz for the effective action that will contain only couplings which are
relevant or marginal according to this counting
scheme.\cite{Amit/Martin-Mayor,Kleinert/Schulte-Frohlinde,Zinn-Justin}

At second order phase transitions, physical quantities exhibit scaling behavior,
which means that they depend on the distance from the phase transition (in our
case this distance is measured by $\kappa_1$) in a power-law fashion $\sim
\kappa_1^{\tau}$, with a generally non-integer exponent $\tau$. In order to
study critical behavior in the RG, we investigate the RG flow starting from the
action fine-tuned to criticality, i.e., with $\kappa_1 = 0$, and approach the
critical point by lowering $k$. Then, scaling behavior of a physical quantity
$g$ shows up as power-law dependence $g \sim k^{\theta}$ on $k$ for $k \to 0$
with a critical exponent $\theta$. In other words, phase transitions are
associated to scaling solutions of the RG flow (not all scaling solutions
correspond to phase transitions\cite{Cardy}), or -- equivalently -- fixed points
of the flow of rescaled couplings $\tilde{g} = k^{-\theta} g$. The dominant
contribution to the exponent $\theta$ associated to a coupling $g$ is determined
by its physical dimension measured in units of momentum $k$, i.e., the canonical
scaling dimension or engineering dimension $[g]$ (we have $[k] =
1$). Anticipating that deviations from canonical scaling will be small (see
Sec.~\ref{sec:scaling-solutions}), let us study the flow of the dimensionless
two-body elastic collision coupling $\tilde{\lambda} = \lambda/k$ (we will see
below that $\tilde{\lambda}$ is indeed dimensionless). In
Sec.~\ref{sec:mean-field-theory} we saw that the flow of $\lambda$ is generated
by the loop diagrams Eq.~\eqref{eq:26}. Then, to the flow of the dimensionless
variable $\tilde{\lambda}$ we have an additional contribution due to the
engineering dimension,
\begin{equation}
  \label{eq:33}
  \partial_t \tilde{\lambda} = - \tilde{\lambda} + \text{loop diagrams},
\end{equation}
where we are taking the derivative with respect to the dimensionless logarithmic
scale $t = \ln(k/\Lambda)$ which is zero for $k = \Lambda$ and goes to $-
\infty$ for $k \to 0$. The loop contribution to the flow of $\tilde{\lambda}$ is
of order $\tilde{\lambda}^2, \tilde{\lambda} \tilde{\kappa}, \tilde{\kappa}^2$
and higher in the dimensionless two-body couplings $\tilde{\lambda}$,
$\tilde{\kappa}$. We find, therefore, a trivial fixed point $\partial_t
\tilde{\lambda} = 0$ for $\tilde{\lambda}_{*} = \tilde{\kappa}_{*} = 0$. The
flow for small $\tilde{\lambda}$ in the vicinity of this Gaussian fixed point is
determined by the canonical scaling contribution on the RHS of Eq.~\eqref{eq:33}
and is directed towards higher values of $\tilde{\lambda}$, i.e., the coupling
$\tilde{\lambda}$ is relevant at the Gaussian fixed point. For increasing
$\tilde{\lambda}$, the loop contributions become important and balance canonical
scaling at a second fixed point. This non-trivial Wilson-Fisher fixed point at
finite $\tilde{\lambda}_{*}, \tilde{\kappa}_{*}$ corresponds to the phase
transition in the interacting system, and for small deviations $\tilde{\lambda}
- \tilde{\lambda}_{*}$ the flow is attracted to $\tilde{\lambda}_{*}$.

The described scenario changes drastically for a coupling with negative
canonical scaling dimension, i.e., when instead of the prefactor $- 1$ for the
first term on the RHS in Eq.~\eqref{eq:33} we had a positive integer. Such a
coupling is irrelevant at the Gaussian fixed point, which means that its flow is
attracted to that fixed point. We can, therefore, as a starting point for a
systematic expansion in the relevance of couplings, set all irrelevant couplings
to zero. Unlike perturbative expansions, the inclusion of irrelevant couplings
in higher orders in the expansion in canonical scaling dimensions results not
only in enhanced quantitative accuracy, but rather refines our picture of the
phase transition, as it involves higher order vertices and a refined treatment
of the momentum dependence of propagators.\cite{tetradis94:_critic}

We proceed by determining the canonical dimensions of the couplings appearing in
the action Eq.~\eqref{eq:16}. They are not uniquely fixed by the requirement
that the action is dimensionless, $[\s] = 0$: Still we have the freedom of
assigning different scaling dimensions to the classical $\phi_c$ and quantum
fields $\phi_q$. We exploit this freedom in order to impose a scaling dimension
upon the Keldysh component of the inverse propagator in Eq.~\eqref{eq:16} that
is the same as in finite-temperature thermodynamic
equilibrium,\cite{kamenev09:_keldy,kamenevbook} i.e., we require $[\gamma] =
0$. While this choice yields a consistent picture of the driven-dissipative Bose
condensation transition as detailed below, it is inappropriate for the
investigation of stationary transport solutions that define genuine
nonequilibrium states with nonvanishing flux which might be contained in our
model.  As already pointed out in Sec. \ref{sec:key-results-physical}, in two dimensions, such a scenario indeed has been recently identified in Ref.~\onlinecite{Altman2013}, showing that the Kardar-Parisi-Zhang non-equilibrium fixed point \cite{kardar86} governs the long wavelength behavior. In three dimensions, a similar scenario is conceivable in principle, however only beyond a certain threshold value for the strength of violation of detailed balance. 

Denoting the dynamical exponent by $[\partial_t] = z$ we find, from the
quadratic part of the action and in $d$ dimensions,
\begin{equation}
  z = [\mu] = [\kappa_1] = 2, \quad [\phi_c] = \frac{d - 2}{2}, \quad [\phi_q]
  = \frac{d + 2}{2}.
\end{equation}
The different scaling dimensions of classical and quantum fields result in
different behavior of the complex couplings associated with the classical and
quantum vertices Eq.~\eqref{eq:19} under renormalization, even though their
values at $k = \Lambda$ are the same. In particular, for a local vertex that
contains $n_c$ classical and $n_q$ quantum fields, the canonical scaling
dimension of the corresponding coupling is
\begin{equation}
  \left[ \lambda_{n_c,n_q} \right] = d + 2 - n_c [\phi_c] - n_q [\phi_q].
\end{equation}
We observe that all couplings $\lambda_{n_c,n_q}$ with $n_q > 2$ ($n_q \geq 1$
is required by
causality\cite{Altland/Simons,kamenev09:_keldy,kamenevbook}) or $n_c
> 5$ are irrelevant. The coupling $\lambda_{3,1}$ associated with the classical
quartic vertex has canonical dimension $4 - d$, i.e., its upper critical
dimension is $d = 4$ and, in the case of interest $d = 3$, it is relevant with
canonical scaling dimension equal to unity. All other quartic couplings are
irrelevant, as are sextic couplings with $n_q > 1$. The classical three-body
coupling $\lambda_{5,1}$ is marginal and we will include it (with both real and
imaginary parts) in our ansatz for the running effective action below, even
though it is not present in the action $\s$. Higher order couplings
$\lambda_{n_c,n_q}$ with $n_c + n_q > 6$ are irrelevant and we will discard
them.

\subsection{Equilibrium symmetry}
\label{sec:equilibrium-symmetry}

According to the canonical power counting scheme outlined in the previous
section, in order to describe critical properties at the driven-dissipative Bose
condensation transition we may disregard quantum vertices in the action
Eq.~\eqref{eq:16} -- this corresponds to a semiclassical
approximation,\cite{Altland/Simons,kamenev09:_keldy,kamenevbook,tauberbook} and
the resulting simplified ``mesoscopic'' action has the same structure as the
classical dynamical models considered in Ref.~\onlinecite{hohenberg77:_theor}
inasmuch as it is linear in the quantum fields apart from the noise term which
is quadratic. Therefore, like in the classical dynamical models, the functional
integral with the mesoscopic action is equivalent to a Langevin equation for the
classical field. This is just the stochastic dissipative Gross-Pitaevskii
equation Eq.~\eqref{eq:1} for a single non-conserved complex field $\psi$ and
bears close resemblance to the equation of motion of MA of HH with $N = 2$ real
components. There are, however, two key differences: First the dynamics in MA is
purely relaxational whereas Eq.~\eqref{eq:1} contains both coherent and
dissipative contributions. Second, and more importantly, dropping all coherent
contributions on the RHS of Eq.~\eqref{eq:1} we find that it is invariant under
the transformation of the
fields:\cite{enz79:_field,aron10:_symmet_langev,canet11:_gener}
\begin{equation}
  \label{eq:212}
  \begin{split}
    \psi(t,\mathbf{x}) & \mapsto \psi^{*}(-t,\mathbf{x}), \\
    \xi(t,\mathbf{x}) & \mapsto - \xi^{*}(-t,\mathbf{x}) - i 2 \partial_t
    \psi^{*}(t,\mathbf{x}).
  \end{split}
\end{equation}
This symmetry of the dynamics implies a FDT for the retarded response and
correlation functions. Its absence in the driven-dissipative model (DDM),
therefore, may be seen as indicating non-equilibrium conditions.  In
Sec.~\ref{sec:relat-equil-dynam} below we discuss a generalized version of the
symmetry transformation Eq.~\eqref{eq:212} and we are led to consider an
extension of MA by coherent dynamics that then differs from the DDM precisely in
the obedience to this generalized symmetry. With regard to critical phenomena,
the difference in symmetries between equilibrium and non-equilibrium situations
renders it possible that novel universal behavior may be found in the latter
case. We proceed to perform an FRG analysis of the critical properties of both
models in the following sections.

\section{Functional Renormalization Group}
\label{sec:funct-renorm-group}

\subsection{FRG approach for the Keldysh effective
  action}
\label{sec:frg-approach-keldysh}

The transition from the action $\s$ to the effective action
$\Gamma$ consists in the inclusion of both statistical and quantum fluctuations
in the latter (see Eq.~\eqref{eq:22}). In the FRG, the functional integral over
fluctuations is carried out stepwise by introducing an infrared regulator which
suppresses fluctuations with momenta less than an infrared cutoff scale
$k$.\cite{berges02:_nonper} This is achieved by adding to the action
in Eq.~\eqref{eq:12} a term
\begin{equation}
  \label{eq:28}
  \Delta \s_k = \int_X \left( \phi_c^{*},\phi_q^{*} \right)  
  \begin{pmatrix}
    0 & R_{k,\bar{K}}(-\Delta) \\
    R_{k,\bar{K}}^{*}(-\Delta) & 0
  \end{pmatrix}  
  \begin{pmatrix}
    \phi_c \\ \phi_q
  \end{pmatrix}
\end{equation}
with a cutoff function $R_{k,\bar{K}}$ which will be specified below in
Sec.~\ref{sec:truncation}. We denote the resulting cutoff-dependent Keldysh
partition function and generating functional for connected correlation functions
by, respectively, $\mathcal{Z}_k$ and $\mathcal{W}_k$. The effective running
action $\Gamma_k$ is then defined as the modified Legendre transform
\begin{multline}
  \Gamma_k[\bar{\Phi}_c,\bar{\Phi}_q] = \mathcal{W}_k[J_c,J_q] \\ + \int_X
  \left( J_c^{\dagger} \bar{\Phi}_c + J_q^{\dagger} \bar{\Phi}_q \right) -
  \Delta \s_k[\bar{\Phi}_c,\bar{\Phi}_q].
\end{multline}
Here the subtraction of $\Delta \s_k$ on the RHS guarantees that the only
difference between the functional integral representations for $\Gamma$ and
$\Gamma_k$ is the inclusion of the cutoff term in the latter,
\begin{multline}
  \label{eq:29}
  e^{i \Gamma_k[\bar{\Phi}_c,\bar{\Phi}_q]} = \int \mathcal{D}[\delta
  \bar{\Phi}_c,\delta \bar{\Phi}_q] \, e^{i \s[\bar{\Phi}_c + \delta
    \bar{\Phi}_c,\bar{\Phi}_q + \delta \bar{\Phi}_q] + i \Delta
    \s_k[\delta \bar{\Phi}_c,\delta \bar{\Phi}_q]}.
\end{multline}
Physically, $\Gamma_k$ can be viewed as the effective action for averages of
fields over a coarse-graining volume with size $\sim k^{- d}$.

We choose the form of the cutoff term $\Delta \s_k$ such that it modifies the
inverse retarded and advanced propagators: Comparing Eqs.~\eqref{eq:16}
and~\eqref{eq:28}, we see that associated with the action $\s + \Delta \s_k$ are
the regularized retarded and advanced inverse propagators $P^R(Q) +
R_{k,\bar{K}}^{*}(q^2)$ and $P^A(Q) + R_{k,\bar{K}}(q^2)$ respectively, whereas
the Keldysh part $P^K$ of the inverse propagator remains unchanged. In other
words, by introducing the cutoff $\Delta \s_k$ we manipulate the spectrum of
single-particle excitations, which is encoded in the zeros of the inverse
propagators $P^{R/A}(Q)$ or, equivalently, in the poles of the propagators
Eq.~\eqref{eq:18}. At the transition, these poles are determined by
Eq.~\eqref{eq:17} with $\kappa_1 = 0$, i.e., we have a pole at $\omega = q = 0$,
and as we have pointed out in the paragraph following Eq.~\eqref{eq:26}, this
leads to infrared divergences that drive critical behavior. For the regularized
propagators, on the other hand, we have $G^R(\omega = 0,q^2 = 0) =
1/R_{k,\bar{K}}^{*}(0)$ and $G^A(\omega = 0,q^2 = 0) = 1/R_{k,\bar{K}}(0)$ which
are finite for
\begin{equation}
  \label{eq:30}
  R_{k,\bar{K}}(q^2) \sim k^2, \quad q \to 0.
\end{equation}
To regulate infrared divergences, it is sufficient to introduce the cutoff
function in the retarded and advanced inverse propagators, as becomes clear from
the discussion following Eq.~\eqref{eq:18}.

We have seen that the effective action $\Gamma_k$ defined by Eq.~\eqref{eq:29}
has an infrared-finite loop expansion. Its main usefulness, however, lies in the
fact that it interpolates between the action $\s$ for $k \to \Lambda$ where
$\Lambda$ is an ultraviolet cutoff scale, and the full effective action $\Gamma$
for $k \to 0$. This is ensured by the requirements on the cutoff
function\cite{berges09:_nonth}
\begin{equation}
  \label{eq:31}
  \begin{aligned}
    R_{k,\bar{K}}(q^2) & \sim \Lambda^2, & k & \to \Lambda, \\
    R_{k,\bar{K}}(q^2) & \to 0, & k & \to 0,
  \end{aligned}
\end{equation}
where under the condition that $\Lambda$ exceeds all energy scales in the action
by far, for $k \to \Lambda$ we may evaluate the functional integral
Eq.~\eqref{eq:29} in a stationary phase approximation. Then, to leading order we
find $\Gamma_{\Lambda} \sim \s$. The evolution of $\Gamma_k$ from this starting
point in the ultraviolet to the full effective action in the infrared for $k \to
0$ is described by the exact Wetterich flow
equation\cite{wetterich93:_exact,berges02:_nonper}
\begin{equation}
  \label{eq:32}
  \partial_k \Gamma_k = \frac{i}{2} \Tr \left[ \left( \Gamma^{(\bar{2})}_k +
      \bar{R}_k \right)^{-1} \partial_k \bar{R}_k \right],
\end{equation}
where $\Gamma_k^{(\bar{2})}$ and $\bar{R}_k$ denote, respectively the second
variations of the effective action and the cutoff $\Delta \s_k$ and will be
specified in Sec.~\ref{sec:truncation}; $\Tr$ denotes summation over internal
field degrees of freedom as well as integration over frequencies and
momenta. The flow equation provides us with an alternative but fully equivalent
formulation of the functional integral Eq.~\eqref{eq:29} as a functional
differential equation. Like the functional integral, the flow equation can not
be solved exactly. It is, however, amenable to various systematic approximation
strategies. Here we perform an expansion of the effective action $\Gamma_k$ in
canonical scaling dimensions as outlined above in
Sec.~\ref{sec:canon-power-count}, keeping only those couplings which are -- in
the sense of the RG -- relevant or marginal at the phase transition.

\subsection{Truncation}
\label{sec:truncation}

In three dimensional classical $O(N)$-symmetric models, already the inclusion of
non-irrelevant couplings gives a satisfactory description of critical
phenomena.\cite{berges02:_nonper} As we will show below, static critical
properties of our non-equilibrium phase transition are described by such a model
with $N = 2$. Therefore, in the following, we will as well restrict ourselves to
the inclusion of relevant and marginal couplings in the ansatz for the effective
action, i.e., we choose a truncation of the form
\begin{equation}
  \label{eq:34}
  \Gamma_k = \int_X \left[ \left( \bar{\phi}_c^{*},\bar{\phi}_q^{*} \right)
    \begin{pmatrix}
      0 & \bar{D}^A \\
      \bar{D}^R & i \bar{\gamma}
    \end{pmatrix}
    \begin{pmatrix}
      \bar{\phi}_c \\ \bar{\phi}_q
    \end{pmatrix}
    - \left( \frac{\partial \bar{U}}{\partial \bar{\phi}_c} \bar{\phi}_q +
      \frac{\partial \bar{U}^{*}}{\partial \bar{\phi}_c^{*}} \bar{\phi}_q^{*}
    \right) \right].
\end{equation}
(Here all couplings depend on the infrared cutoff scale $k$. However, for the
sake of keeping the notation simple, we will not state this dependence
explicitly.) All terms involving derivatives are contained in $\bar{D}^R = i
Z^{*} \partial_t + \bar{K}^{*} \Delta$ and $\bar{D}^A = \bar{D}^{R \dagger}$. In
contrast to the action Eq.~\eqref{eq:16}, however, we allow for complex coefficients
$Z = Z_R + i Z_I$ and $\bar{K} = \bar{A} + i \bar{D}$: Due to the presence of
complex couplings $\lambda + i \kappa$ in the classical action, imaginary parts
of $Z$ and $\bar{K}$ will be generated in the RG flow as indicated at the end of
Sec.~\ref{sec:mean-field-theory}, even though they are zero initially at $k \to
\Lambda$.

A complex prefactor $Z$ of the time derivative -- often referred to as
wave-function renormalization -- obscures the physical interpretation of the
other complex couplings: The field equation $\delta \Gamma_k/\delta
\bar{\phi}_q^{*} = 0$ contains $i Z^{*} \partial_t \bar{\phi}_c = - \bar{K}^{*}
\Delta \bar{\phi}_c + \dotsb$. The physical meaning of the gradient coefficient
$\bar{K}$ becomes clear only after division by $Z^{*}$, i.e., in the form
$i \partial_t \bar{\phi}_c = - \left( A - i D \right) \Delta \bar{\phi}_c +
\dotsb$ where we introduced the decomposition $K = \bar{K}/Z = A + i D$ into
real and imaginary parts. In this form, the interpretation of $A$ and $D$ as
encoding coherent propagation and diffusive behavior of particles is
apparent. Similar considerations hold for the other couplings in
Eq.~\eqref{eq:34}, and we will elaborate on this point in
Sec.~\ref{sec:geom-interpr-equil}.

In our truncation containing only non-irrelevant contributions, the only
momentum-independent couplings we keep are the Keldysh and spectral masses,
$\bar{\gamma}$ and $\bar{u}_1 = - \bar{\mu} + i \bar{\kappa}_1$, as well as the
classical quartic and sextic couplings (i.e., those vertices containing only one
quantum field but three and five classical field variables respectively). These
are included in the part in Eq.~\eqref{eq:34} that involves the potential
$\bar{U}$, which is a function of the $U(1)$ invariant $\bar{\rho}_c =
\abs{\bar{\phi}_c}^2$ and given by
\begin{equation}
  \label{eq:35}
  \bar{U}(\bar{\rho}_c) = \bar{u}_1 \left( \bar{\rho}_c - \bar{\rho}_0
  \right) + \frac{1}{2} \bar{u}_2 \left( \bar{\rho}_c - \bar{\rho}_0
  \right)^2 + \frac{1}{6} \bar{u}_3 \left( \bar{\rho}_c - \bar{\rho}_0
  \right)^3,
\end{equation}
where both $\bar{u}_2 = \bar{\lambda} + i \bar{\kappa}$ and $\bar{u}_3 =
\bar{\lambda}_3 + i \bar{\kappa}_3$ are complex. In the symmetric phase, we keep
$\bar{u}_1 \neq 0$ as a running coupling and set $\bar{\rho}_0 = 0$, whereas in
the ordered phase we set the masses to zero, $\bar{u}_1 = 0$, and regard the
condensate amplitude as a running coupling, $\bar{\rho}_0 \neq 0$. Then, the
parameterization Eq.~\eqref{eq:35} corresponds to an expansion of the potential
around its minimum in both the symmetric and ordered phases. It ensures that the
field equations $\delta \Gamma_k/\delta \bar{\phi}_c^{*} = 0, \delta
\Gamma_k/\delta \bar{\phi}_q^{*} = 0$ are solved by $\bar{\rho}_c = 0$ and
$\bar{\rho}_c = \bar{\rho}_0$ in the symmetric and ordered phases respectively
(in both cases we require $\bar{\phi}_q = \bar{\phi}_q^{*} = 0$) for all values
of $k$.

In what follows we will find it advantageous to introduce renormalized fields
$\phi_c = \bar{\phi}_c, \phi_q = Z \bar{\phi}_q$ (the various symbols for
bare/renormalized fields etc.\ are summarized in Tab.~\ref{tab:notation}). With
this choice the complex wave-function renormalization $Z$ that multiplies the
time derivative in Eq.~\eqref{eq:34} is absorbed in the field variables and we
can write the effective action in the form ($\sigma_z$ denotes the Pauli matrix)
\begin{equation}
  \label{eq:36}
  \Gamma_k = \int_X \Phi_q^{\dagger} \left[ i \sigma_z \left( \partial_t
      \Phi_c + \frac{\delta \mathcal{U}_D}{\delta \Phi_c^{*}} \right) -
    \frac{\delta \mathcal{U}_H}{\delta \Phi_c^{*}} + i
    \frac{\gamma}{2} \Phi_q \right].
\end{equation}
The renormalized Keldysh mass is $\gamma = \bar{\gamma}/\abs{Z}^2$. For the
variational derivatives with respect to the classical fields we are using the
notation $\delta/\delta \Phi_c^{*} = \left( \delta/\delta
  \phi_c^{*},\delta/\delta \phi_c \right)^T$, and the renormalized potential
functionals that encode unitary and dissipative terms respectively, read
\begin{equation}
  \begin{split}
    \frac{\delta \mathcal{U}_H}{\delta \Phi_c^{*}} & = \left( - A \Delta +
      U_H' \right) \Phi_c, \\ \frac{\delta \mathcal{U}_D}{\delta
      \Phi_c^{*}} & = \left( - D \Delta + U_D' \right) \Phi_c,
  \end{split}
\end{equation}
where $A$ and $D$ are the real and imaginary parts of the renormalized gradient
coefficient $K = \bar{K}/Z = A + i D$. Primes denote derivatives with respect to
$\rho_c = \abs{\phi_c}^2$ of the real and imaginary parts of the renormalized
potential $U = \bar{U}/Z = U_H + i U_D$, which is given by
\begin{equation}
  U(\rho_c) = u_1 \left( \rho_c - \rho_0 \right) + \frac{1}{2} u_2 \left(
    \rho_c - \rho_0 \right)^2 + \frac{1}{6} u_3 \left( \rho_c - \rho_0 \right)^3
\end{equation}
with renormalized couplings $u_1 = \bar{u}_1/Z = - \mu + i \kappa_1, u_2 =
\bar{u}_2/Z = \lambda + i \kappa$, and $u_3 = \bar{u}/Z = \lambda_3 + i
\kappa_3$. The inclusion of the classical three-body coupling $u_3$ adds the
vertex
\begin{equation}
  \parbox{1.12in}{\includegraphics{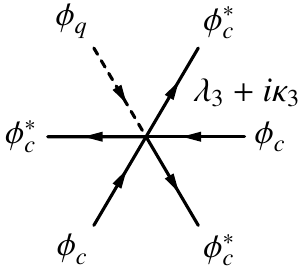}}
\end{equation}
to the building blocks Eqs.~\eqref{eq:18} and~\eqref{eq:26}.
\begin{table}
  \centering
  \begin{tabular}{|c|c|c|c|}
    \hline $\hat{\psi}$ & field operator & \ref{sec:master-equation} \\
    $\psi_{\sigma}, \sigma = \pm$ & fields on Keldysh contour &
    \ref{sec:keldysh-path-integr} \\ $\Psi_{\sigma} = \left( \psi_{\sigma},\psi_{\sigma}^{*} \right)^T$ & spinor of $\pm$-fields &
    \ref{sec:keldysh-path-integr} \\ $\phi_{\nu}, \nu = c, q$ & classical and
    quantum fields & \ref{sec:keldysh-path-integr} \\
    $\Phi_{\nu} = \left( \phi_{\nu}, \phi_{\nu}^{*} \right)^T$ & spinor of $c$ and
    $q$ fields & \ref{sec:effective-action} \\ $\bar{\Phi}_{\nu}$ & 
    field expectation values/bare fields & \ref{sec:effective-action} \\ $\Gamma_k^{(\bar{2})}, \bar{R}_k$ & derivatives WRT
    bare fields & \ref{sec:frg-approach-keldysh} \\
    $\phi_c = \bar{\phi}_c, \phi_q = Z \bar{\phi}_q$ & renormalized fields &
    \ref{sec:truncation} \\ $\chi_{\nu,n}, n = 1,2$ & real fields &
    \ref{sec:truncation} \\ $\Gamma_k^{(2)}, R_k$ &
    derivatives WRT renormalized fields & \ref{sec:truncation} \\ $Z, \bar{K}, \bar{u}_1, \bar{u}_2, \dotsc$ &
    bare couplings & \ref{sec:truncation} \\ $K, u_1, u_2, \dotsc$ &
    renormalized couplings & \ref{sec:truncation} \\
    $\tilde{\Phi}_{\nu}, \hat{\Phi}_{\nu}$ & transformed bare fields &
    \ref{sec:relat-equil-dynam} \\ $\tilde{u}_1,
    \tilde{u}_3, \dotsc$ & dimensionless couplings & \ref{sec:resc-flow-equat}
    \\ \hline
  \end{tabular}
  \caption{Summary of notation. The columns are symbols, their meaning, and the
    section in which they are introduced.}
  \label{tab:notation}
\end{table}

\begin{figure}
  \centering
  \includegraphics{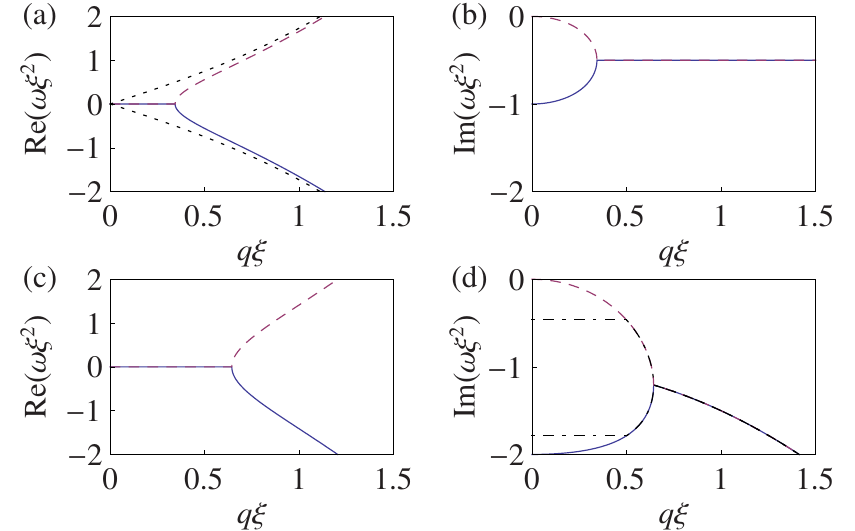}  
  \caption{(Color online) Dispersion relation of single-particle excitations in
    the ordered phase. Frequencies and momenta are measured in units of the
    healing length $\xi = 1/\sqrt{\lambda \rho_0}$. (a) and (b): The Goldstone
    and massive modes Eq.~\eqref{eq:25}, obtained in mean-field approximation,
    are shown as, respectively, dashed and solid lines for $\kappa =
    \lambda/2$. For small momenta both modes are purely diffusive and
    non-propagating. The dotted lines in (a) correspond to the usual Bogoliubov
    dispersion relations for $\kappa = 0$. (c) and (d): Dispersion relations
    Eq.~\eqref{eq:42} with gradient coefficients $A, D$ that are generated upon
    renormalization. (d) For finite $D$, the damping rate grows $\propto q^2$
    for large $q$. The regularized dispersion relations, where $a q^2$ is
    replaced by $p_a(q^2)$ for $a = A, D$ (cf.~Eq.~\eqref{eq:45}), are shown as
    a dash-dotted lines. Here we chose parameters $A = 1, D = 1/2, \kappa =
    \lambda, k = 1/(2 \xi)$.}
  \label{fig:dispersion_relation}
\end{figure}

As we have already indicated, the first variational derivative of the effective
action yields field equations that determine the stationary state values of the
classical and quantum fields. In the ordered phase, these are constant in space
and time and read $\phi_c \rvert_{\mathrm{ss}} = \phi_c^{*} \rvert_{\mathrm{ss}}
= \sqrt{\rho_0}$ (our choice of a real condensate amplitude does not cause a
loss of generality) and $\phi_q \rvert_{\mathrm{ss}} = \phi_q^{*}
\rvert_{\mathrm{ss}} = 0$. Then, the scale-dependent inverse connected
propagator is given by the second variational derivative of the effective
action,\cite{Amit/Martin-Mayor,Kleinert/Schulte-Frohlinde,Zinn-Justin} evaluated
in stationary state. We will carry out this variational derivative in a basis of
real fields, which we introduce by decomposing the classical and quantum fields
into real and imaginary parts according to $\phi_{\nu} = \frac{1}{\sqrt{2}}
\left( \chi_{\nu,1} + i \chi_{\nu,2} \right)$ for $\nu = c, q$. The inverse
propagator at the scale $k$ is then given by
\begin{equation}
  \label{eq:37}
  P_{ij}(Q) \delta(Q - Q') = \frac{\delta^2 \Gamma_k}{\delta \chi_i(-Q) \delta
    \chi_j(Q')} \biggr\rvert_{\mathrm{ss}}.
\end{equation}
Here the indices $i, j$ enumerate the four components of the field vector
\begin{equation}
  \label{eq:38}
  \chi(Q) = \left( \chi_{c,1}(Q), \chi_{c,2}(Q), \chi_{q,1}(Q),
    \chi_{q,2}(Q) \right)^T.
\end{equation}
Analogous to the inverse propagator in the action Eq.~\eqref{eq:16}, the inverse
propagator at the scale $k$ is structured into retarded, advanced, and Keldysh
blocks,
\begin{equation}
  \label{eq:39}
  P(Q) =
  \begin{pmatrix}
    0 & P^A(Q) \\
    P^R(Q) & P^K
  \end{pmatrix}.
\end{equation}
However, here these blocks are themselves $2 \times 2$ matrices. (This
additional Nambu structure emerges in the ordered phase.) We have explicitly
\begin{equation}
  \label{eq:40}
  \begin{split}
    P^R(Q) & =
    \begin{pmatrix}
      - A q^2 - 2 \lambda  \rho_0 & i \omega - D q^2  \\
      - i \omega + D q^2 + 2 \kappa \rho_0 & - A q^2
    \end{pmatrix}
    = P^A(Q)^{\dagger}, \\ P^K & = i \gamma \id.
  \end{split}
\end{equation}
These expressions can be used to deduce the dispersion relation for
single-particle excitations. It is determined by solving
\begin{equation}
  \label{eq:41}
  \det P(Q) = \det \left( P^R(Q) \right) \det \left( P^A(Q) \right) = 0
\end{equation}
for $\omega$. Due to the second relation Eq.~\eqref{eq:40}, two of the four
solutions to Eq.~\eqref{eq:41} are complex conjugate. The zeros of the
determinant of the retarded inverse propagator encode the two branches
\begin{equation}
  \label{eq:42}
  \omega_{1,2}^R = - i D q^2 - i \kappa  \rho_0 \pm \sqrt{A q^2 \left( A q^2
      + 2 \lambda  \rho_0 \right) - \left( \kappa \rho_0 \right)^2},
\end{equation}
which differ from the mean-field expression Eq.~\eqref{eq:25} by the
contribution $- i D q^2$ due to the explicit inclusion of a dissipative kinetic
term in our truncation, and by the appearance of the scale dependent gradient
coefficient $A$. The dissipative Goldstone mode is now characterized by the
low-momentum behavior $\omega_1^R \sim - i \left( D + A \frac{\lambda}{\kappa}
\right) q^2$, whereas for the ``massive'' (the mass is purely imaginary) mode we
reproduce the form of the mean-field expression $\omega_2^R \sim - i 2 \kappa
\rho_0$ -- however, in a scale-dependent version with all couplings running in
the course of the RG. In this way, structural properties such as Goldstone's
theorem are preserved during the flow. The dispersion relation Eq.~\eqref{eq:42}
is depicted in Fig.~\ref{fig:dispersion_relation}.

We proceed by specifying the cutoff function $R_{k,\bar{K}}$ which appears in
Eq.~\eqref{eq:28}. We will work with an optimized cutoff\cite{litim00:_optim}
\begin{equation}
  \label{eq:43}
  R_{k,\bar{K}}(q^2) = - \bar{K} \left( k^2 - q^2 \right) \theta(k^2 - q^2),
\end{equation}
which obviously meets the requirements Eqs.~\eqref{eq:30} and~\eqref{eq:31}. The
regularized propagator, which appears in the loop diagrams that generate the RG
flow, reads
\begin{equation}
  \label{eq:44}
  G_k(Q) = \left( P(Q) + R_k(Q) \right)^{-1},
\end{equation}
where the $4 \times 4$ matrix $R_k(Q)$ is defined in analogy to the inverse
propagator Eq.~\eqref{eq:37} as the second variational derivative of the
cutoff Eq.~\eqref{eq:28} with respect to the real fields Eq.~\eqref{eq:38},
\begin{equation}
  R_{k,ij}(q^2) \delta(Q - Q') = \frac{\delta^2 \Delta \s_k}{\delta \chi_i(- Q)
    \delta \chi_j(Q')}.
\end{equation}
Due to the cutoff $R_k(Q)$ in the denominator in Eq.~\eqref{eq:44}, the poles of
$G_k(Q)$ are given by Eq.~\eqref{eq:42}, however, with $A q^2$ and $D q^2$
replaced by $p_A(q^2)$ and $p_D(q^2)$ respectively, where the function
$p_a(q^2)$ for $a = A, D$ reads
\begin{equation}
  \label{eq:45}
  p_a(q^2) = a q^2 - R_{k,a}(q^2) =
  \begin{cases}
    a k^2 & \text{for } q^2 < k^2, \\
    a q^2 & \text{for } q^2 \geq k^2.
  \end{cases}
\end{equation}
The thus modified dispersion relations are finite for $q \to 0$, i.e., infrared
divergences of loop diagrams are regularized. In panel (d) in
Fig.~\ref{fig:dispersion_relation} the regularized dispersion relations are
shown as dashed-dotted lines.

In Sec.~\ref{sec:funct-renorm-group} we introduced most of the ingredients for a
FRG investigation of the steady state driven-dissipative Bose condensation
transition. Before we present the explicit flow equations in
Sec.~\ref{sec:non-eq-frg-flow}, we will now provide a detailed discussion of the
relation between our non-equilibrium model and the classical equilibrium
dynamical MA of HH.\cite{hohenberg77:_theor}

\section{Relation to equilibrium dynamical models}
\label{sec:relat-equil-dynam}

Here we extend the discussion of Sec.~\ref{sec:equilibrium-symmetry} and work
out the precise relation of the DDM to MA with $N = 2$ components. We
reemphasize that these considerations rely on the power counting introduced in
Sec.~\ref{sec:canon-power-count}, which implies that we may omit quantum
vertices from an effective long-wavelength description close to criticality; The
resulting action Eq.~\eqref{eq:36} is equivalent to a Langevin equation of the
form of
Eq.~\eqref{eq:1}.\cite{Tauber20127,tauberbook,kamenev09:_keldy,kamenevbook}

Originally, MA was formulated in terms of such a Langevin equation for a
non-conserved, coarse-grained order parameter. It provides for a
phenomenological description of the relaxational dynamics of the order parameter
subject to stochastic fluctuations, which are introduced necessarily as a
consequence of the coarse-graining over a volume of extent $\kcg^{-d}$: The
effects of fluctuations with momenta $q$ greater than the coarse-graining scale
$\kcg$ are included by introducing random noise sources in the evolution
equation.

For our model, coarse-graining amounts to integrating out fluctuations with
momenta $q$ greater than $\kcg$ in the functional integral
Eq.~\eqref{eq:29},\cite{berges02:_nonper} which results in an effective action
$\Gamma_{\cg}$ that can be regarded as the starting point of a phenomenological
description in the spirit of HH, i.e., we may interpret it as the action
$\s_{\cg} = \Gamma_{\cg}$ for slow modes with momenta $q < \kcg$.

The equation of motion of MA is constructed such that its stationary state is
thermodynamic equilibrium, which manifests itself in a
FDT\cite{hohenberg77:_theor} relating the order parameter retarded response and
correlation functions. The FDT can be derived as a consequence of a specific
equilibrium symmetry of the dynamics which is related to time reversal and
expresses detailed
balance.\cite{enz79:_field,aron10:_symmet_langev,canet11:_gener} This symmetry,
however, does not restrict the dynamics to be purely relaxational as is the case
in MA. In fact, one can conceive an extension of MA by reversible mode couplings
(MAR) which differs from the DDM \emph{only} in the obedience of the
symmetry. (Note that the DDM generically features both coherent and dissipative
contributions). As universality classes are fully characterized by the spatial
dimensionality and symmetries of a system, however, this opens up the
possibility of novel critical behavior in the DDM.

In the remainder of this section we illuminate the consequences of the
equilibrium symmetry through a detailed comparison between MAR in which it is
present at the outset and the DDM model, where it is only emergent at long
scales.  We give a simple geometric interpretation of the restriction that the
symmetry imposes on the couplings that parameterize the effective action and
specify the submanifolds in the coupling space for the DDM that correspond to MA
and MAR.

While these considerations demonstrate formally the non-equilibrium character of
the DDM, the equilibrium MAR constructed in the above way may seem a bit
academic. In fact, as we will see in Sec.~\ref{sec:geom-interpr-equil} it
amounts to an unrealistic fine-tuning of the ratios of all coherent vs.\
dissipative couplings. The physically relevant model which the DDM should be
compared to is model E, which describes the equilibrium Bose condensation
transition. An important difference between the DDM and model E is the presence
of an exact particle number conservation in the latter case which can be seen to
rule out a finite $\kappa_1$ mass term.\footnote{It may -- and does -- occur as
  a regularization, meaning however that it has to be sent to zero in such a way
  that it does not affect any physical result.}  Therefore, according to the
arguments given in Sec.~\ref{sec:key-results-physical}, the standard equilibrium
Bose condensation transition exhibits only three independent exponents (as
opposed to four in the DDM) and, in particular, no counterpart to
$\eta_r$. Moreover, as a consequence of the exact particle number conservation
an additional slow mode occurs at criticality and modifies the \emph{dynamical}
exponent.

\subsection{Model A with $N = 2$ and reversible mode couplings (MAR)}
\label{sec:mar}

We specify the equilibrium symmetry in terms of fields $\tilde{\Phi}_{\nu}$
which are related to the bare fields $\bar{\Phi}_{\nu}$ of Eq.~\eqref{eq:34}
via
\begin{equation}
  \label{eq:46}
  \tilde{\Phi}_c = \bar{\Phi}_c, \quad \tilde{\Phi}_q = \frac{\zrcg - \bar{r}
    \zicg}{1 + \bar{r}^2} \left( \bar{r} \id + i \sigma_z \right) \bar{\Phi}_q,
\end{equation}
where $\zrcg$ and $\zicg$ denote the real and imaginary parts of the
wave-function renormalization at the coarse-graining scale $\kcg$ and $\bar{r}$
is a real parameter, the physical meaning of which will become clear in the
following. The symmetry transformation is denoted by $\mathcal{T}$ and
reads\cite{enz79:_field,aron10:_symmet_langev,canet11:_gener}
\begin{equation}
  \label{eq:47}
  \begin{split}
    \mathcal{T} \tilde{\Phi}_c(t,\mathbf{x}) & = \sigma_x \tilde{\Phi}_c(-t,\mathbf{x}), \\
    \mathcal{T} \tilde{\Phi}_q(t,\mathbf{x}) & = \sigma_x \left(
      \tilde{\Phi}_q(-t,\mathbf{x}) + \frac{i}{2 T} \partial_t \tilde{\Phi}_c(-t,\mathbf{x})
    \right),
  \end{split}
\end{equation}
cf.\ the implementation in the Langevin formulation Eq.~\eqref{eq:212}. It
includes complex conjugation (in the form of multiplication with the Pauli
matrix $\sigma_x$) and time reversal; $T$ is the temperature. As outlined above,
we now construct the action for MAR as follows: We identify the effective action
Eq.~\eqref{eq:34} at the coarse-graining scale $\kcg$ with the action for
low-momentum modes, $\s_{\cg} = \Gamma_{\cg}$, and enforce thermodynamic
equilibrium by requiring invariance of $\s_{\cg}$ under the transformation
$\mathcal{T}$, which results in
\begin{multline}
  \label{eq:48}
  \scgmar = \int_X \bar{\Phi}_q^{\dagger} \left[ \left( \zrcg \sigma_z - i \zicg
      \id \right) i \partial_t \bar{\Phi}_c \vphantom{\frac{\delta
        \bar{\mathcal{U}}_{D,\cg}}{\delta \bar{\Phi}_c^{*}}} \right. \\ \left. +
    \left( i \sigma_z - \bar{r} \id \right) \frac{\delta
      \bar{\mathcal{U}}_{D,\cg}}{\delta \bar{\Phi}_c^{*}} + i
    \frac{\bar{\gamma}_{\cg}}{2} \bar{\Phi}_q \right].  
\end{multline}
(See App.~\ref{sec:symm-constr-acti} for details of the derivation.) The action
$\scgmar$ contains coherent dynamics in the form of $\bar{\mathcal{U}}_{H,\cg} =
\bar{r} \bar{\mathcal{U}}_{D,\cg}$, i.e., the parameter $\bar{r}$ plays the role
of the common fixed ratio between coherent and dissipative couplings. This
relation ensures compatibility of coherent dynamics with the equilibrium
symmetry. We note that here, crucially, both the irreversible and the reversible
dynamics have the same physical origin, being generated by the same functional
$\bar{\mathcal{U}}_{D,\cg}$. This is motivated in the frame of a
phenomenological, effective model for relaxation dynamics in the absence of
explicit drive.

However, not only the values of the couplings encoding coherent dynamics are
restricted by the symmetry, but also the Keldysh mass $\bar{\gamma}_{\cg}$ is
determined by the temperature that appears in the symmetry transformation as
\begin{equation}
  \label{eq:49}
  \bar{\gamma}_{\cg} = \frac{4}{1 +
    \bar{r}^2} \left( \zrcg - \bar{r} \zicg \right)^2 T.
\end{equation}
Finally we note that Eq.~\eqref{eq:48} includes MA with effectively purely
dissipative dynamics as a special case: Indeed we can derive the action for MA
in the same way as we derived the action for MAR from the truncation for the
DDM, i.e., by enforcing an additional symmetry. Requiring invariance of
$\scgmar$ under complex conjugation of the fields,
\begin{equation}
  \label{eq:50}
  \mathcal{C} \tilde{\Phi}_{\nu} = \sigma_x \tilde{\Phi}_{\nu},
\end{equation}
we find the additional constraint $\bar{r} = - \zicg/\zrcg$ (see
App.~\ref{sec:symm-constr-acti}), reducing the number of independent parameters
further. Then, after rescaling the quantum fields with $Z_{\cg}$ it becomes
apparent that this model describes purely dissipative dynamics as we will show
in Sec.~\ref{sec:geom-interpr-equil}.

\subsection{Truncation for MAR}
\label{sec:truncation-mar}

We proceed by specifying the truncation for a FRG analysis of MAR. Here it is
crucial to note that the transformation $\mathcal{T}$ Eq.~\eqref{eq:47} not only
leaves the action Eq.~\eqref{eq:48} invariant, but is actually a symmetry of the
full theory,\cite{aron10:_symmet_langev} i.e., of the effective action. Then, if
the cutoff $\Delta \s_k$ in Eq.~\eqref{eq:29} is $\mathcal{T}$-invariant as well
(this is indeed the case for the choice Eq.~\eqref{eq:28}), also the
scale-dependent effective action $\Gamma^{\mar}_k$ must obey the symmetry. This
requirement implies restrictions on the RG flow: Invariance of the effective
action on all scales is guaranteed by the ansatz
\begin{multline}
  \label{eq:51}
  \Gamma^{\mar}_k = \int_X \bar{\Phi}_q^{\dagger} \left[ \left( Z_R \sigma_z - i
      Z_I \id \right) i \partial_t \bar{\Phi}_c
    \vphantom{\frac{\bar{\mathcal{U}}_H}{\delta \bar{\Phi}_c^{*}}} \right. \\
  \left. + \left( i \sigma_z - \bar{r} \id \right) \frac{\delta
      \bar{\mathcal{U}}_D}{\delta \bar{\Phi}_c^{*}} + i \frac{\bar{\gamma}}{2}
    \bar{\Phi}_q \right],
\end{multline}
which follows by enforcing the symmetry on the truncation Eq.~\eqref{eq:34} (see
App.~\ref{sec:symm-constr-acti} for details). We note in particular that
compatibility of coherent and dissipative dynamics is conserved in the RG
flow. In contrast to the DDM, here the Keldysh mass is not an independent
running coupling, as it is linked to the wave-function renormalization $Z = Z_R
+ i Z_I$ by the Ward identity of the symmetry Eq.~\eqref{eq:47},
\begin{equation}
  \label{eq:52}
  \bar{\gamma} = \frac{Z_R - \bar{r} Z_I}{\zrcg - \bar{r} \zicg} \bar{\gamma}_{\cg}.
\end{equation}
In comparison to the DDM, therefore, MAR is described by a reduced number of
couplings: Our truncation Eq.~\eqref{eq:34} for the DDM is parameterized by a vector
of couplings
\begin{equation}
  \label{eq:53}
  \bar{\mathbf{g}} = \left( Z, \bar{K}, \bar{\rho}_0, \bar{u}_1, \bar{u}_2,
    \bar{u}_3, \bar{\gamma} \right)^T,
\end{equation}
where $Z, \bar{K}, \bar{u}, \bar{u}_3$ are complex whereas $\bar{\rho}_0,
\bar{\gamma}$ are positive real numbers. In MAR, the real parts of the complex
couplings in the functional $\bar{\mathcal{U}}$ are determined by imaginary ones
and the ratio $\bar{r}$ which appears as a fixed parameter in the action at the
coarse-graining scale $\kcg$. Additionally the Keldysh mass is related to the
wave-function renormalization via Eq.~\eqref{eq:52}, so that a reduced set of
running couplings,
\begin{equation}
  \label{eq:54}
  \bar{\mathbf{g}}_{\mar} = \left(
    Z, \bar{D}, \bar{\rho}_0, \bar{\kappa}_1, \bar{\kappa},
    \bar{\kappa}_3 \right)^T,
\end{equation}
is sufficient to fully specify the truncation Eq.~\eqref{eq:51}. In the purely
dissipative MA, finally, the symmetry Eq.~\eqref{eq:50} determines the ratio of
imaginary to reals parts of the wave-function renormalization $Z$ as $\bar{r} =
- Z_I/Z_R$ (see App.~\ref{sec:symm-constr-acti}), so that $Z$ can be
parametrized in terms of a single real running coupling. The truncation for MA,
therefore, is described by the couplings:
\begin{equation}
  \label{eq:55}
  \bar{\mathbf{g}}_{\ma} = \left( Z_R, \bar{D},
    \bar{\rho}_0, \bar{\kappa}_1, \bar{\kappa}, \bar{\kappa}_3 \right)^T.
\end{equation}

\subsection{Fluctuation-dissipation theorem}
\label{sec:fluct-diss-theor}

In the following we will show that the symmetry Eq.~\eqref{eq:47} implies a
classical FDT for MAR.\cite{enz79:_field,aron10:_symmet_langev,canet11:_gener}
If we regard the full propagators of the theory as the $k \to 0$ limits of the
RG flow of scale-dependent propagators, we may say that the FDT holds for MAR
(and, \textit{a fortiori}, for MA) for all $0 < k < \kcg$. In addition we will
see that this is not the case for the driven-dissipative system we
consider. There the equilibrium symmetry is not present at mesoscopic scales but
rather emergent for the system at criticality in the infrared for $k \to 0$. As
a result, thermalization sets in only at low frequencies and long wavelengths.

As indicated at the beginning of the preceding section, the transformation
$\mathcal{T}$ Eq.~\eqref{eq:47} is a symmetry of the full theory. In particular,
for two-point correlation functions we have
\begin{equation}
  \label{eq:56}
  \langle \tilde{\phi}_{\nu}(t,\mathbf{x}) \tilde{\phi}_{\nu'}^{*}(t',\mathbf{x}') \rangle = \langle
  \mathcal{T} \tilde{\phi}_{\nu}(t,\mathbf{x}) \mathcal{T}
  \tilde{\phi}_{\nu'}^{*}(t',\mathbf{x}') \rangle,
\end{equation}
and corresponding relations hold for higher correlation functions. Here
expectation values are defined as
\begin{equation}
  \langle \dotsb \rangle = \int \mathcal{D} [\Phi_c,\Phi_q] \dotsb \, e^{i \scgmar[\Phi_c,\Phi_q]}.
\end{equation}
The relation Eq.~\eqref{eq:56} implies a FDT: For the particular choice of
correlations between quantum fields $\nu = \nu' = q$ which vanish by
construction of the Keldysh functional
integral,\cite{Altland/Simons,kamenev09:_keldy,kamenevbook} we find
\begin{equation}
  \label{eq:57}
  0 = \langle \tilde{\phi}_q(t,\mathbf{x}) \tilde{\phi}_q^{*}(t',\mathbf{x}') \rangle = \langle
  \mathcal{T} \tilde{\phi}_q(t,\mathbf{x}) \mathcal{T} \tilde{\phi}_q^{*}(t',\mathbf{x}') \rangle.
\end{equation}
Inserting here explicit expressions for the $\mathcal{T}$-transformed fields and
performing a Fourier transformation, we obtain the classical FDT
\begin{equation}
  \label{eq:58}
  \tilde{G}^K\oq = \frac{2 T}{\omega} \left( \tilde{G}^R\oq - \tilde{G}^A\oq \right).
\end{equation}

Such a relation is in general not valid in the DDM. It is, however, emergent for
the critical system in the long-wavelength limit: In the basis $\hat{\phi}_c =
\bar{\phi}_c, \hat{\phi}_q = i \left( Z/\abs{Z} \right) \bar{\phi}_q$ we have
for the inverse propagators at the scale $k$ (for convenience we are working
here in the symmetric phase; the scale-dependent inverse propagators are
determined by the quadratic part of the effective action Eq.~\eqref{eq:34})
$\hat{P}^R\oq = i \abs{Z} \left( \omega - \xi^{*}(q) \right) =
\hat{P}^A\oq^{\dagger}$ where $\xi(q) = K q^2 + u_1$ (note that here the
renormalized quantities appear) and $\hat{P}^K = \bar{P}^K$. With these inverse
propagators we form the ratio
\begin{equation}
  \label{eq:59}
  \frac{\omega}{2} \frac{\hat{P}^K}{\hat{P}^R(Q) - \hat{P}^A(Q)} = \frac{\bar{\gamma}}{4
    \abs{Z}} \frac{\omega}{\omega - \Re \xi(q)},
\end{equation}
which would equal the temperature if a FDT were valid.\footnote{Due to the
  relation $\hat{G}^K\oq = - \hat{G}^R\oq \hat{P}^K \hat{G}^A\oq$, in
  Eq.~\eqref{eq:58} the propagators can be replaced by the inverse propagators.}
As we will see in Sec.~\ref{sec:wilson-fisher-fixed}, the effective action for
the critical system becomes purely dissipative for $k \to 0$. In particular we
have $\Re \xi(q) \to 0$ so that Eq.~\eqref{eq:59} indeed reduces to an FDT with
an effective temperature
\begin{equation}
  \label{eq:60}
  \teff = \frac{\bar{\gamma}}{4 \abs{Z}}.
\end{equation}
Note that for purely dissipative dynamics Eq.~\eqref{eq:52} implies that the
ratio $\bar{\gamma}/\abs{Z}$ is a constant of the RG flow. For the DDM the
emergence of an FDT with $\teff$ manifests itself in the relation Eq.~\eqref{eq:97}
between the anomalous dimensions of $\bar{\gamma}$ and $Z$ valid at the fixed
point. The flow of $\bar{\gamma}/\left( 4 \abs{Z} \right)$ is shown in
Fig.~\ref{fig:teff}.

\subsection{Geometric interpretation of the equilibrium symmetry}
\label{sec:geom-interpr-equil}

\begin{figure}
  \centering
  \includegraphics{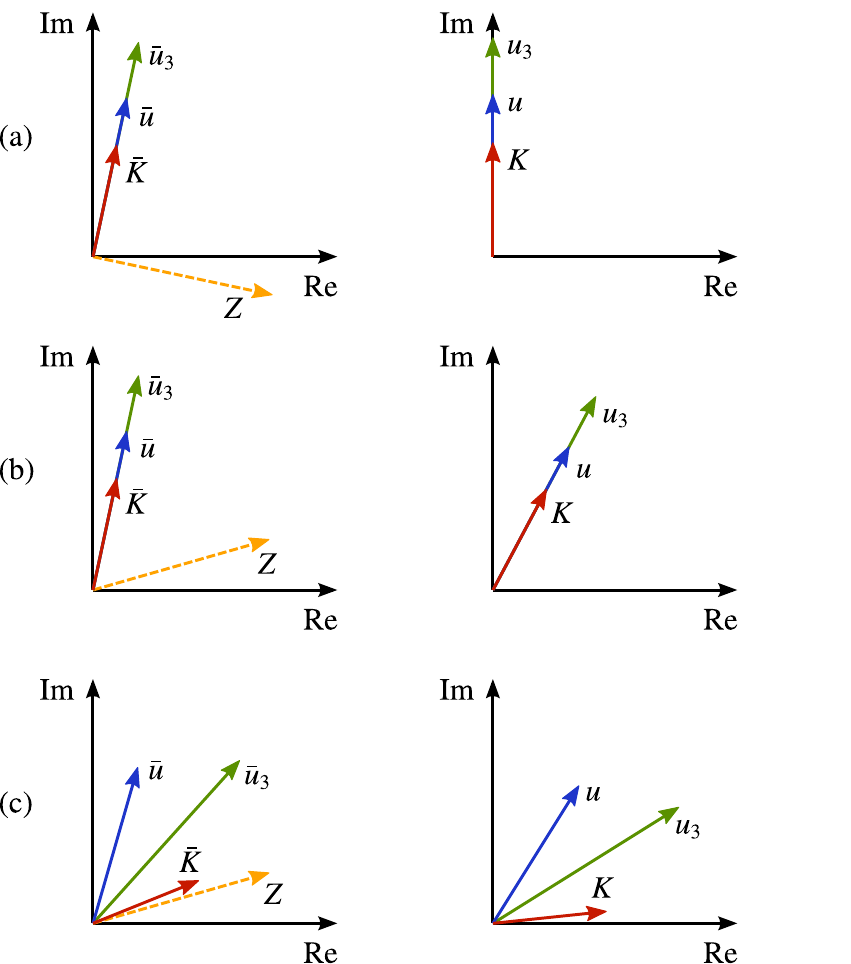}
  \caption{(Color online) Visualization of the renormalization with $Z$. Left
    column: Bare couplings. Right column: Renormalized couplings. The
    renormalization of a complex coupling $\bar{g}$ corresponds to a rescaling
    $\abs{g} = \abs{\bar{g}}/\abs{Z}$ of the modulus and a rotation of the phase
    by the argument of $Z$, $\arg g = \arg \bar{g} - \arg Z$. (a) When all bare
    couplings lie on a single ray that is perpendicular to $Z$, the renormalized
    couplings are purely imaginary as in MA. (b) Deviations from the right angle
    incorporate MA with compatible reversible mode couplings. (c) In a generic
    non-equilibrium situation there is no fixed relation between the arguments
    of the various couplings.}
  \label{fig:Z_renormalization}
\end{figure}

For our truncation of the effective action $\Gamma_k^{\mar}$, the relation
$\bar{\mathcal{U}}_H = \bar{r} \bar{\mathcal{U}}_D$ between the real and
imaginary parts of the functional $\bar{\mathcal{U}} = \bar{\mathcal{U}}_H + i
\bar{\mathcal{U}}_D$ implies that the couplings parameterizing
$\bar{\mathcal{U}}_H$ and $\bar{\mathcal{U}}_D$ share a common ratio $\bar{r}$
of real to imaginary parts
\begin{equation}
  \label{eq:61}
  \bar{r} = \frac{\bar{A}}{\bar{D}} = \frac{\bar{\lambda}_1}{\bar{\kappa}_1} =
  \frac{\bar{\lambda}}{\bar{\kappa}} = \frac{\bar{\lambda}_3}{\bar{\kappa}_3}.
\end{equation}
The same applies to the renormalized couplings, however, with a different value
$r$: With $z = - Z_I/Z_R$ we have
\begin{equation}
  \label{eq:62}
  r = \frac{A}{D} = \frac{\lambda_1}{\kappa_1} = \frac{\lambda}{\kappa} =
  \frac{\lambda_3}{\kappa_3} = \frac{\bar{r} - z}{1 + \bar{r} z}.
\end{equation}
This can be visualized conveniently in the complex plane, where the ratio of
real to imaginary parts contains the same information as the argument of a
complex number (the argument is $\tan(1/r)$): The renormalization of a complex
coupling $\bar{g}$ with $Z$ corresponds to a rescaling $\abs{g} =
\abs{\bar{g}}/\abs{Z}$ of the modulus and a rotation of the phase by the
argument of $Z$, $\arg g = \arg \bar{g} - \arg Z$. The condition
Eq.~\eqref{eq:61} corresponding to MAR is depicted in
Fig.~\ref{fig:Z_renormalization} (b): All bare\footnote{Here we denote the
  couplings that are not divided by $Z$ as bare.} couplings lie on a single
ray. In the purely dissipative case with $r = 0$ and $\bar{r} = z$, which is
shown in Fig.~\ref{fig:Z_renormalization} (a), this ray is perpendicular to
$Z$. As a result, in this case the renormalized couplings are purely
imaginary. Generally, only the renormalized quantities allow for an immediate
physical interpretation: $A$ and $D$ describe propagation and diffusive behavior
of particles, respectively, while $\lambda$ ($\lambda_3$) and $\kappa$
($\kappa_3$) are two-body (three-body) elastic collisions and loss. In the
generic driven-dissipative case, we have no \textit{a priori} relations between
these couplings because they are due to different physical mechanisms:
Dissipative couplings describe local incoherent single particle pump and loss,
as well as local two-body loss. On the other hand, unitary dynamics is given by
coherent propagation and elastic collisions. Geometrically, the physical
couplings point in different directions in the first quadrant of the complex
plane (see Fig.~\ref{fig:Z_renormalization} (c)), the latter restriction being
due to the physical stability of the system (see
Sec.~\ref{sec:master-equation}).

This concludes our discussion of the relation of the DDM to dynamical
equilibrium models. In the following section we will proceed to derive explicit
flow equations for the couplings Eq.~\eqref{eq:53}.

\section{Non-Equilibrium FRG flow equations}
\label{sec:non-eq-frg-flow}

In the following we discuss how the functional differential equation
Eq.~\eqref{eq:32} for the effective action is reduced to a set of ordinary
differential equations by virtue of the ansatz Eq.~\eqref{eq:34} for
$\Gamma_k$. First we derive the flow equation for the effective potential, i.e.,
the part of the effective action that involves all momentum-independent
couplings. Then we proceed to specify the flow of the inverse propagator which
determines flow equations for the wave-function renormalization $Z$ and the
gradient coefficient $\bar{K}$. In the FRG, we approach the critical point from
the ordered (symmetry-broken) side of the transition. This allows us to capture
the leading divergences of two-loop effects in a calculation that is formally
one-loop\cite{berges02:_nonper} by means of diagrams like the second one in
Eq.~\eqref{eq:27} in the spirit of the background field method in gauge
theories.\cite{dashen81:_relat}

We denote the truncation Eq.~\eqref{eq:34}, evaluated for homogeneous, i.e., space-
and time-independent ``background fields'' by
\begin{equation}
  \label{eq:63}
  \Gamma_{k,cq} = - \Omega \left( \bar{U}' \bar{\rho}_{cq} + \bar{U}^{\prime *}
    \bar{\rho}_{qc} - i \bar{\gamma} \bar{\rho}_q \right),
\end{equation}
(the subscript $cq$ indicates that we have classical and quantum background
fields) where $\Omega$ is the quantization volume and the $U(1)$ invariant
combinations of fields are $\bar{\rho}_{cq} = \bar{\phi}_c^{*} \bar{\phi}_q =
\bar{\rho}_{qc}^{*}$ and $\bar{\rho}_q = \abs{\bar{\phi}_q}^2$. This
representation of $\Gamma_{k,cq}$ implies that the flow equation for the
potential $\bar{U}'$ can be obtained from Eq.~\eqref{eq:32} by taking the
derivative with respect to $\bar{\rho}_{cq}$ and setting the quantum background
fields to their stationary value (which is zero) afterwards,
\begin{equation}
  \label{eq:64}
  \partial_t \bar{U}' = - \frac{1}{\Omega} \left[ \partial_{\bar{\rho}_{cq}} \partial_t
    \Gamma_{k,cq} \right]_{\qnb},
\end{equation}
where the dimensionless RG flow parameter $t$ is related to the cutoff scale $k$
via $t = \ln(k/\Lambda)$. The flow equation for the renormalized potential
follows straightforwardly by taking the scale derivative of the relation
$\bar{U} = Z U$, which results in
\begin{equation}
  \partial_t \bar{U}' = Z \left( - \eta_Z U' + \partial_t U' \right),
\end{equation}
where we introduced the anomalous dimension of the wave-function
renormalization,
\begin{equation}
  \label{eq:65}
  \eta_Z = -\partial_t Z/Z.
\end{equation}
Then, using $\partial_{\bar{\rho}_{cq}} = Z \partial_{\rho_{cq}}$, the flow equation
for the renormalized potential can be written as
\begin{equation}
  \label{eq:66}
  \partial_t U' = \eta_Z U' + \zeta', \quad \zeta' = - \frac{1}{\Omega}
  \left[ \partial_{\rho_{cq}} \partial_t
    \Gamma_{k,cq} \right]_{\qn}.
\end{equation}

We proceed by specifying the projection prescriptions that allow us to derive
the flow of the couplings $u_n$ in the ordered phase from the flow
equation~\eqref{eq:66}. Taking the scale derivatives of the relation $u_n =
U^{(n)}(\rho_0)$ we find
\begin{equation}
  \label{eq:67}
  \partial_t u_n = \left( \partial_t U^{(n)} \right)(\rho_0) + U^{(n
    + 1)}(\rho_0) \partial_t \rho_0.
\end{equation}
Based on the power-counting arguments of Sec.~\ref{sec:canon-power-count}, our
truncation includes terms up to cubic order in the $U(1)$ invariants, i.e., for
derivatives of the effective potential of the order of $n \geq 4$ we have
$U^{(n)} = 0$. The flow equations for the quartic and sextic couplings are then
given by (the RHS of these equations determine the so-called $\beta$-functions)
\begin{align}
  \label{eq:68}
  \partial_t u_2 & = \beta_{u_2} = \eta_Z u_2 + u_3 \partial_t \rho_0
  + \partial_{\rho_c} \zeta' \bigr\rvert_{\mathrm{ss}},
  \\ \label{eq:69} \partial_t u_3 & = \beta_{u_3} = \eta_Z u_3
  + \partial_{\rho_c}^2 \zeta' \bigr\rvert_{\mathrm{ss}},
\end{align}
where according to Eq.~\eqref{eq:67} in $\zeta'$ we specify the classical
background field $\rho_c$ it to its stationary value $\rho_c
\rvert_{\mathrm{ss}} = \rho_0$. As we have seen above (cf.\
Secs.~\ref{sec:master-equation} and~\ref{sec:mean-field-theory}, the latter is
determined by the dissipative part of the field equation, i.e., by the condition
$\Im U'(\rho_0) = 0$. Taking here the derivative with respect to the RG
parameter $t$, we find
\begin{equation}
  \label{eq:70}  
  \partial_t \rho_0 = - \left( \Im \partial_t U' \right)(\rho_0)/\Im U''(\rho_0)
  = - \Im \zeta' \bigr\rvert_{\mathrm{ss}}/\kappa.
\end{equation}
Having thus specified the flow equations for the couplings that parameterize the
potential $U$, we proceed to the Keldysh mass $\bar{\gamma}$, which is the
coefficient of the term that is proportional to the quantum $U(1)$ invariant
$\bar{\rho}_q$ in Eq.~\eqref{eq:63}. We obtain the flow equation for
$\bar{\gamma}$ as
\begin{equation}
  \label{eq:71}
  \partial_t \bar{\gamma} = - \frac{i}{\Omega}
  \left[ \partial_{\bar{\rho}_q} \partial_t \Gamma_{k,cq} \right]_{\mathrm{ss}}.
\end{equation}
For the renormalized Keldysh mass, which is related to the bare one via $\gamma
= \bar{\gamma}/\abs{Z}^2$, we have (the transformation from bare to renormalized
fields implies $\partial_{\bar{\rho}_q} = \abs{Z}^2 \partial_{\rho_q}$)
\begin{equation}
  \label{eq:72}
  \partial_t \gamma = \beta_{\gamma} = 2 \ezr \gamma + \zeta_{\gamma}, \quad \zeta_{\gamma} = -
  \frac{i}{\Omega} \left[ \partial_{\rho_q} \partial_t \Gamma_{k,cq} \right]_{\mathrm{ss}}.
\end{equation}

While the flow of $\Gamma_{k,cq}$ (i.e., the flow equation evaluated at
homogeneous background fields) yields flow equations for all
momentum-independent couplings, we have to consider the flow of the inverse
propagator
\begin{equation}
  \label{eq:73}
  \partial_t \bar{P}_{ij}(Q) \delta(Q - Q') = \left[ \frac{\delta^2 \partial_t \Gamma_k}{\delta
      \bar{\chi}_i(-Q) \delta \bar{\chi}_j(Q')} \right]_{\mathrm{ss}},
\end{equation}
in order to derive flow equations for the wave-function renormalization $Z$ and
the gradient coefficient $\bar{K}$. The retarded component of the inverse
propagator in the presence of real stationary background fields $\bar{\phi}_c =
\bar{\phi}_c^{*} = \bar{\phi}_0$ reads
\begin{equation}
  \label{eq:74}    
  \bar{P}^R(Q) =
  \begin{pmatrix}
    - i Z_I \omega - \bar{K}_R q^2 - 2 \bar{\lambda} \bar{\rho}_0 & i Z_R \omega - \bar{K}_I q^2 \\
    - i Z_R \omega + \bar{K}_I q^2 + 2 \bar{\kappa} \bar{\rho}_0 & - i Z_I
    \omega - \bar{K}_R q^2
  \end{pmatrix},
\end{equation}
Then, for the kinetic coefficient $\bar{K}$ we choose from the flow
equation~\eqref{eq:73} the elements of the inverse propagator that do not have
mass-like contributions\cite{berges02:_nonper} $2 \bar{\lambda} \bar{\rho}_0$
and $2 \bar{\kappa} \bar{\rho}_0$,
\begin{equation}  
  \label{eq:75}
  \partial_t \bar{K} = - \partial_{q^2} \left( \partial_t \bar{P}^R_{22}(Q)
    + i \partial_t \bar{P}^R_{12}(Q) \right) \Bigr\rvert_{Q = 0}.
\end{equation}
The flow equation for the wave-function renormalization $Z$ as specified below,
on the other hand, mixes massive and massless components symmetrically
\begin{equation}
  \label{eq:76}
  \partial_t Z = - \frac{1}{2} \partial_{\omega} \tr \left[ \left( \id + \sigma_y
    \right) \partial_t \bar{P}^R(Q) \right] \Bigr\rvert_{Q = 0}.
\end{equation}
This choice allows for the locking of the flows of the Keldysh mass and $Z$ as
implied by the emergence of the symmetry Eq.~\eqref{eq:47} in the purely
dissipative IR regime (see Sec.~\ref{sec:scaling-solutions}). Finally, the flow
equation for the renormalized coefficient $K$ follows by straightforward
differentiation of its definition $K = \bar{K}/Z$ in terms of bare
quantities. We find
\begin{equation}
  \label{eq:77}
  \partial_t K = \beta_K = \eta_Z K + \partial_t \bar{K}/Z.
\end{equation}

The truncation Eq.~\eqref{eq:34} is parameterized in terms of the couplings
Eq.~\eqref{eq:53}. Renormalization of the fields with $Z$ leads to a description
in terms of $\mathbf{g} = \left( K, \rho_0, u_2, u_3, \gamma \right)^T$ (where
we omit the mass $u_1$: as indicated above we approach the critical point from
the ordered phase, i.e., we parameterize the effective action in terms of the
stationary condensate density $\rho_0$ instead of the mass $u_1$). In this
section we derived the $\beta$-functions for these renormalized couplings, i.e.,
we have specified a closed set of flow equations $\partial_t \mathbf{g} =
\beta_{\mathbf{g}}(\mathbf{g})$ from which $Z$ can be completely eliminated (the
anomalous dimension $\eta_Z$ entering the $\beta$-functions can again be
expressed in terms of the couplings $\mathbf{g}$ alone). More explicit
expressions for the $\beta$-functions are provided in
App.~\ref{sec:flow-inverse-prop}).

\section{Scaling solutions}
\label{sec:scaling-solutions}

As one considers an effective description of a system at a continuous phase
transition at longer and longer scales (which is equivalent to following the RG
flow to smaller values of $k$), physical observables and the couplings that
describe the system exhibit scaling behavior. The search for such scaling
solutions to the flow equations is facilitated by introducing rescaled
dimensionless (in the sense of the canonical power counting introduced in
Sec.~\ref{sec:canon-power-count}) couplings which remain constant, i.e., by
searching for a fixed point of the flow equations of these rescaled couplings
instead. In the following section we introduce such rescaled couplings and derive
the corresponding flow equations.

\subsection{Scaling form of the flow equations}
\label{sec:resc-flow-equat}

As a first step we trade the real parts of $K, u_2,$ and $u_3$ for the ratios of
real to imaginary parts
\begin{equation}
  \label{eq:78}
  r_K = A/D, \quad r_{u_2} = \lambda/\kappa, \quad r_{u_3} = \lambda_3/\kappa_3,
\end{equation}
which measure the relative strength of coherent and dissipative dynamics. As we
will show below, at criticality all these ratios flow to zero signaling
decoherence. Their flow is given by
\begin{align}
  \label{eq:79}
  \partial_t r_K & = \beta_{r_K} = \frac{1}{D} \left( \beta_A - r_K \beta_D
  \right), \\ \label{eq:80} \partial_t r_{u_2} & = \beta_{r_{u_2}} =
  \frac{1}{\kappa} \left( \beta_{\lambda} - r_{u_2} \beta_{\kappa} \right),
  \\ \label{eq:81} \partial_t r_{u_3} & = \beta_{r_{u_3}} = \frac{1}{\kappa_3}
  \left( \beta_{\lambda_3} - r_{u_3} \beta_{\kappa_3} \right).
\end{align}
(The $\beta$-functions for the real and imaginary parts of $K, u_2,$ and $u_3$
are specified in App.~\ref{sec:flow-eff-pot}, see Eq.~\eqref{eq:164}.) We
proceed by introducing a dimensionless mass term
\begin{equation}
  \label{eq:82}
  w = \frac{2 \kappa \rho_0}{k^2 D},
\end{equation}
the flow equation of which mixes contributions from the $\beta$-functions of
$\rho_0, \kappa,$ and $D$, and reads
\begin{equation}
  \label{eq:83}
  \partial_t w = \beta_w = - \left( 2 - \eta_D \right) w + \frac{w}{\kappa}
  \beta_{\kappa} + \frac{2 \kappa}{k^2 D} \beta_{\rho_0},
\end{equation}
where the anomalous dimension of $D$ is defined as
\begin{equation}
  \label{eq:84}
  \eta_D = - \partial_t D/D.
\end{equation}
Finally we replace the quartic and sextic couplings by dimensionless ones. For a
momentum-independent $n$-body coupling $u_n$ we can construct a corresponding
dimensionless coupling by means of the relation
\begin{equation}
  \label{eq:85}
  \tilde{u}_n = \frac{k^{\left( d - 2 \right) n - d}}{D^n} \left( \frac{\gamma}{2} \right)^{n - 1} u_n.
\end{equation}
The flow equations for the imaginary parts $\tilde{\kappa}$ and
$\tilde{\kappa}_3$ of the dimensionless quartic and sextic couplings, therefore,
are given by
\begin{gather}
  \partial_t \tilde{\kappa} = \beta_{\tilde{\kappa}} = - \left( 4 - d - 2 \eta_D
    + \eta_{\gamma} \right) \tilde{\kappa} + \frac{k^{-4 + d} \gamma}{2 D^2}
  \beta_{\kappa}, \\ \partial_t \tilde{\kappa}_3 = \beta_{\tilde{\kappa}_3} = -
  \left( 6 - 2 d - 3 \eta_D + 2 \eta_{\gamma} \right) \tilde{\kappa}_3 +
  \frac{k^{- 6 + 2 d} \gamma^2}{4 D^3} \beta_{\kappa_3},
\end{gather}
and include contributions from the anomalous dimension
\begin{equation}
  \label{eq:86}
  \eta_{\gamma} = - \partial_t \gamma/\gamma.
\end{equation}

Thus we are left with six dimensionless running couplings, which we collect in
vectors $\mathbf{r} = \left( r_K,r_{u_2},r_{u_3} \right)^T$ and $\mathbf{s} =
\left( w,\tilde{\kappa},\tilde{\kappa}_3 \right)^T$. Their flow equations form a
closed set,
\begin{equation}
  \label{eq:87}
  \partial_t \mathbf{r} = \beta_{\mathbf{r}}(\mathbf{r},\mathbf{s}), \quad
  \partial_t \mathbf{s} = \beta_{\mathbf{s}}(\mathbf{r},\mathbf{s}).
\end{equation}
The $\beta$-functions on the RHS of these equations contain the anomalous
dimensions $\eta_Z, \eta_D$, and $\eta_{\gamma}$, which in turn can be expressed
as functions of the running couplings $\mathbf{r}$ and $\mathbf{s}$ alone. We
note in passing that according to the discussion of
Sec.~\ref{sec:geom-interpr-equil}, the equilibrium model MAR is described by
$r_K = r_{u_2} = r_{u_3} = r$, i.e.,
\begin{equation}
  \label{eq:88}
  \mathbf{r}_{\mar} = r \left( 1,1,1 \right)^T
\end{equation}
(MA is realized for the special case $r = 0$). Inserting the same value $r$ for
all three ratios in the respective $\beta$-functions we find $\beta_{r_K} =
\beta_{r_{u_2}} = \beta_{r_{u_3}}$, which shows that for MAR the common ratio is
preserved by the flow as it should be.

Our analysis of the flow equations~\eqref{eq:87} will proceed in two steps:
First we will search for fixed points $\mathbf{r}_{*}$ and $\mathbf{s}_{*}$,
which are solutions to the algebraic equations
\begin{equation}
  \label{eq:89}
  \beta_{\mathbf{r}}(\mathbf{r}_{*},\mathbf{s}_{*}) =
  \beta_{\mathbf{s}}(\mathbf{r}_{*},\mathbf{s}_{*}) = \mathbf{0}.
\end{equation}
In Sec.~\ref{sec:gaussian-fixed-point} we briefly discuss the trivial Gaussian
fixed point and then turn to the Wilson-Fisher fixed point that describes the
critical system in~\ref{sec:wilson-fisher-fixed}). Second we will solve the full
flow equations numerically and provide our results in
Sec.~\ref{sec:numer-integr-flow}. While already the linearized flow equations in
the vicinity of the Wilson-Fisher fixed point encode universal physics at the
phase transition and determine the asymptotic flow of the system for $k \to 0$
(or $t \to - \infty$), the numerical integration of the full flow equations
provides us with information on non-universal aspects such as the extent of the
scaling regime.

\subsection{Gaussian fixed point}
\label{sec:gaussian-fixed-point}

All $\beta$-functions vanish on the manifold of Gaussian fixed points which is
parameterized by $\mathbf{s}_{*} = 0$ and $\mathbf{r}_{*} \in \R^3$. We note
that the combination of vanishing imaginary parts $\tilde{\kappa}_{*}$ and
$\tilde{\kappa}_{3*}$ of the quartic and sextic couplings and arbitrary finite
ratios of real to imaginary parts implies that also the real parts of
$\tilde{u}_{2*}$ and $\tilde{u}_{3*}$ are zero on this fixed point manifold. In
a linearization of the flow equations around $\mathbf{s}_{*} = 0$, the
fluctuation contributions vanish and the scaling behavior is determined solely
by the canonical scaling dimensions, implying in particular that the Gaussian
fixed point is unstable (for small values $\mathbf{s} \neq \mathbf{s}_{*}$ the
flow is directed away from the fixed point) and, therefore, physically not
relevant. Non-trivial scaling behavior at criticality is governed by the
Wilson-Fisher fixed point which we will discuss in the next section.

\subsection{Wilson-Fisher fixed point: critical behavior}
\label{sec:wilson-fisher-fixed}

As discussed in Sec.~\ref{sec:relat-equil-dynam}, our driven-dissipative model
reduces to MA when we set the real parts of all renormalized couplings to zero,
cf.\ Fig.~\ref{fig:Z_renormalization}, i.e., for $\mathbf{r} = \mathbf{0}$. It
is well-known that MA exhibits a non-trivial Wilson-Fisher fixed
point,\cite{hohenberg77:_theor} and indeed we find this fixed point at
\begin{equation}
  \label{eq:90}
  \begin{split}
    \mathbf{r}_{*} & = \left( r_{K *},r_{u *},r_{u' *} \right) = \mathbf{0},\\
    \mathbf{s}_{*} & = \left( w_{*},\tilde{\kappa}_{*},\tilde{\kappa}'_{*}
    \right)= \left( 0.475,5.308,51.383 \right).
  \end{split}
\end{equation}
The values of the coupling $\mathbf{s}_{*}$ are identical to those obtained in
an equilibrium classical $O(2)$ model from functional RG calculations at the
same level of truncation.\cite{berges02:_nonper} We note that this fixed point
is also contained in the subspace of couplings corresponding to MAR, which is
characterized by Eq.~\eqref{eq:88}, i.e., the phase transitions in both the
equilibrium and non-equilibrium models are described by the same fixed
point. Critical behavior, however, is determined by the RG flow in the vicinity
of the fixed point. Here the non-equilibrium setting adds two more independent
directions, thereby opening up the possibility for deviations from equilibrium
criticality as we will now show.

The asymptotic flow for $k \to 0$ of the critical system is determined by a
linearization of the flow equations in the deviations $\delta \mathbf{s} =
\mathbf{s} - \mathbf{s}_{*}, \delta \mathbf{r} =\mathbf{r}$ from the fixed
point. In the linear regime the two sectors corresponding to $\mathbf{s}$ and
$\mathbf{r}$ decouple as described by the block diagonal stability matrix
\begin{equation}
  \label{eq:91}
  \partial_t
  \begin{pmatrix}
    \delta \mathbf{r} \\ \delta \mathbf{s}
  \end{pmatrix}
  =
  \begin{pmatrix}    
    N & 0 \\
    0 & S
  \end{pmatrix}
  \begin{pmatrix}
    \delta \mathbf{r} \\ \delta \mathbf{s}
  \end{pmatrix},
\end{equation}
where the $3 \times 3$ submatrices $S$ and $N$ are given by
\begin{align}
  \label{eq:92}  
  S & = \nabla_{\mathbf{s}}^T \beta_{\mathbf{s}} \bigr\rvert_{\mathbf{r} =
    \mathbf{r}_{*}, \mathbf{s} = \mathbf{s}_{*}} =
  \begin{pmatrix}
    -1.620 & 0.088 & 0.005 \\
    -3.183 & 0.290 & 0.036 \\
    -15.374 & -42.249 & 2.183
  \end{pmatrix}, \\
  \label{eq:93}
  N & = \nabla_{\mathbf{r}}^T \beta_{\mathbf{r}} \bigr\rvert_{\mathbf{r} =
    \mathbf{r}_{*}, \mathbf{s} = \mathbf{s}_{*}} =
  \begin{pmatrix}
    0.053 & 0.059 & 0.032 \\
    0 & -0.053 & 0.196 \\    
    0.498 & -2.327 & 1.973
  \end{pmatrix}.
\end{align}
The matrix $N$ would be identically zero in the absence of anomalous additions
to the canonical scaling dimensions (note that the ratios $\mathbf{r}$ have
canonical scaling dimension zero), or even if coherent and dissipative couplings
would exhibit identical anomalous scaling. The non-vanishing of this block thus
indicates a different universal behavior of these two types of couplings. Due to
the decoupling of the flows of $\mathbf{r}$ and $\mathbf{s}$ we may discuss the
linearized flow of each set of couplings separately.

In the matrix $S$ we find one negative eigenvalue $s_1$ corresponding to the
correlation length exponent $\nu = - 1/s_1 = 0.716$ (our findings for critical
exponents are summarized in Tab.~\ref{tab:critical_exponents}). Considering that
we are restricting ourselves to relevant and marginal terms in our truncation,
the agreement of the numerical value of $\nu$ with results from more
sophisticated calculations\cite{guida98:_critic_n} is reasonable. Furthermore
there are two complex conjugate eigenvalues $s_{2,3} = 1.124 \pm i 0.622$ with
positive real parts (indicating that these directions are stable). The imaginary
parts are known artifacts of this level of truncation for the $O(2)$ model and
vanish upon inclusion of higher order terms in the effective
potential.\cite{litim02:_critic}
\begin{table}
  \centering
  \begin{tabular}{|c|c|c|c|c|}
    \hline
    & $\nu$ & $\eta$ & $z$ & $\eta_r$ \\
    \hline
    $O(2)$ & 0.716 & 0.039 & & \\
    MA & 0.716 & 0.039 & 2.121 & \\    
    MAR & 0.716 & 0.039 & 2.121 & - 0.143 \\
    DDM & 0.716 & 0.039 & 2.121 & - 0.101 \\
    \hline
  \end{tabular}
  \caption{Results for the correlation length exponent $\nu$, the anomalous
    dimension $\eta$, the dynamical critical exponent $z$, and the decoherence
    exponent $\eta_r$ in our truncation.}
  \label{tab:critical_exponents}
\end{table}

The scaling behavior of the couplings $Z, D,$ and $\gamma$ is determined by the
values of the respective anomalous dimensions at the fixed point. In addition we
define the anomalous dimension for the bare kinetic coefficient $\bar{K}$ as
\begin{equation}
  \label{eq:94}
  \eta = - \partial_t \bar{K}/\bar{K} = \frac{1}{1 + r_K^2} \left[
    r_K^2 \bar{\eta}_A + \bar{\eta}_D - i r_K \left( \bar{\eta}_A - \bar{\eta}_D \right) \right],
\end{equation}
where the representation in terms of $\bar{\eta}_A$ and $\bar{\eta}_D$ follows
from the definition of these quantities in Eq.~\eqref{eq:162}. At the fixed point
$\eta$ takes the value
\begin{equation}
  \label{eq:95}
  \eta = 0.039,
\end{equation}
which is again the result for the anomalous dimension of the classical $O(2)$
model in $d = 3$ dimensions at the same level of
truncation\cite{berges02:_nonper} and agrees well with results from more
accurate calculations.\cite{guida98:_critic_n} In summary, the static critical
behavior coincides precisely with the one of the classical $O(2)$ model,
implying that the dynamical anomalous dimension $\eta_Z$ effectively does not
enter the corresponding equations. This can be seen as follows: Inserting
$\mathbf{r} = \mathbf{0}$ in the expressions for the anomalous dimensions, we
find
\begin{equation}
  \label{eq:96}
  \ezr = - \eta_{\gamma}, \quad \ezi = 0.
\end{equation}
(We note that this holds for all values of the static couplings $\mathbf{s}$,
i.e., it is always realized in MA.) These relations ensure that $\ezr$ and
$\eta_{\gamma}$ compensate each other in all flow equations.\footnote{The
  cancellation of $\ezr$ and $\eta_{\gamma}$ can be made explicit by inserting
  the $\beta$-functions for $\kappa$ and $\kappa_3$, Eqs.~\eqref{eq:185}
  and~\eqref{eq:186} respectively, as well as the expression for $\eta_D$ that
  follows from Eq.~\eqref{eq:181}, in the flow equations for $\tilde{\kappa}$
  and $\tilde{\kappa}_3$. In the resulting expressions the anomalous dimensions
  $\ezr$ and $\eta_{\gamma}$ appear only as the sum $\ezr + \eta_{\gamma}$.}
Moreover they imply that the ratio $\bar{\gamma}/\abs{Z}$ appearing on the RHS
of the fluctuation-dissipation relation Eq.~\eqref{eq:59} approaches a constant
value at the fixed point: According to the definition of the anomalous
dimensions Eqs.~\eqref{eq:65} and~\eqref{eq:86}, close to the fixed point the
flow of $Z$ and $\gamma$ is described by $Z \sim k^{-\eta_Z}$ (note that
$\eta_Z$ is real so that this behavior indeed describes algebraic scaling and
does not contain oscillatory parts) and $\gamma \sim k^{-\eta_{\gamma}}$ with
$\eta_Z$ and $\eta_{\gamma}$ evaluated at $\mathbf{r}_{*}$ and
$\mathbf{s}_{*}$. Thus we find $\bar{\gamma}/\abs{Z} = \abs{Z} \gamma \sim
k^{-\eta_Z - \eta_{\gamma}} = \mathrm{const.}$, i.e., the symmetry
Eq.~\eqref{eq:47}, which manifests itself in this quantity approaching a
constant value (cf.\ Eq.~\eqref{eq:60}), emerges in the IR without imposing it
in the microscopic model. In other words, the driven-dissipative system obeys a
classical FDT in the long-wavelength limit (see Fig.~\ref{fig:teff}). At the
fixed point we find the value
\begin{equation}
  \label{eq:97}
  \eta_Z = - \eta_{\gamma} = 0.161.
\end{equation}

Let us now consider the upper left block $N$ of the stability matrix. It has
three positive eigenvalues,
\begin{equation}
  \label{eq:98}
  n_1 = 0.101, \quad  n_2 = 0.143, \quad n_3 = 1.728,
\end{equation}
which indicates that the ratios $\mathbf{r}$ are attracted to their fixed point
value zero. The corresponding eigenvectors are
\begin{equation}
  \label{eq:99}
  \mathbf{u}_1 = 
  \begin{pmatrix}
    0.022 \\
    0.109 \\
    0.994
  \end{pmatrix},
  \quad \mathbf{u}_2 = \frac{1}{\sqrt{3}}
  \begin{pmatrix}
    1 \\ 1 \\ 1
  \end{pmatrix},
  \quad \mathbf{u}_3 = 
  \begin{pmatrix}
    0.802 \\
    0.469 \\
    0.370
  \end{pmatrix}.
\end{equation}
The smallest of the eigenvalues determines the scaling behavior of $\mathbf{r}$
in the deep IR. In order to see this let us expand $\mathbf{r}$ in the basis of
eigenvectors of the matrix $N$,
\begin{equation}
  \label{eq:100}
  \mathbf{r} = \sum_{i = 1}^3 \mathbf{u}_i c_i.
\end{equation}
The coefficients in this expansion are given by $c_i = \mathbf{v}_i \cdot
\mathbf{r}$, where $\mathbf{v}_i$ are the left eigenvectors of $N$ (the latter
is not symmetric and its left and right eigenvectors, therefore, are not equal),
normalized such that $\mathbf{u}_i \cdot \mathbf{v}_j = \delta_{ij}$. The
asymptotic behavior of the flow of the so-called scaling
fields\cite{Altland/Simons} $c_i$ is given by $c_i \sim e^{n_i t} = k^{n_i}$,
which implies that for $\mathbf{r}$ we indeed find
\begin{equation}
  \label{eq:101}
  \mathbf{r} \sim  \mathbf{u}_1 k^{n_1} = \mathbf{u}_1 k^{- \eta_r},
\end{equation}
with only subdominant contributions in the directions of $\mathbf{u}_2$ and
$\mathbf{u}_3$. This leads us to identify the decoherence exponent
\begin{equation}
  \label{eq:102}
  \eta_r = - n_1 = - 0.101.
\end{equation}

From the scaling behavior of the ratios $\mathbf{r}$ we may infer the one of the
coherent couplings. For the coefficient of coherent propagation $A$, in
particular, we have
\begin{equation}
  \label{eq:103}
  A = r_K D \sim k^{n_1 - \eta_D} = k^{- \eta_A}.
\end{equation}
Then, with the anomalous dimension of the dissipative kinetic coefficient $D$ at
the fixed point,
\begin{equation}
  \label{eq:104}  
  \eta_D = - 0.121,
\end{equation}
we obtain the value
\begin{equation}
  \label{eq:105}
  \eta_A = - 0.223.
\end{equation}

Let us discuss the consequences of this result for the effective dispersion
relation of long-wavelength excitations, which is encoded in the running inverse
propagator Eq.~\eqref{eq:37}. Once the cutoff scale $k$ becomes smaller than the
external momentum $q$, the effective infrared cutoff is given by $q$ instead of
$k$.\cite{wetterich08:_funct} Then, in the dispersion relation Eq.~\eqref{eq:42},
which we rewrite here in terms of the scaling variables introduced in
Sec.~\ref{sec:resc-flow-equat} as
\begin{equation}
  \label{eq:106}
  \omega_{1,2}^R = D q^2 \left[ -i \left( 1 + w/2 \right) \pm \sqrt{r_K^2 + r_K
      r_{u_2} w - \left( w/2 \right)^2} \right],
\end{equation}
we may insert the scaling forms $w \sim w_{*}, r_K \sim r_{K 0} q^{-\eta_r},
r_{u_2} \sim r_{u_2 0} q^{-\eta_r},$ and $D \sim D_0 q^{-\eta_D}$. For $q \to 0$ both
modes are purely diffusive with $\omega_1^R \sim - i D_0 q^{2 - \eta_D}$ and
$\omega_2^R \sim - i D_0 q^{2 - \eta_D} \left( 1 + w_{*} \right)$, and for the
dynamical critical exponent $z$ which is defined via the relation $\omega^R \sim
-i q^z$ we find the value
\begin{equation}
  \label{eq:107}
  z = 2 - \eta_D = 2.121.
\end{equation}

\begin{figure}  
  \includegraphics{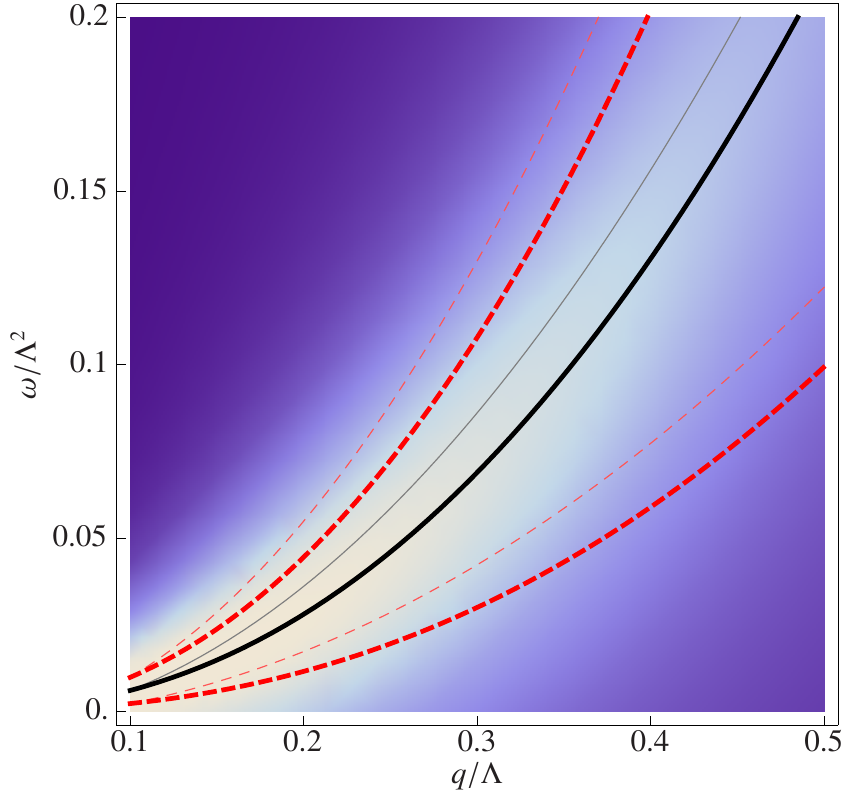}
  \caption{(Color online) Spectral density Eq.~\eqref{eq:218} in the scaling
    regime. The solid line corresponds to the position $\sim A_0 q^{2 - \eta_A}$
    of the peak of the Lorentzian curve while its width $\sim D_0 q^{2 -
      \eta_D}$ is indicated by dashed lines. For comparison we also show peak
    position and width for canonical scaling $\sim q^2$ as thin solid and dashed
    lines, respectively. (Canonical and anomalous scaling forms are chosen to
    coincide at $q \Lambda = 0.1$.) Parameters are $A_0 \Lambda^{\eta_A} = Z_0
    \Lambda^{\eta_Z} = 1$ and $D_0 \Lambda^{\eta_D} = 1/2$.}
  \label{fig:spectral_fct_density}
\end{figure}

Above the purely diffusive IR regime, when $w \ll r_K, r_{u_2}$, the dispersion
relation simplifies to
\begin{equation}
  \label{eq:108}
  \omega_{1,2}^R \sim \left( -i D \pm D r_K \right) q^2 \sim -i D_0 q^{2 - \eta_D} \pm
  A_0 q^{2 - \eta_A},
\end{equation}
i.e., coherent propagation and diffusive contributions scale differently with
the momentum $q$. In the symmetric phase the branch $\omega^R_2$ is absent and
the bare retarded response function is dominated by a single pole at $\omega =
\omega^R_1$, i.e., we have (the bare scale-dependent propagators are determined
by the quadratic part of the effective action Eq.~\eqref{eq:34})
\begin{equation}
  \label{eq:216}
  \bar{G}^R(Q) = \frac{1}{Z^{*} \left( \omega - \omega^R_1 \right)}.
\end{equation}
As explained at the end of Sec.~\ref{sec:key-results-physical} this quantity
and, in particular, the spectral density which is related to its imaginary
part,\cite{Altland/Simons}
\begin{equation}
  \label{eq:217}
  A(Q) = - 2 \Im \bar{G}^R(Q),
\end{equation}
are direct experimental observables. For $\omega \approx \omega^R_1$ the
spectral density has the shape of a Lorentzian centered at $\Re \omega^R_1$ and
with width determined by $\Im \omega^R_1$,
\begin{equation}
  \label{eq:218}
  A(Q) = \frac{2}{\abs{Z}^2} \frac{Z_R \Im \omega^R_1}{\left( \omega - \Re
      \omega^R_1 \right)^2 + \left( \Im \omega^R_1 \right)^2}.
\end{equation}
Inserting here the scaling forms $Z \sim Z_0 k^{-\eta_Z}$ and Eq.~\eqref{eq:108}
for $\omega^R_1$ with different scaling behavior of real and imaginary parts,
the structure sketched in Fig.~\ref{fig:sketch} emerges. For the specific values
of the anomalous dimensions obtained in this section the spectral density is
shown in Fig.~\ref{fig:spectral_fct_density}, where a pronounced feature is
clearly visible.

Before moving on to a numerical integration of the flow equations in the next
section, we briefly contrast our findings for the DDM with the equilibrium case
of MAR. There, analyzing the stability of the fixed point Eq.~\eqref{eq:90} we
have to take into account only one direction $r = r_K = r_{u_2} = r_{u_3}$, and
we find (as $r_{*} = 0$ we have $\delta r = r$)
\begin{equation}
  \label{eq:109}
  \partial_t
  \begin{pmatrix}
    \delta r \\ \delta \mathbf{s}
  \end{pmatrix}
  =
  \begin{pmatrix}
    R & 0 \\
    0 & S
  \end{pmatrix}
  \begin{pmatrix}
    \delta r \\ \delta \mathbf{s},
  \end{pmatrix}
\end{equation}
where the matrix $S$ is the same as above and the element $R$ is given by the
``middle'' eigenvalue Eq.~\eqref{eq:98} of the stability matrix $N$ in the
non-equilibrium problem,
\begin{equation}
  \label{eq:110}
  R = \partial_r \beta_r \bigr\rvert_{r = r_{*}, \mathbf{s} = \mathbf{s}_{*}} =
  n_2,
\end{equation}
i.e., also in the equilibrium setting we find decoherence at the longest scales,
however, with a value of the decoherence exponent that is different from the one
in non-equilibrium. Let us finally remark that in the linearized regime, the
fact that MAR is contained as a special case in the non-equilibrium problem,
becomes visible in the form of the second eigenvector Eq.~\eqref{eq:99} which
realizes the constraint Eq.~\eqref{eq:88}.

We finally comment on the relation of the critical exponents obtained here with other approaches. The static critical exponent $\eta$ shows very good agreement with sophisticated high order perturbative calculations. For the dynamical critical exponent  $z$, to the best of our knowledge, currently there are no high-precision results for MA with $N = 2$ components available. (The
situation is different for the Ising-symmetric case with $N = 1$, where the
dynamical critical exponent has been calculated with high accuracy, see
Ref.~\onlinecite{CanetLeonie;Chate2007} and references therein.) Thus the value $z= 2.121$
obtained here has to be compared to $z = 2.026$ which corresponds to the third
order in $\epsilon$-expansion\cite{Zinn-Justin,Antonov1984}. The decoherence exponent ($\eta_r =-0.101$ here) has been computed in a recent complementary perturbative field theoretical study to second order in $\epsilon$-expansion, where it takes the value $\eta_r = -0.003$, see Ref.~\onlinecite{Tauber2013}. The
discrepancy between these values can only be resolved by extending the truncation advocated here, by including higher order
corrections in pertrubative field theoretical approaches, or by means of large-scale computer
simulations.

\section{Numerical integration of flow equations}
\label{sec:numer-integr-flow}

In the previous section we have seen that the flow equations Eq.~\eqref{eq:87}
entail non-trivial critical behavior governed by the Wilson-Fisher fixed point
Eq.~\eqref{eq:90}. While these results were based on an analysis of the
linearized flow equations in the vicinity of the fixed point, we will now turn
to a numerical integration of the full non-linear equations. One the one hand,
this serves to illustrate the concept of universality: Independently from the
initial values $\mathbf{r}_{\Lambda},\tilde{\kappa}_{\Lambda}$, and
$\tilde{\kappa}_{3 \Lambda}$ at the mesoscopic starting point of the RG flow,
critical behavior can be induced by a proper fine-tuning of $w_{\Lambda}$ and
becomes apparent in the approach of the RG flow to the scaling solution. Apart
from that, the availability of the full flow in the framework of the FRG allows
us to extract non-universal aspects. In particular, we will give an estimate of
the Ginzburg scale, i.e., the scale that separates the region of non-universal
flow from the universal scaling regime and thus is important for determining
experimental requirements on the necessary frequency resolution.

Our approach for finding numerical solutions to the flow equations that exhibit
critical behavior is as follows: We choose initial values $\mathbf{r}_{\Lambda},
\tilde{\kappa}_{\Lambda},$ and $\tilde{\kappa}_{3 \Lambda}$ at the mesoscopic
scale $k = \Lambda$ ($t = 0$), which are appropriate for the description of the
model introduced in Sec.~\ref{sec:model}. This model contains two-body elastic
interactions and loss, while three-body terms are contained only in an effective
low-momentum description, implying $\tilde{\kappa}_{\Lambda} \approx 1$ and
$\tilde{\kappa}_{3 \Lambda} \ll 1$. The dissipative kinetic coefficient $D$ is
very small in the microscopic description, so that $r_{K \Lambda} \gg 1$
initially, while for the two-body terms we have $r_{u_2 \Lambda} \approx 1$. The
latter generate the three-body couplings and we assume that $r_{u_3 \Lambda}
\approx 1$ as well.  For such a choice of initial values, there is a critical
value $w_{\Lambda} = w_c$ so that the resulting RG trajectories $\mathbf{r}(t)$
and $\mathbf{s}(t)$ approach the scaling solution, i.e., the fixed point, for $k
\to 0$ ($t \to -\infty$). Any solution obtained by \emph{numerically}
integrating the flow equations with $w_{\Lambda}$ fine-tuned to $w_c$, however,
eventually always flows away from the fixed point, as due to limited accuracy
the solution develops a non-zero component in the unstable direction of the
fixed point at some stage. For all solutions shown in the figures we choose
$w_{\Lambda}$ slightly below $w_c$, so that the trajectory at large RG ``times''
$t$ flows to the symmetric phase with $w = 0$.
\begin{figure}
  \centering
  \includegraphics{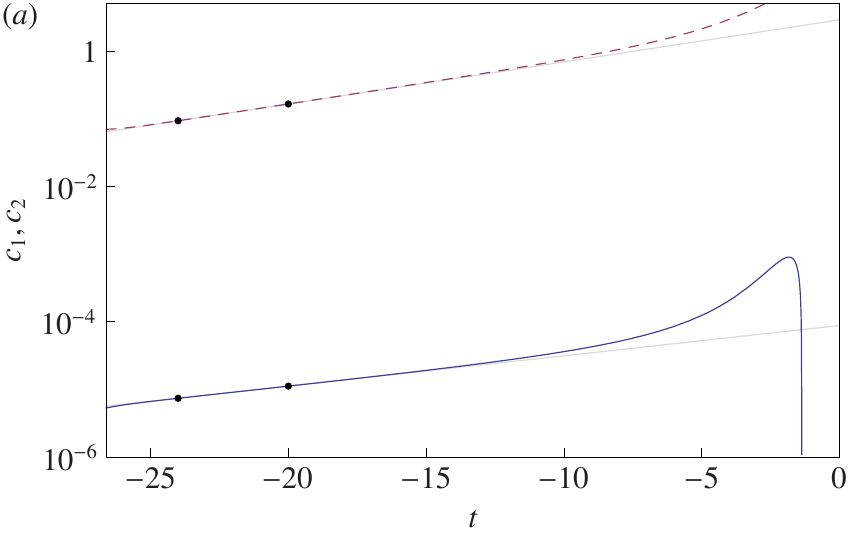}
  \includegraphics{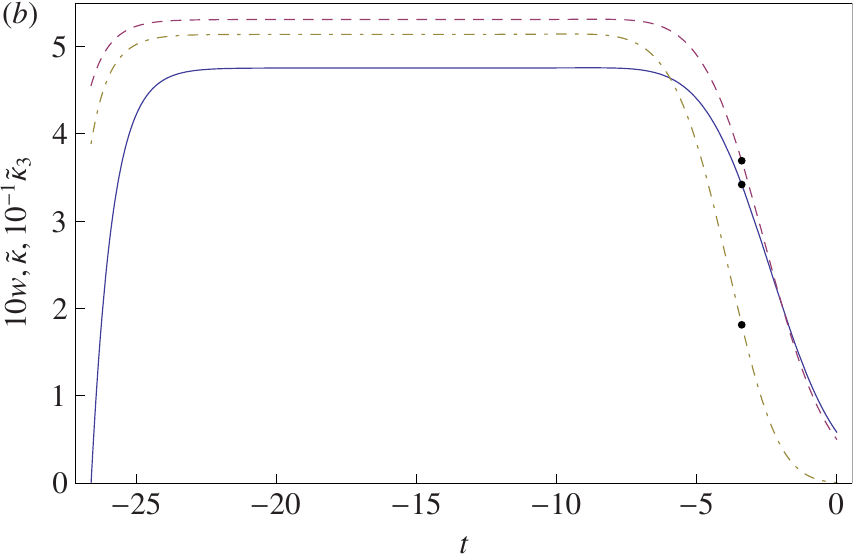}  
  \caption{(Color online) (a) The flow of $c_1$ (solid line) describes the
    vanishing of coherent dynamics. A fit with $\ln c_1 = a_1 t + b_1$ in the
    region $t \in [-24,-20]$ (the points $t = -24$ and $t = -20$ are highlighted
    by dots on the trajectory) yields the slope $a_1 = 0.10$ in agreement with
    smallest eigenvalue $n_1 = - \eta_r$ of the stability matrix
    Eq.~\eqref{eq:93}. We also show the evolution of the coefficient $c_2$
    (dashed line). For the evolution of $c_2$, the slope of a linear fit is $a_2
    = 0.14$ and reproduces the eigenvalue $n_2$. In the scaling region, the
    coefficient $c_3$ drops to very small values $\lesssim 10^{-11}$ on a scale
    $1/n_3 \approx 0.6$. The exponential decay of the components of $\mathbf{r}$
    is in this range still dominated by the contribution stemming from
    $c_2$. (b) The couplings $10 w$ (solid), $\tilde{\kappa}$ (dashed), and
    $10^{-1} \tilde{\kappa}_3$ (dot-dashed) are close to their fixed point
    values in the range from $t \approx - 5$ to $t \approx -25$. A measure for
    the extent of the universal domain is given by the Ginzburg scale
    Eq.~\eqref{eq:112} which here takes the value $t_G \approx -3.4$. Initial
    conditions for both (a) and (b) are $r_{K \Lambda} = r_{u \Lambda} = 10$,
    $r_{u_3 \Lambda} = 1$, $w_{\Lambda} \approx 0.05810, \tilde{\kappa} = 0.5,$
    and $\tilde{\kappa}_3 = 0.01$.}
  \label{fig:c1_c2_w_kappaT_kappa3T_flow}
\end{figure}

When such a near-critical trajectory approaches the scaling solution, the
couplings $\mathbf{s}$ flow towards their fixed point values $\mathbf{s}_{*}$ on
a scale $1/\Re s_{2,3} \approx 1$ determined by the eigenvalues $s_{2,3}$ of the
stability matrix $S$, cf.\ Fig.~\ref{fig:c1_c2_w_kappaT_kappa3T_flow}, and stay
there for a long ``time'' $t_s$. Depending on how close $w_{\Lambda}$ is to
$w_c$, this duration is typically $t_s = 10$ to 20 which corresponds to several
orders of magnitude in $k/\Lambda$. During $t_s$ the ratios $\mathbf{r}$ decay
according to Eq.~\eqref{eq:100}, i.e., as the sum of three exponentials, with
decay rates given by the eigenvalues Eq.~\eqref{eq:98} of the stability matrix
$N$. In order to extract these eigenvalues from the numerical solution, we
consider the flow of the coefficients $c_i \sim e^{n_i t}$ in the expansion of
$\mathbf{r}$ in the basis of eigenvectors of $N$
Eq.~\eqref{eq:100}. Figure~\ref{fig:c1_c2_w_kappaT_kappa3T_flow} shows $c_{1,2}$
along with exponential fits, which reproduce the eigenvalues $n_{1,2}$ to
satisfactory accuracy.
\begin{figure}
  \centering
  \includegraphics{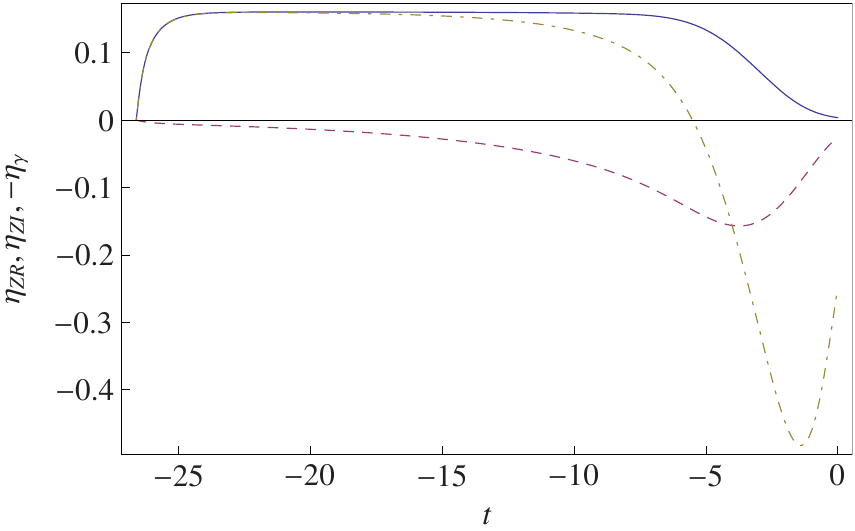}  
  \caption{(Color online) Anomalous dimensions $\ezr$ (solid), $\ezi$ (dashed),
    and $-\eta_{\gamma}$ (dot-dashed) for the solution of
    Fig.~\ref{fig:c1_c2_w_kappaT_kappa3T_flow}. From $t \approx -5$ to $t
    \approx -25$, where the values of $\mathbf{s}$ are close to the scaling
    solution, $\ezr$ takes the constant value Eq.~\eqref{eq:97}, while $\ezi$
    decays to zero. The value of $-\eta_{\gamma}$ approaches the one of $\ezr$
    so that Eq.~\eqref{eq:97} is satisfied at late ``times'' $t$. Eventually, as
    the trajectory is driven away from the fixed point and enters the symmetric
    phase with $w = 0$, the anomalous dimensions drop to zero.}
  \label{fig:flow_int_anom_dims}
\end{figure}

An important result of the previous section is the scaling relation
Eq.~\eqref{eq:97} between the anomalous dimensions $\eta_Z$ and $\eta_{\gamma}$
of the wave-function renormalization and the Keldysh mass respectively, which
implies that when evaluated along a critical trajectory, the value of $-
\eta_{\gamma}$ approaches the one of the real part $\ezr$ of $\eta_Z$, while the
imaginary part $\ezi$ flows to zero. This prediction -- physically implying
asymptotic thermalization -- is verified numerically in
Fig.~\ref{fig:flow_int_anom_dims}.

\begin{figure}
  \centering
  \includegraphics{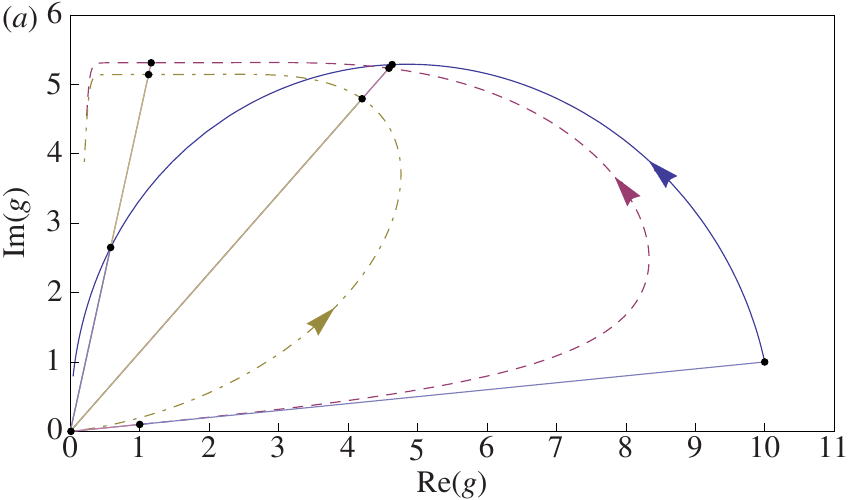}
  \includegraphics{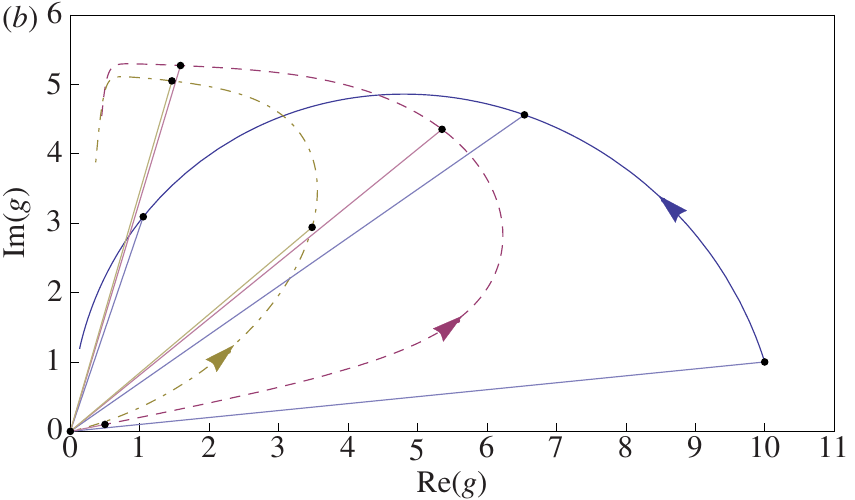}    
  \caption{(Color online) Equilibrium vs.\ non-equilibrium flow: (a) As
    discussed in Sec.~\ref{sec:geom-interpr-equil}, in thermodynamic equilibrium
    all couplings lie on a single ray in the complex plane. (b) This geometric
    constraint is absent out-of-equilibrium. We show $g = 10 K, \tilde{u},$ and
    $10^{-1} \tilde{u}_3$ as solid, dashed, and dot-dashed lines
    respectively. Stages of the flow at $t = 0, -8,$ and $-16$ are indicated
    with points on the trajectories. Initial values are (a) $r_{\Lambda} = 10,
    w_{\Lambda} = 0.01281$ and (b) $r_{K \Lambda} = 10, r_{u \Lambda} = 5,
    r_{u_3 \Lambda} = 1, w_{\Lambda} = 0.01264$. In both cases we have
    $\tilde{\kappa}_{\Lambda} = 0.1, \tilde{\kappa}_{3 \Lambda} = 0.01,$ and
    $K_{\Lambda} = 1 + i 0.1$.}
  \label{fig:mechanism}
\end{figure}

As the anomalous dimensions $\eta_a$ of $a = Z,D,$ and $\gamma$ are
functions of the renormalized dimensionless couplings $\mathbf{r}$ and
$\mathbf{s}$ alone and not the quantities $a$ themselves, we get the
solutions to the flow equations $\partial_t a = - \eta_a a$
simply by exponentiating the integrals of the anomalous dimensions along RG
trajectories $\mathbf{r}(t)$ and $\mathbf{s}(t)$, i.e.,
\begin{equation}
  \label{eq:111}
  a(t) = a_{\Lambda} e^{-\int_0^t d t' \eta_a}.
\end{equation}
In this way we obtain the trajectories of $K$ shown in Fig.~\ref{fig:mechanism}
and the flow of the effective temperature $\teff = \bar{\gamma}/\left( 4 \abs{Z}
\right) = \gamma \abs{Z}/4$ which according to the discussion in
Sec.~\ref{sec:fluct-diss-theor} at low frequencies saturates to a constant
value as illustrated in Fig.~\ref{fig:teff}. While this asymptotic
value depends on the initial values of $\bar{\gamma}$ and $Z$ at the scale
$\Lambda$ and is, therefore, non-universal, the manner in which it is
approached \emph{is} universal as it is determined by the exponent $\eta_r$:
According to Eq.~\eqref{eq:111} the flow of $\teff$ is given by
\begin{equation}
  \label{eq:211}
  \teff(t) = T_{\mathrm{eff} \Lambda} e^{-\int_0^t dt' \left(\ezr+ \eta_{\gamma} \right)}.
\end{equation}
Close to the fixed point we may expand the anomalous dimensions in the
exponential in powers of $\delta \mathbf{r} = \mathbf{r}$ and $\delta
\mathbf{s}$. As both $\ezr$ and $\eta_{\gamma}$ are even functions of
$\mathbf{r}$ there is no linear term in the expansion and we may write for $a =
\mathit{ZR}$ and $\gamma$ (here we
are indicating the anomalous dimension evaluated at the fixed point explicitly
as $\eta_{a *}$):
\begin{equation}
  \label{eq:213}
  \eta_a = \eta_{a *} + \frac{1}{2} \mathbf{r} \cdot \left[
    \nabla_{\mathbf{r}}^T \nabla_{\mathbf{r}} \eta_a \right]_{\mathbf{r} = \delta
    \mathbf{s} = \mathbf{0}} \mathbf{r},
\end{equation}
where we are neglecting corrections that are quartic in $\abs{\mathbf{r}}$ or
contain mixed powers of $\abs{\mathbf{r}}$ and $\abs{\delta \mathbf{s}}$. Both
types of corrections are small as compared to the leading contribution that is
quadratic in $\abs{\mathbf{r}}$: In the scaling regime we have $\abs{\mathbf{r}}
\sim e^{-\eta_{r *} t} \gg \abs{\delta \mathbf{s}} \sim e^{\Re s_{2,3} t}$,
where $s_{2,3}$ are eigenvalues with positive real parts $\Re s_2 = \Re s_3$ of
the stability matrix $S$ Eq.~\eqref{eq:92} and determine the leading corrections
to scaling in the static sector. Therefore, neglecting exponentially small
corrections, we have
\begin{equation}
  \label{eq:214}
  \eta_a = \eta_{a *} + \eta_a'' e^{-2 \eta_{r *} t}.
\end{equation}
Note that the quantities $\eta_a''$ depend on the precise prefactor in the
scaling form Eq.~\eqref{eq:101} of $\mathbf{r}$, i.e., they depend on
microscopic parameters and are thus non-universal. Then, using
Eq.~\eqref{eq:214} and keeping in mind that $\eta_{\gamma *} = -
\eta_{\mathit{ZR} *}$, we find the asymptotic behavior
\begin{equation}
  \label{eq:215}
  \teff(t) = T_{\mathrm{eff} 0} \left( 1 + \frac{\ezr'' + \eta_{\gamma}''}{2 \eta_{r*}} e^{-2
      \eta_{r*} t} \right),
\end{equation}
where in the last line we are again dropping exponentially small
corrections. This form confirms the physical intuition that long-wavelength
thermalization of the DDM is governed by the exponent that is unique to this
model and manifestly witnesses the microscopic non-equilibrium nature of this
model. We finally note that the effective temperature defined in
Eq.~\eqref{eq:60} is not the one that enters the FDT Eq.~\eqref{eq:58} for
MAR. The latter can be established by means of the basis transformation
Eq.~\eqref{eq:46} which involves the parameter $\bar{r}$. This parameter,
however, is characteristic the of MAR and has no counterpart in the DDM.

The near-critical trajectories we consider in this section illustrate the
concept of universality in that they show how details of the microscopic model,
which determine the initial conditions of the RG flow, are lost as we lower $k
\to 0$, where all of these trajectories converge towards the scaling solution,
cf.\ Fig.~\ref{fig:universality}. However, a distinctly non-universal feature of
these trajectories is the point where the crossover to the universal regime
takes place, which is known as the Ginzburg
scale.\cite{Amit/Martin-Mayor,Kleinert/Schulte-Frohlinde,Zinn-Justin}
Physically, the Ginzburg scale marks the breakdown of mean-field theory as we
approach the fluctuation-dominated critical region. In a perturbative estimate
in the symmetric phase, we compare the bare distance from the phase transition
$\kappa_1$ to the corresponding one-loop correction. Demanding these quantities
to be of the same order of magnitude
yields\cite{sieberer13:_dynam_critic_phenom_driven_dissip_system}
\begin{equation}
  \label{eq:112}
  \kappa_{1G} = \frac{1}{D_{\Lambda}^3} \left( \frac{\gamma_{\Lambda}
      \kappa_{\Lambda}}{2 C} \right)^2,
\end{equation}
where $C$ is a numerical constant (we find $C = 2 \pi$ if we set the bare value
$\kappa_{1G}$ exactly equal to its one-loop correction). Expressing
$\kappa_{1G}$ through a momentum scale as $\kappa_{1G} = D_{\Lambda} k_G^2$ we
find Eq.~\eqref{eq:3}, and for the dimensionless RG ``time'' $t_G = \ln \left(
  k_G/\Lambda \right)$, in terms of the dimensionless two-body loss rate
$\tilde{\kappa}$ introduced in Sec.~\ref{sec:resc-flow-equat}, we have
\begin{equation}
  \label{eq:113}
  t_G = \ln \left( \tilde{\kappa}_{\Lambda}/C \right).
\end{equation}
Fitting this logarithmic dependence to numerically obtained trajectories in
Fig.~\ref{fig:universality}, we find $C \approx 14.8$. The Ginzburg scale
delimits also the region where the driven-dissipative system obeys a FDT and the
ratio $\teff = \bar{\gamma}/\left( 4 \abs{Z} \right)$ saturates to a constant
value as shown in Fig.~\ref{fig:teff}.

\section{Conclusions}
\label{sec:conclusions}

We have studied the nature of Bose criticality in driven open systems. To this
end, starting from a description of the microscopic physics in terms of a
many-body quantum master equation, we have developed and put into practice a FRG
approach based on a Keldysh functional integral reformulation of the quantum
master equation for the quantitative determination of the universality
class. The absence of both an exact particle number conservation and the
detailed balance condition were seen to underly the existence of a new and
independent critical exponent governing universal decoherence, while the
distribution function shows asymptotic thermalization despite the microscopic
driven nature of the system.

This work is just a first step in the exploration of non-equilibrium critical
behavior. Key questions for future studies concern the status of critical points
in lower dimensionality as, e.g., relevant for current exciton-polariton systems. In particular, in Ref.~\onlinecite{Altman2013} it has been shown that the thermal fixed point is unstable in two dimensions, and instead is replaced by the non-equilibrium Kardar-Parisi-Zhang \cite{kardar86} fixed point. It is also a key issue to investigate different
symmetries beyond the $O(2)$ case. For example, Heisenberg models realized with
ensembles of trapped ions may exhibit $O(3)$ symmetry.\cite{porras04}
Furthermore, given the fact that many light-matter systems are pumped coherently
as opposed to the incoherent pump considered here, it will be important to
understand the impact of the coherent drive on potential criticality in these
classes of systems. Finally, it is an intriguing question whether driven open
systems which realize non-equilibrium counterparts of quantum criticality can be
identified. In the long run, it remains to be seen whether a classification of
non-equilibrium criticality with similarly clear structure as familiar from
equilibrium dynamical criticality\cite{hohenberg77:_theor} can be reached.

\section*{Acknowledgments}

We thank J. Berges, I. Boettcher, M. Buchhold, I. Carusotto, T. Gasenzer,
S.\,G. Hofer, A. Imamoglu, J.\,M. Pawlowski, A. Rosch, U.\,C. T\"auber,
A. Tomadin, C. Wetterich, and P. Zoller for stimulating discussions. This work
was supported by the Austrian Science Fund (FWF) through the START Grant No. Y
581-N16, the SFB FoQuS (FWF Project No. F4006- N16) (LMS, SD), and the ISF under
Grant No. 1594/11 (EA).

\appendix
\section{Markovian dissipative action}
\label{sec:mark-diss-acti}

\subsection{Translation table: Master equation
  vs. Keldysh functional integral }
\label{sec:transl-tabl-mast}

Here we specify the relation between second quantized master equation and the
equivalent Keldysh functional integral, defined with a markovian dissipative
action. In particular, we review how the presence of external driving underlies
the validity of the master equation and markovian dissipative action. We start
from a master equation governing the time evolution of a system density matrix,
\begin{equation}
  \label{eq:mastergen}
  \partial_t \hat{\rho} = -i \left[ \hat{H}_s,\hat{\rho} \right] +
  \kappa \left( \hat{L} \hat{\rho}
    \hat{L}^{\dagger} - \frac{1}{2} \left\{ \hat{L}^{\dagger}
      \hat{L}, \hat{\rho} \right\} \right).
\end{equation}
Here, $\hat{H}_s$ is a system Hamiltonian generating the unitary evolution and
$\hat{L}$ is a Lindblad operator making up the dissipative part of the
Liouvillian. For simplicity we consider only a single dissipative channel. The
generalization to several channels as in Eq.~\eqref{eq:6}, realized through the
coupling to several baths, is straightforward. Equation~\eqref{eq:mastergen}
results from a more general system-bath setting, $\hat{H}_{\mathrm{tot}} =
\hat{H}_s + \hat{H}_b + \hat{H}_{sb}$ ($\hat{H}_b$ and $\hat{H}_{sb}$ are a
quadratic bath Hamiltonian with a continuum of frequencies and a system-bath
Hamiltonian linear in the bath operators, respectively) under the following
three assumptions: (i) the system-bath coupling $\sqrt{\gamma(\omega)}$ is weak
compared to a typical scale $\omega_0$ in the system (or, by energy
conservation, in the bath) indicating, e.g., the level spacing in an atom (Born
approximation $\gamma(\omega)/\omega_0\ll 1$) (ii) the frequency dependence of
the system-bath coupling is negligible over the bandwidth $\vartheta$ of the
bath centered around $\omega_0$, implying $\delta$-correlations in the time
domain (Markov approximation $\gamma(\omega) \approx \text{const.}$), and (iii)
the system is \emph{driven with an external field with frequency $\nu$} to
bridge the large energy separation of the levels, $\left( \nu - \omega_0
\right)/\left( \nu + \omega_0 \right) \ll 1$. This makes it possible to work in
the rotating wave approximation, in which only the detuning $\Delta = \nu -
\omega_0$ occurs as a physical scale, while all fast terms involving $\nu +
\omega_0$ are dropped. From this consideration, it is clear that the master
equation is an accurate description of strongly driven systems coupled to an
environment. A typical realization in quantum optics is an atom with two
relevant levels separated by $\omega_0$, connected by an external laser drive
with frequency $\nu$, which is detuned from resonance by $\Delta = \nu -
\omega_0$. Only the laser drive makes the excited level accessible and gives
rise to two-level dynamics such as Rabi oscillations, with frequency determined
by the laser intensity. The excited level is unstable and can undergo
spontaneous emission by coupling to the radiation field, providing for the
reservoir -- this mechanism is physically completely independent of the coherent
dynamics. Alternatively but fully equivalent to the operator formalism, the
above approximations can be performed in a Keldysh path integral setting (see
below). In this way, the physics of a given quantum master equation becomes
amenable to quantum field theoretical approaches, which is particularly useful
for bosonic and fermionic driven-dissipative many-body systems. Here the
starting point is the Keldysh partition function
\begin{equation}
  \label{eq:114}
  \mathcal{Z} = \int \mathcal{D}[a^{*},a,b^{*},b] e^{- i \s_{\mathrm{tot}}[a^*,a,b^*,b]},
\end{equation}
which results from a ``Trotterization'' of the Hamiltonian dynamics (after
normal ordering) acting on the density matrix in the integrated form of the von
Neumann equation in the basis of coherent states; in this process, the second
quantized system and bath field operators, $\hat{a}_i$ (the index $i$ denotes
both position and internal indices, such as different particle species) and
$\hat{b}_{\mu}$ ($\mu$ labels the bath modes and will be chosen a continuous
index below) respectively, are mapped to time-dependent complex valued fields in
the action
\begin{multline}
  \label{eq:115}
  \s_{\mathrm{tot}} = \sum_{\sigma = \pm} \sigma \int d t \left( \sum_i
    a^*_{i,\sigma}(t) i \partial_t a_{i,\sigma}(t) \right. \\
  \left. \vphantom{\sum_i} + \sum_{\mu} b^*_{\mu,\sigma}(t) i \partial_t
    b_{\mu,\sigma}(t) - H_{\mathrm{tot},\sigma}(t) \right),
\end{multline}
where $H_{\mathrm{tot},\sigma}(t)$ is a quasilocal polynomial of these
fields. The relative minus sign for the evolution on the forward (+) and
backward (-) contours clearly reflects the commutator structure in the von
Neumann equation of motion for the system-bath density operator above. We have
omitted an imaginary regularization term ensuring convergence of the functional
integral\cite{Altland/Simons,kamenev09:_keldy,kamenevbook} for simplicity, as it
does not affect any of the next steps. Integrating out the harmonic bath
variables using approximations (i) -- (iii) and considering for the moment
Lindblad operators $\hat{L}$ which are linear in the system field operators, we
arrive at the following effective Markovian dissipative action:
\begin{multline}
  \label{eq:116}
  \s = \sum_{\sigma} \sigma \int d t \left( \sum_i a^*_{i,\sigma}(t)
    i \partial_t a_{i,\sigma}(t) - H_{s,\sigma}(t) \right) \\ - i \kappa \left[
    L_{+}(t) L_{-}^{*}(t) - \frac{1}{2} \left( L_{+}^{*}(t) L_{+}(t) +
      L_{-}^{*}(t) L_{-}(t) \right) \right].
\end{multline}
While the relative minus sign for the system Hamiltonian $H_s$ on the + and -
contours preserve the commutator structure, the dissipative terms clearly
reflect the temporally local Lindblad structure of Eq.~\eqref{eq:mastergen}. We
thus arrive at a simple translation rule for bosonic\footnote{For fermions,
  additional signs arise due to the Grassmann nature of the fermion field, but a
  similar translation table exists.} master equations into the corresponding
Keldysh functional integral: (i) the temporal derivative terms can be read off
from the last equation; (ii) for all (normal ordered) operators on the right
(left) of the density matrix, introduce a contour index + (-) and write down the
Markovian dissipative action. The linear Lindblad operators we consider here are
not affected by normal ordering. For the more general case of Lindblad operators
that are quasilocal polynomials in the system field operators, operator ordering
can be tracked by a suitable temporal regularization procedure as elaborated in
the next section.

\subsection{Derivation in the Keldysh setting}
\label{sec:deriv-keldysh-sett}

Here we present a derivation of the Markovian dissipative action in the $\pm$
basis for arbitrary (non-linear) Lindblad jump operators, which allows for the
most direct comparison with the master equation. In particular, we pay special
attention to the question how the operator ordering in the master equation is
reflected in the path integral formulation. We leave the system action
unspecified, requiring only the property that after proper rotating frame
transformation the evolution of the system is much slower than the correlation
time of the bath $\tau_c = 1/\vartheta$ (broadband bath). The action of the bath
is, in the $\pm$ basis,
\begin{multline}
  \s_b = \sum_{\mu} \int dt dt' \left( b_{\mu,+}^{*}(t),
    b_{\mu,-}^{*}(t) \right) \\ \times
  \begin{pmatrix}
    ~G_{\mu}^{++}(t,t') &  G_{\mu}^{+-}(t,t') \\
    ~G_{\mu}^{-+}(t,t') & G_{\mu}^{--}(t,t')
  \end{pmatrix}^{-1}
  \begin{pmatrix}
    b_{\mu,+}(t') \\ b_{\mu,-}(t')
  \end{pmatrix}.
\end{multline}
The Green's functions for the oscillators of the bath are assumed to be in
thermal equilibrium and read
\begin{equation}
  \label{eq:117}
  \begin{split}
    G_{\mu}^{+-}(t,t') & = -i\bar{n}(\omega_{\mu}) e^{-i\omega_{\mu} \left( t-t'
      \right)}, \\
    G_{\mu}^{-+}(t,t') & = -i \left( \bar{n}(\omega_{\mu}) +1 \right)
    e^{-i\omega_{\mu}\left( t-t' \right)}, \\ G_{\mu}^{++}(t,t') & = \theta (t -
    t')
    G_{\mu}^{-+}(t,t') + \theta (t' - t) G_{\mu}^{+-}(t,t'),\\
    G_{\mu}^{--}(t,t') & = \theta (t' - t) G_{\mu}^{-+}(t,t') + \theta (t - t')
    G_{\mu}^{+-}(t,t').
  \end{split}
\end{equation}
The linear coupling between system and the bath is (note that the case of
several dissipative channels and local baths as in Eq.~\eqref{eq:6} can be
implemented by adding appropriate indices to the $L_{\sigma}$ and
$b_{\mu,\sigma}$ and summing over these indices)
\begin{multline}
  \label{eq:118}
  \s_{sb} = \sum_{\mu} \sqrt{\gamma_{\mu}} \int dt \left( L_{+}^{*}(t)
    b_{\mu,+}(t) \right. \\ \left. + L_{+}(t) b_{\mu,+}^{*}(t) - L_{-}^{*}(t)
    b_{\mu,-}(t) - L_{-}(t) b_{\mu,-}^{*}(t) \right),
\end{multline}
where $L_{\pm}$ correspond to the quantum jump operators which are typically
quasilocal polynomials of the system's creation and annihilation operators. To
be consistent with the derivation of the path integral, we require the jump
operators to have been normal ordered before the Trotter decomposition giving
rise to the path integral. The partition function is of the general form
\begin{equation}
  \label{eq:119}
  \mathcal{Z} = \int \mathcal{D}[a^{*},a,b^{*},b] e^{i \left( \s_s[a^{*},a] + \s_b[b^{*},b] +
      \s_{sb}[a^{*},a,b^{*},b] \right)},
\end{equation}
Now we integrate out the bath via completion of the square which results in an
effective action $\s_{\mathrm{eff}}$ for the system degrees of freedom. The
contribution $\s_{{\mathrm{eff}},\mu} $ of the $\mu$th mode to the effective
action reads
\begin{multline}
  \label{eq:120}
  \s_{\mathrm{eff},\mu} = \gamma_{\mu} \int dt dt' \left( L_{+}^{*}(t), -
    L_{-}^{*}(t) \right) \\ \times
  \begin{pmatrix}
    G_{\mu}^{++}(t,t') & G_{\mu}^{+-}(t,t') \\
    G_{\mu}^{-+}(t,t') & G_{\mu}^{--}(t,t')
  \end{pmatrix}
  \begin{pmatrix}
    L_{+}(t')\\ - L_{-}(t')
  \end{pmatrix}.
\end{multline}
The signs for the operators on the $-$ contour comes from the backward
integration in time. Thus the mixed terms will occur with an overall $-$ sign,
while the $++$ and $--$ terms come with an overall $+$. Summing over all the
modes $\mu$ we obtain the effective action for the field variables of the
subsystem due to the coupling to the bath. We now take the continuum limit of
densely lying bath modes, centered around some central frequency $\omega_0$ and
with bandwidth $\vartheta$. That is, we substitute the sum over the modes with
an integral in the energy $\Omega$ weighted by a (phenomenologically introduced)
density of states $\nu(\Omega)$ of the bath
$  \sum_{\mu}\gamma_\mu \simeq \int_{0}^{\infty} d\Omega \gamma(\Omega)
\nu(\Omega)$, and obtain
\begin{multline}
  \label{eq:121}
  \s_{\mathrm{eff}} = - \int_{\omega_0-\vartheta}^{\omega_0 + \vartheta} d\Omega
  \gamma(\Omega) \nu(\Omega)\int dt d\tau \left( L_{+}^{*}(t), - L_{-}^{*}(t) \right) \\
  \times 
  \begin{pmatrix}
    G_{\Omega}^{++}(\tau) & G_{\Omega}^{+-}(\tau) \\
    G_{\Omega}^{-+}(\tau) & G_{\Omega}^{--}(\tau)
  \end{pmatrix}
  \begin{pmatrix}
    L_{+}(t-\tau )\\ -L_{-}(t-\tau)
  \end{pmatrix},
\end{multline}
where in addition we have used the translation invariance of the bath Green's
function, $G_\Omega^{\alpha\beta}(t,t') = G_\Omega^{\alpha\beta}(t - t')$ to
suitably shift the integration variables. We consider the various terms
separately. In doing the Markov approximation, we use (a) that by assumption it
is possible to choose a rotating frame in which the evolution of the system is
slow compared to the scales in the bath, $\omega_{\mathrm{sys}} \ll
\omega_0,\vartheta$. In this case, a zeroth order temporal derivative
approximation for the jump operators is appropriate. This gives rise to a
\emph{temporally local} form of the markovian dissipative action. However, for
the evaluation of tadpole diagrams for this action, ambiguities due to a
temporally local vertex arises. In these diagrams -- and only in these -- it is
then important to specify the proper regularization of the system's Green's
function at equal time arguments. To keep track of this, we indicate the sign of
the next time step in the approximated jump operators by $t_{\pm\delta} = t \pm
\delta t$. In step (b) below, we assume that the density of states and the
coupling of the system to bath are well approximated as constant over the
relevant reservoir width,
\begin{widetext}
  \begin{multline}
    \label{eq:markovcalc}
    - \int dt L_{+}^{*}(t) \int d\tau \int_{\omega_0-\vartheta}^{\omega_0 +
      \vartheta} \frac{d\Omega}{2\pi} \gamma(\Omega) \nu(\Omega)
    G_{\Omega}^{+-}(\tau) L_{-}(t-\tau) = i \int dt L_{+}^{*}(t) \int d\tau
    \int_{\omega_0-\vartheta}^{\omega_0 + \vartheta} \frac{d\Omega}{2\pi}
    \gamma(\Omega) \nu(\Omega) \bar n(\Omega)e^{-i\Omega \tau} L_{-}(t-\tau) \\
    \begin{aligned}
      & \stackrel{\text{(a)}}{\approx} i \int dt L_{+}^{*}(t) \int d\tau
      \int_{\omega_0-\vartheta}^{\omega_0 + \vartheta} \frac{d\Omega}{2\pi}
      \gamma(\Omega) \nu(\Omega) \bar{n}(\Omega)e^{-i\Omega \tau}
      L_{-}(t_{-\delta}) \stackrel{\text{(b)}}{\approx} i \int dtL_{+}^{*}(t)
      \gamma \nu \int d\tau \int_{\omega_0-\vartheta}^{\omega_0 + \vartheta}
      \frac{d\Omega}{2\pi} \bar{n}(\Omega)e^{-i\Omega \tau} L_{-}(t_{-\delta})\\
      & \approx i \int dtL_{+}^{*}(t) \gamma \nu\int_{-\infty}^{\infty} d\Omega
      \bar{n}(\Omega)\delta (\Omega - \omega_0) L_{-}(t_{-\delta}) = i \kappa
      \bar{n}\int dtL_{+}^{*}(t) L_{-}(t_{-\delta}),
    \end{aligned}
  \end{multline}
  where we have shifted the frequency integration domain by $-\omega_0$ and taken
  the limit $\vartheta \to \infty$, as well as $\kappa = \gamma \nu$ and $\bar{n}
  = \bar{n} (\omega_0)$. Further note the relation to the operator formalism
  $\int_{-\infty}^{\infty} \frac{d\Omega}{2\pi} \bar{n}(\Omega)e^{-i\Omega \tau} =
  \langle \hat b^{\dagger}(\tau) \hat b (0)\rangle $. Similarly,
  \begin{equation}
    \label{eq:122}
    - \int dtL_{-}^{*}(t) \int d\tau \int_{\omega_0-\vartheta}^{\omega_0 +
      \vartheta} \frac{d\Omega}{2\pi} \gamma(\Omega) \nu(\Omega)
    G_{\Omega}^{-+}(\tau) L_{+}(t-\tau) \approx i \kappa (\bar{n} + 1)\int dt
    L_{-}^{*}(t) L_{+}(t_{-\delta})
  \end{equation}
  and $\int_{-\infty}^{\infty}
  \frac{d\Omega}{2\pi} (\bar{n}(\Omega) + 1) e^{-i\Omega \tau} = \langle \hat
  b(\tau) \hat b^{\dagger} (0) \rangle $. For the terms on the forward
  contour, we obtain
  \begin{multline}
    \label{eq:123}
    \int dt L_{+}^{*}(t) \int d\tau \int_{\omega_0-\vartheta}^{\omega_0 +
      \vartheta} \frac{d\Omega}{2\pi} \gamma(\Omega) \nu(\Omega)
    G_{\Omega}^{++}(\tau) L_{+}(t-\tau) \\
    \begin{aligned}
      & = - i \int dt L_{+}^{*}(t) \int d\tau
      \int_{\omega_0-\vartheta}^{\omega_0 + \vartheta} \frac{d\Omega}{2\pi}
      \gamma(\Omega) \nu(\Omega) \left[ \theta (\tau) \left( \bar{n}(\Omega) + 1
        \right) +
        \theta (-\tau) \bar{n}(\Omega) \right] e^{-i\Omega \tau} L_{+}(t-\tau) \\
      & \stackrel{\text{(a)}}{\approx} - i \int dt L_{+}^{*}(t) \int d\tau
      \int_{\omega_0-\vartheta}^{\omega_0 + \vartheta} \frac{d\Omega}{2\pi}
      \gamma(\Omega) \nu(\Omega) \left[ \theta (\tau) \left( \bar{n}(\Omega) + 1
        \right) + \theta (-\tau) \bar{n}(\Omega) \right] e^{-i\Omega \tau}
      L_{+}(t_{-\delta}) \\ & \stackrel{\text{(b)}}{\approx} - i \int
      dtL_{+}^{*}(t) \gamma \nu \left[ \int d\tau \theta(\tau)
        \int_{\omega_0-\vartheta}^{\omega_0 + \vartheta} \frac{d\Omega}{2\pi}
        \left( \bar{n}(\Omega) + 1 \right) e^{-i\Omega \tau} + \int d\tau
        \theta(-\tau) \int_{\omega_0-\vartheta}^{\omega_0 + \vartheta}
        \frac{d\Omega}{2\pi} \bar{n}(\Omega) e^{-i\Omega \tau} \right]
      L_{+}(t_{-\delta}) \\ & \approx - i \int dt \left\{ \left[ \frac{1}{2}
          \kappa \left( \bar{n} +1 \right) - i \delta E_1 \right] L_{+}^{*}(t)
        L_{+}(t_{-\delta}) + \left( \frac{1}{2} \kappa \bar{n} + i \delta E_2
        \right) L_{+}^{*}(t) L_{+}(t_{+\delta}) \right\}.
    \end{aligned}
  \end{multline}
\end{widetext}
In the last line we have used
\begin{multline}
  \label{eq:124}
  \int d\tau \theta(\tau)
  \int_{\omega_0-\vartheta}^{\omega_0 + \vartheta} \frac{d\Omega}{2\pi} ( \bar
  n(\Omega) + 1) e^{-i\Omega \tau} L_{+}(t_{-\delta}) \\
  \begin{aligned}
    & \approx \int_{-\infty}^{\infty} \frac{d\Omega}{2\pi} \left(
      \bar{n}(\Omega) + 1 \right) \left( \pi \delta(\Omega- \omega_0) - i
      \mathcal P \frac{1}{\Omega- \omega_0 } \right) L_{+}(t_{-\delta}) \\ & =
    \left[ \frac{1}{2} \kappa \left(\bar{n} +1 \right) - i \delta E_1 \right]
    L_{+}(t_{-\delta})
  \end{aligned}
\end{multline}
and
\begin{multline}
  \label{eq:125}
  \int d\tau \theta( - \tau)
  \int_{\omega_0-\vartheta}^{\omega_0 + \vartheta} \frac{d\Omega}{2\pi} \bar
  n(\Omega) e^{-i\Omega \tau} L_{+}(t_{-\delta}) \\
  \begin{aligned}
    & = \int d\tau \theta( \tau) \int_{\omega_0-\vartheta}^{\omega_0 +
      \vartheta} \frac{d\Omega}{2\pi} \bar{n}(\Omega) e^{+ i\Omega \tau}
    L_{+}(t_{+\delta}) \\ & \approx \int_{-\infty}^{\infty} \frac{d\Omega}{2\pi}
    \bar n(\Omega) \left( \pi \delta(\Omega - \omega_0) - i \mathcal P
      \frac{1}{\Omega - \omega_0} \right) L_{+}(t_{+\delta})\\ & = \left(
      \frac{1}{2} \kappa \bar{n} + i \delta E_2 \right) L_{+}(t_{+\delta}).
  \end{aligned}
\end{multline}
Importantly, note the sign change in the regularization of the time argument
upon reversal of integration direction. This gives a hint which operator ``comes
first'' in the coarse grained evolution where the bath has been integrated out,
and reflects the fact that in the corresponding master equation, the ``cooling''
dissipation terms $\sim \left( \bar{n} + 1 \right)$ are normal ordered in the
jump operators ($\sim \hat L^{\dagger} \hat L$), while the ``heating'' terms
$\sim \bar{n} $ are anti-normal ordered ($\sim \hat L \hat
L^{\dagger}$). Similarly, we obtain on the backward contour,
\begin{multline}
  \label{eq:126}
  \int dt L_{-}^{*}(t) \int d\tau \int_{\omega_0-\vartheta}^{\omega_0 +
    \vartheta} \frac{d\Omega}{2\pi} \gamma(\Omega) \nu(\Omega)
  G_{\Omega}^{--}(\tau) L_{-}(t-\tau) \\ \approx - i \int dt \left\{
    \left[\frac{1}{2} \kappa \left( \bar{n} +1 \right) + i \delta E_1 \right]
    L_{-}^{*}(t) L_{-}(t_{+\delta}) \right. \\ \left. + \left( \frac{1}{2}
      \kappa \bar{n} - i \delta E_2 \right) L_{-}^{*}(t) L_{-}(t_{-\delta})
  \right\},
\end{multline}
where the changes in the signs relative to the forward term emerge from the
reverse signs in the $\theta$-functions. In summary, we obtain the following
dissipative contribution to the action:
\begin{multline}
  \label{eq:dissact}
  \s_d = i \kappa \int dt \left\{ \left( \bar{n} + 1
    \right) \left[ \vphantom{\frac{1}{2}} L_{-}^{*}(t) L_{+}(t_{-\delta}) \right. \right. \\
  \left. \left. - \frac{1}{2} \left( L_{+}^{*}(t) L_{+}(t_{-\delta}) +
        L_{-}^{*}(t) L_{-}(t_{+\delta}) \right) \right] \right. \\ \left. +
    \bar{n} \left[ L_{+}^{*}(t) L_{-}(t_{-\delta}) - \frac{1}{2} \left(
        L_{+}^{*}(t) L_{+}(t_{+\delta}) + L_{-}^{*}(t) L_{-}(t_{-\delta})
      \right) \right] \right\}.
\end{multline}
In addition, there is a ``Lamb shift'' which reads
\begin{multline}
  \label{eq:127}
  \s_L = - \int dt \left[ \delta E_1 \left(- L_{+}^{*}(t) L_{+}(t_{-\delta}) +
      L_{-}^{*}(t) L_{-}(t_{+\delta}) \right) \right. \\ \left. + \delta E_2 \left( L_{+}^{*}(t)
      L_{+}(t_{+\delta}) - L_{-}^{*}(t) L_{-}(t_{-\delta}) \right) \right].
\end{multline}
This gives a contribution to the coherent dynamics which has the same physical
origin as the dissipative dynamics. However, typically there is a dominant
independent Hamiltonian contribution, such that the effective Hamiltonian
parameters after the Lamb shift renormalization are properly regarded as
independent of the Liouvillian ones.

\section{Symmetry constraints on the action and truncation for MAR}
\label{sec:symm-constr-acti}

In this section we derive the action Eq.~\eqref{eq:48} and the truncation
Eq.~\eqref{eq:51} for MAR. Our starting point is the truncation
Eq.~\eqref{eq:34} appropriate for the driven-dissipative model on which we
impose invariance under the equilibrium symmetry transformation
Eq.~\eqref{eq:47}. This leads to Eq.~\eqref{eq:51} which reduces to the action
Eq.~\eqref{eq:48} when we set $k = k_{\cg}$.

In terms of the bare spinors $\bar{\Phi}_{\nu}$ the truncation for the DDM can
be written as
\begin{multline}
  \label{eq:128}
  \Gamma_k = \int_X \bar{\Phi}_q^{\dagger} \left[ \left( Z_R \sigma_z - i Z_I
      \id \right) i \partial_t \bar{\Phi}_c
    \vphantom{\frac{\bar{\mathcal{U}}_H}{\delta \bar{\Phi}_c^{*}}} \right. \\
  \left. - \frac{\delta \bar{\mathcal{U}}_H}{\delta \bar{\Phi}_c^{*}} + i
    \sigma_z \frac{\delta \bar{\mathcal{U}}_D}{\delta \bar{\Phi}_c^{*}} + i
    \frac{\bar{\gamma}}{2} \bar{\Phi}_q \right].
\end{multline}
We perform the change of basis Eq.~\eqref{eq:46} and obtain for the
contributions in the sum $\Gamma_k = \Gamma_{\mathrm{dyn},k} + \Gamma_{H,k} +
\Gamma_{D,k} + \Gamma_{\mathrm{reg},k}$ the expressions
\begin{align}
  \label{eq:129}
  \Gamma_{\mathrm{dyn},k} & = i \frac{\bar{r} Z_R + Z_I}{\zrcg - \bar{r} \zicg}
  \int_X \tilde{\Phi}_q^{\dagger} \sigma_z \partial_t \tilde{\Phi}_c,
  \\ \label{eq:130} \Gamma_{H,k} & = \frac{i}{\zrcg - \bar{r} \zicg} \int_X
  \tilde{\Phi}_q^{\dagger} \sigma_z \left( \bar{r} \frac{\delta
      \bar{\mathcal{U}}_D}{\delta \tilde{\Phi}_c^{*}} -
    \frac{\delta \bar{\mathcal{U}}_H}{\delta \tilde{\Phi}_c^{*}} \right), \\
  \Gamma_{D,k} & = - \frac{1}{\zrcg - \bar{r} \zicg} \int_X
  \tilde{\Phi}_q^{\dagger} \left( \frac{\delta \bar{\mathcal{U}}_D}{\delta
      \tilde{\Phi}_c^{*}} + \bar{r} \frac{\delta
      \bar{\mathcal{U}}_H}{\delta \tilde{\Phi}_c^{*}} \right),
\end{align}
and
\begin{multline}
  \label{eq:131}
  \Gamma_{\mathrm{reg},k} = \frac{i}{\zrcg - \bar{r} \zicg} \int_X
  \tilde{\Phi}_q^{\dagger} \left( \left( Z_R - \bar{r} Z_I \right) i \partial_t
    \tilde{\Phi}_c \vphantom{\frac{1 + \bar{r}^2}{\zrcg - \bar{r} \zicg}}
  \right. \\ \left. + \frac{1 + \bar{r}^2}{\zrcg - \bar{r} \zicg}
    \frac{\bar{\gamma}}{2} \tilde{\Phi}_q \right).
\end{multline}
Both $\Gamma_{\mathrm{dyn},k}$ and $\Gamma_{D,k}$ are symmetric under the
transformation Eq.~\eqref{eq:47}. Demanding the remaining contributions
$\Gamma_{H,k}$ and $\Gamma_{\mathrm{reg},k}$ to be invariant we find that a term
of the form of Eq.~\eqref{eq:130} is actually forbidden by the symmetry, i.e.,
we must have $\Gamma_{H,k} = 0$, which is satisfied for $\bar{\mathcal{U}}_H =
\bar{r} \bar{\mathcal{U}}_D$. For the regularization term
$\Gamma_{\mathrm{reg},k}$ we obtain the additional constraint
Eq.~\eqref{eq:52}. All these requirements are implemented in the truncation
Eq.~\eqref{eq:51} which is easily seen to reduce to Eq.~\eqref{eq:48} for $k =
k_{\cg}$.

If in addition to the equilibrium symmetry we demand invariance under complex
conjugation of the fields Eq.~\eqref{eq:50} as is the case for MA, we find the
condition $\Gamma_{\mathrm{dyn},k} = 0$. This is met for all $0 < k < k_{\cg}$
if $\bar{r} = - Z_I/Z_R$.

\section{Non-Equilibrium FRG flow equations}
\label{sec:non-eq-frg-flow-app}

Here we present details of the derivation of the non-equilibrium FRG flow
equations in Sec.~\ref{sec:non-eq-frg-flow}. To start with, we rewrite the flow
equation~\eqref{eq:32} such that only renormalized quantities appear on the RHS,
\begin{equation}
  \label{eq:132}
  \partial_t \Gamma_k = \frac{i}{2} \Tr \left[ \left( \Gamma_k^{(2)} +
      R_k \right)^{-1} \dtt R_k \right].
\end{equation}
The second functional derivatives appearing under the trace on the RHS are taken
with respect to renormalized real fields Eq.~\eqref{eq:38}. These can be written in
terms of the bare ones as $\chi(Q) = z \bar{\chi}(Q)$, where the matrix $z$ is
given by
\begin{equation}
  \label{eq:133}
  z = \id \oplus 
  \begin{pmatrix}    
    Z_R & - Z_I \\
    Z_I & Z_R
  \end{pmatrix}.
\end{equation}
The linear transformation from bare to renormalized fields implies for
functional derivatives the relations
\begin{gather}
  \label{eq:134}
  \Gamma_k^{(\bar{2})}= z^T \Gamma_k^{(2)} z, \quad \bar{R}_k = z^T R_k z,
\end{gather}
and inserting these in the flow equation~\eqref{eq:32} yields Eq.~\eqref{eq:132} if
in addition we replace the derivative with respect to $t$ by the differential
operator $\dtt$ which is defined as
\begin{equation}
  \label{eq:135}
  \dtt \equiv \partial_t R_{k,\bar{K}} \partial_{R_{k,\bar{K}}}
  + \partial_t R_{k,\bar{K}}^{*} \partial_{R_{k,\bar{K}}^{*}}.
\end{equation}
With this definition we may write $\partial_t \bar{R}_k = \dtt \bar{R}_k$, which
has the advantage that $\dtt$ commutes with the multiplicative renormalization
with $Z$ (note that also $Z$ is a running coupling and depends on $t$), i.e., we
have
\begin{equation}
  \label{eq:136}
  \dtt \bar{R}_k = \dtt \left( z^T R_k z \right) = z^T \left( \dtt R_k \right) z.
\end{equation}
Furthermore, since $\dtt$ acts only on the cutoff and not the inverse propagator
$\Gamma_k^{(2)}$, we may rewrite the exact flow equation~\eqref{eq:132} in the
simple form
\begin{equation}
  \label{eq:137}  
  \partial_t \Gamma_k = \frac{i}{2} \Tr \dtt \ln \left( \Gamma_k^{(2)} + R_k \right).
\end{equation}

\subsection{Expansion in fluctuations}
\label{sec:expans-fluct}

According to its definition in Sec.~\ref{sec:effective-action}, the effective
action is a functional of the field expectation values, and also the flow
equation~\eqref{eq:137} can be evaluated for arbitrary field configurations. A
particularly useful form of the flow equation can be obtained by decomposing the
fields into homogeneous and frequency- and momentum-dependent fluctuation parts
as $\chi(Q) = \chi \delta(Q) + \delta \chi(Q)$ and expanding the logarithm on
the RHS of Eq.~\eqref{eq:137} to second order in the fluctuations $\delta
\chi(Q)$. Then, the zeroth order term determines the flow of the
momentum-independent couplings whereas the $\beta$-functions for the
wave-function renormalization and the gradient coefficient can be obtained from
the second order contribution.

We begin by deriving an explicit expression for the full inverse propagator
$\Gamma_k^{(2)}$ up to second order in $\delta \chi$. To this end we rewrite the
effective action Eq.~\eqref{eq:36} in the form
\begin{equation}
  \label{eq:138}
  \Gamma_k = \frac{1}{2} \int_Q \chi(-Q)^T D(Q) \chi(Q) - \int_X V,
\end{equation}
where $\int_Q = \int \frac{d \omega d^d \mathbf{q}}{\left( 2 \pi \right)^{d +
    1}}$. The frequency- and momentum-dependent part of the inverse propagator
Eq.~\eqref{eq:37} is denoted by $D(Q) = P(Q) - P(0)$, and the effective
potential $V$ that contains all momentum-independent couplings is given by
\begin{equation}
  \label{eq:139}
  V = U' \rho_{cq} + U^{\prime *} \rho_{qc} - i \gamma \rho_q.
\end{equation}
The second functional derivative of the effective action can then be expressed
as the sum of two contributions,
\begin{equation}
  \label{eq:140}
  \Gamma_k^{(2)}(Q,Q') = D(Q) \delta(Q - Q') - \mathcal{V}^{(2)}(Q,Q'),
\end{equation}
where the second term is just the functional derivative of the effective
potential,
\begin{equation}
  \label{eq:141}
  \mathcal{V}^{(2)}_{ij}(Q,Q') = \frac{\delta^2}{\delta \chi_i(-Q) \delta
    \chi_j(Q')} \int_X V = \int_X e^{i \left( Q - Q' \right) X} V^{(2)}_{ij},
\end{equation}
which can be reduced to ordinary (i.e., not functional) partial derivatives with
respect to the fields in the time domain and real space,
\begin{equation}
  \label{eq:142}
  V^{(2)}_{ij} = \frac{\partial^2}{\partial \chi_i \partial \chi_j} V.
\end{equation}
Setting the fluctuation components of the fields to zero in Eq.~\eqref{eq:140}
we obtain the inverse propagator in the presence of homogeneous classical and
quantum background fields,
\begin{equation}
  \label{eq:143}
  P_{cq}(Q) \delta(Q - Q') = \Gamma_k^{(2)}(Q,Q') \bigr\rvert_{\delta \chi = 0} =
  \left( D(Q) - V^{(2)}_{cq} \right) \delta(Q - Q').
\end{equation}
Note that the difference between $P_{cq}(Q)$ and the inverse propagator
Eq.~\eqref{eq:37} is that in the latter the background fields are set to their
stationary values while in the former they remain unspecified. The background
fields are all contained in the second contribution $V_{cq}^{(2)}$ which we
split into $2 \times 2$ blocks according to
\begin{equation}
  \label{eq:144}
  V_{cq}^{(2)} =
  \begin{pmatrix}
    V_{cq}^{(2) H} & V_c^{(2) A} \\
    V_c^{(2) R} & V^{(2) K}
  \end{pmatrix}.
\end{equation}
While the upper left block $V_{cq}^{(2) H}$ is linear in the quantum fields
(and, therefore, vanishes when we set these to zero, giving rise to the
causality structure of the inverse propagator Eq.~\eqref{eq:37}),
\begin{widetext}
  \begin{equation}
    \label{eq:145}
    \begin{split}
      V_{cq,11}^{(2) H} & = \left[ \left( \rho_{cq} + \rho_{qc} \right) U_H^{(3)}
        + i \left( \rho_{cq} - \rho_{qc} \right) U_D^{(3)} \right] \chi_{c,1}^2 +
      \left( \rho_{cq} + \rho_{qc} + 2
        \chi_{c,1} \chi_{q,1} \right) U_H'' + i \left( \rho_{cq} - \rho_{qc} + i 2
        \chi_{c,1} \chi_{q,2} \right) U_D'', \\
      V_{cq,12}^{(2) H} & = V_{cq,21}^{(2) H} = \left( \chi_{c,2} \chi_{q,1} +
        \chi_{c,1} \chi_{q,2} \right) U_H'' + \left( \chi_{c,1} \chi_{q,1} -
        \chi_{c,2} \chi_{q,2} \right) U_D'' + \chi_{c,1} \chi_{c,2} \left[ \left(
          \rho_{cq} + \rho_{qc} \right) U_H^{(3)} + i \left( \rho_{cq} - \rho_{qc}
        \right) U_D^{(3)}
      \right], \\
      V_{cq,22}^{(2) H} & = \left[ \left( \rho_{cq} + \rho_{qc} \right) U_H^{(3)}
        + i \left( \rho_{cq} - \rho_{qc} \right) U_D^{(3)} \right] \chi_{c,2}^2 +
      \left( \rho_{cq} + \rho_{qc} + 2 \chi_{c,2} \chi_{q,2} \right) U_H'' + i
      \left( \rho_{cq} - \rho_{qc} - i 2 \chi_{c,2} \chi_{q,1} \right) U_D'',
    \end{split}
  \end{equation}
  the retarded and advanced components only contain classical background fields
  (hence we omit the index $q$),
  \begin{equation}
    \label{eq:146}
    V_c^{(2) R} = 
    \begin{pmatrix}
      U_H' + \chi_{c,1} \left( \chi_{c,2} U_D'' + \chi_{c,1} U_H''\right) &
      U_D' + \chi_{c,2} \left( \chi_{c,2} U_D'' + \chi_{c,1} U_H''\right) \\
      \chi_{c,1} \left( \chi_{c,2} U_H'' - \chi_{c,1} U_D''\right) - U_D' &
      U_H' + \chi_{c,2} \left( \chi_{c,2} U_H'' - \chi_{c,1} U_D''\right) \\
    \end{pmatrix}, \quad V_c^{(2) A} = \left( V_c^{(2) R} \right)^{\dagger},
  \end{equation}
\end{widetext}
and the Keldysh component is field-independent and given by $V^{(2) K} = - i
\gamma \id$. In Eq.~\eqref{eq:137}, the inverse propagator is supplemented by
the cutoff to yield the regularized propagator
\begin{equation}
  \label{eq:147}
  P_{k,cq}(Q) = P_{cq}(Q) + R_k(q^2),
\end{equation}
which determines the zeroth order contribution in the fluctuation expansion of
the flow equation.

We proceed by expanding the inverse propagator Eq.~\eqref{eq:140} to second order in
the fluctuations $\delta \chi$. With Eq.~\eqref{eq:143} we may write
\begin{equation}
  \label{eq:148}
  \Gamma_k^{(2)}(Q,Q') = P_{cq}(Q) \delta(Q - Q') + \mathcal{F}(Q,Q') + O \left(
    \delta \chi^3 \right),
\end{equation}
where the matrix $\mathcal{F}$ is given by the sum $\mathcal{F} = \mathcal{F}_1
+ \mathcal{F}_2$ with $\mathcal{F}_{1,2}$ being of first and second order in
$\delta \chi$. The explicit dependence of these matrices on the fluctuations
reads
\begin{align}
  \label{eq:149}  
  \mathcal{F}_1(Q,Q') & = -
  \sum_i V^{(3)}_i \delta \chi_i(Q - Q'), \\ \label{eq:150} \mathcal{F}_2(Q,Q') & = -
  \frac{1}{2} \sum_{ij} V^{(4)}_{ij} \int_P \delta \chi_i(-P) \delta \chi_j(P
  + Q - Q').
\end{align}
Here, for given values of $i$ and $j$ the quantities $V^{(3)}_i$ and $V^{(4)}_{ij}$ are $4 \times 4$
matrices defined as the partial derivatives of $V^{(2)}$,
\begin{equation}
  \label{eq:151}
  V^{(3)}_i = \frac{\partial V^{(2)}}{\partial \chi_i}, \quad  V^{(4)}_{ij} =
  \frac{\partial V^{(2)}}{\partial \chi_i \partial \chi_j}.
\end{equation}
Inserting the decomposition Eq.~\eqref{eq:148} in Eq.~\eqref{eq:137} and expanding
the logarithm in the fluctuations $\delta \chi$ yields
\begin{equation}
  \label{eq:152}
  \partial_t \Gamma_k = \frac{i}{2} \left[ \Tr \dtt \ln
    P_{k,cq} - \frac{1}{2} \dtt \Tr \left( G_{k,cq} \mathcal{F}_1 \right)^2 \right],
\end{equation}
where $G_{k,cq}(Q) = P_{k,cq}(Q)^{-1}$ is the propagator in the presence of
classical and quantum background fields. Note that the appearance of
$G_{k,cq}^2$ makes the trace in the last term UV-convergent and thereby allowed
us to commute $\dtt$ with $\Tr$. In the expansion Eq.~\eqref{eq:152} we are keeping
only terms of zeroth and second order, as these determine, respectively, the
flow of the effective potential and the frequency- and momentum-dependent
contributions to the inverse propagator. We also omit a term $\dtt \Tr G_{k,cq}
\mathcal{F}_2$ which in our truncation with momentum-independent vertices does
not contribute to the flow of $Z$ and $\bar{K}$.

\subsection{Flow equation for the effective potential}
\label{sec:flow-eff-pot}

Equation~\eqref{eq:152} reduces to the flow equation for the effective potential
if we set the fluctuations $\delta \chi$ to zero. Then the second term on the
RHS vanishes and we have
\begin{equation}
  \label{eq:153}
  \frac{1}{\Omega} \partial_t \Gamma_{k,cq} = \frac{i}{2} \int_Q \dtt \ln \dip_{cq}(\omega,q^2)
\end{equation}
where $\dip_{cq}(\omega,q^2) = \det P_{k,cq}(Q)$ denotes the determinant of the
regularized inverse propagator Eq.~\eqref{eq:147} in the presence of classical and
quantum background fields. Since our model is symmetric under simultaneous phase
rotations $\phi_{\nu} \to e^{i \alpha} \phi_{\nu}$ of the classical and quantum
fields, the determinant $\dip_{cq}(\omega,q^2)$ can be expressed as a function
of the $U(1)$-invariant field combinations $\rho_c, \rho_{cq}, \rho_{qc},$ and
$\rho_q$. It can not be written as a function of these invariants without
ambiguity though, as can be seen by noting that the product of four fields
$\phi_c^{*} \phi_q^{*} \phi_c \phi_q$ equals both $\rho_c \rho_q$ and $\rho_{cq}
\rho_{qc}$. However, the form of the field-dependent contribution Eq.~\eqref{eq:144}
to the inverse propagator implies that $\dip_{cq}(\omega,q^2)$ contains terms
that are at most quadratic in the quantum fields and that there is no
contribution that contains $\phi_q^{*} \phi_q$ but no classical fields. All
contributions containing quantum \emph{and} classical fields can be expressed in
powers of $\rho_c, \rho_{cq}$, and $\rho_{qc}$. Therefore, in the following we
will consider $\dip_{cq}(\omega,q^2)$ to be a function of this reduced set of
invariants. Then, inserting Eq.~\eqref{eq:153} in the definition of $\zeta'$ in
Eq.~\eqref{eq:66} we find
\begin{equation}
  \label{eq:154}
  \zeta' = - \frac{i}{2} \int_Q \dtt
  \left\{ \frac{1}{\dip_c(\omega,q^2)} \left[ \partial_{\rho_{cq}}
      \dip_{cq}(\omega,q^2) \right]_{\rqn} \right\},
\end{equation}
where $\dip_c(\omega,q^2) = \det P_{k,c}(Q)$ is the determinant of the
regularized propagator with only classical background fields,
\begin{equation}
  P_{k,c}(Q) = P_{k,cq}(Q) \bigr\rvert_{\qn},
\end{equation}
which differs from $P_{k,cq}(Q)$ only in the block $V_{cq}^{(2) H}$ (note that
the other blocks in Eq.~\eqref{eq:144} do not contain quantum fields) which
vanishes for $\phi_q = \phi_q^{*} = 0$. Accordingly the inverse propagator
$P_{k,c}(Q)$ acquires the causality structure Eq.~\eqref{eq:39} which implies that
the determinant $\dip_c(\omega,q^2)$ factorizes into retarded and advanced
contributions,
\begin{equation}
  \label{eq:155}
  \dip_c(\omega,q^2) = \dip_c^R(\omega,q^2) \dip_c^A(\omega,q^2).
\end{equation}
These are simply related by a change of the sign of the frequency variable,
$\dip_c^R(\omega,q^2) = \dip_c^A(- \omega,q^2).$ Inserting Eq.~\eqref{eq:155} in
Eq.~\eqref{eq:154} we can rewrite the latter as
\begin{equation}
  \label{eq:156}
  \zeta' = 2 v_d \int_0^{\infty} d x x^{d/2 - 1} \dtt \zeta'(x),
\end{equation}
where $v_d = \left( 2^{d + 1} \pi^{d/2} \Gamma(d/2) \right)^{-1}$ and we
introduced a new integration variable $x = q^2$; the function appearing in the
integrand is given by the integral over frequencies
\begin{equation}
  \label{eq:157}
  \zeta'(q^2) = - \frac{i}{4 \pi} 
  \int_{- \infty}^{\infty} d \omega \frac{\left[ \partial_{\rho_{cq}} \dip_{cq}(\omega,q^2)
    \right]_{\rqn}}{\dip_c^A(\omega,q^2) \dip_c^A(-\omega,q^2)},
\end{equation}
which can be performed with the aid of
Ref.~\onlinecite{Prudnikov/Brychkov/Marichev:I}, p.~308, 18.\ (where a factor of
$(-1)^{n + 1}$ is missing\cite{hofer13}). We omit the rather lengthy result.

Let us proceed by specifying the action of $\dtt$ in Eq.~\eqref{eq:156}. The
function $\zeta'(x)$ depends on the cutoff via its dependence on $p_a(x)$ for
which we have $\dtt p_a(x) = - \dtt R_{k,a}(x)$, see
Eq.~\eqref{eq:45}, and thus
\begin{equation}
  \label{eq:158}
  \dtt \zeta'(x) = -\sum_{a = A,D} \dtt R_{k,a}(x) \partial_{p_a(x)} \zeta'(x).
\end{equation}
Recalling the definition Eq.~\eqref{eq:135} of the differential operator $\dtt$
according to which it effectively acts as a scale derivative of the bare cutoff,
we find
\begin{equation}
  \label{eq:159}
  \begin{split}
    \dtt R_{k,A}(x) & = \Re \left( \partial_t R_{k,\bar{K}}(x)/Z \right), \\
    \dtt R_{k,D}(x) & = \Im \left( \partial_t R_{k,\bar{K}}(x)/Z \right).
  \end{split}
\end{equation}
Inserting here the expression
\begin{equation}
  \label{eq:160}
  \partial_t R_{k,\bar{K}}(x) = - \left[ \left( 2 \bar{K} + \partial_t \bar{K}
    \right) k^2 - \partial_t \bar{K} x \right] \theta(k^2 - x),
\end{equation}
we end up with
\begin{equation}
  \label{eq:161}
  \dtt R_{k,a}(x) = - \left[ \left( 2 - \bar{\eta}_a \right) k^2 +
    \bar{\eta}_a x \right] a \theta(k^2 - x),
\end{equation}
where we defined
\begin{equation}
  \label{eq:162}   
  \bar{\eta}_A = - \frac{1}{A} \Re \left( \partial_t \bar{K}/Z \right), \quad
  \bar{\eta}_D = - \frac{1}{D} \Im \left( \partial_t \bar{K}/Z \right).
\end{equation}
Plugging these results in Eq.~\eqref{eq:156} and using that the
$\theta$-function restricts the range of integration over $x$ to the interval
$[0,k^2]$, where $p_a(x) = a k^2$ (cf.\ Eq.~\eqref{eq:45}) and therefore
$\zeta'(x) = \zeta'(k^2)$ does not depend on $x$, we get
\begin{equation}
  \label{eq:163}
  \zeta' = \frac{8 v_d k^{d + 2}}{d} \sum_a \left( 1 - \frac{\bar{\eta}_a}{d + 2}
  \right) a \left[ \partial_{p_a(x)} \zeta'(x) \right]_{p_A(x) = A k^2, p_D(x) = D k^2}.
\end{equation}
The further evaluation of this expression is most conveniently performed on the
computer using \mathematica{}.

In Sec.~\ref{sec:non-eq-frg-flow} we specified prescriptions that allow us to
obtain flow equations for the complex two- and three-body couplings from the
flow equation for the effective potential, cf.\ Eqs.~\eqref{eq:68}
and~\eqref{eq:69}. When we switch to \mathematica{} for an explicit evaluation
of the flow equations, however, it is more convenient to work with real
couplings. The flow equations for the quartic and sextic couplings are then
given by
\begin{equation}
  \label{eq:164}
  \begin{split}    
    \partial_t \lambda & = \beta_{\lambda} = \ezr \lambda - \ezi \kappa +
    \lambda_3 \partial_t \rho_0 + \partial_{\rho_c} \zeta_H'
    \bigr\rvert_{\mathrm{ss}}, \\ \partial_t \kappa & = \beta_{\kappa} = \ezr
    \kappa + \ezi \lambda + \kappa_3 \partial_t \rho_0 + \partial_{\rho_c}
    \zeta_D' \bigr\rvert_{\mathrm{ss}}, \\ \partial_t \lambda_3 & =
    \beta_{\lambda_3} = \ezr \lambda_3 - \ezi \kappa_3 + \partial_{\rho_c}^2
    \zeta_H' \bigr\rvert_{\mathrm{ss}}, \\ \partial_t \kappa_3 & =
    \beta_{\kappa_3} = \ezr \kappa_3 + \ezi \lambda_3 + \partial_{\rho_c}^2
    \zeta_D' \bigr\rvert_{\mathrm{ss}},
  \end{split}
\end{equation}
where we decompose $\zeta' = \zeta_H' + i \zeta_D'$ and $\eta_Z = \ezr + i \ezi$
into real and imaginary parts. For completeness we also state the flow equation
of $\rho_0$ in terms of these quantities:
\begin{equation}
  \label{eq:165}
  \partial_t \rho_0 = \beta_{\rho_0} = - \zeta_D'\bigr\rvert_{\mathrm{ss}}/\kappa.
\end{equation}

To conclude this section let us specify the flow equation for $\gamma$. Similar
to Eq.~\eqref{eq:156} we can express the quantity $\zeta_{\gamma}$ defined in
Eq.~\eqref{eq:72} as
\begin{equation}
  \label{eq:166}
  \zeta_{\gamma} = 2 v_d \int_0^{\infty} dx x^{d/2 - 1} \dtt \zeta_{\gamma}(x).
\end{equation}
As anticipated in the paragraph following Eq.~\eqref{eq:153}, the determinant
$\dip_{cq}(\omega,q^2)$ can be expressed in terms of $\rho_c, \rho_{cq}$, and
$\rho_{qc}$ solely. Therefore, the term that is proportional to $\phi_q^{*}
\phi_q$ and determines the flow of $\gamma$ can then be found taking the
derivative
\begin{equation}
  \label{eq:167}
  \frac{\partial^2}{\partial \phi_q^{*} \partial \phi_q} = \frac{\partial
    \rho_{cq}}{\partial \phi_q} \frac{\partial \rho_{qc}}{\partial \phi_q^{*}} 
  \frac{\partial^2}{\partial \rho_{cq} \partial \rho_{qc}} = \rho_c
  \frac{\partial^2}{\partial \rho_{cq} \partial \rho_{qc}},
\end{equation}
and we find for the integrand in Eq.~\eqref{eq:166} the expression
\begin{multline}
  \label{eq:168}
  \zeta_{\gamma}(q^2) = \frac{\rho_0}{4 \pi} \int_{-\infty}^{\infty} d \omega
  \left[ \frac{\partial_{\rho_{cq},\rho_{qc}}^2 \dip_{cq}(\omega,q^2)
    }{\dip_c(\omega,q^2)} \right. \\ \left. - \frac{\partial_{\rho_{cq}}
      \dip_{cq}(\omega,q^2) \partial_{\rho_{qc}}
      \dip_{cq}(\omega,q^2)}{\dip_c^2(\omega,q^2)} \right]_{\mathrm{ss}}.
\end{multline}
This can be treated in the same way as Eq.~\eqref{eq:157} above.

\subsection{Flow equation for the inverse propagator}
\label{sec:flow-inverse-prop}

The second term on the RHS of Eq.~\eqref{eq:152} determines the flow of both the
wave-function renormalization and the gradient coefficient. It is quadratic in
the fluctuations $\delta \chi$, hence we can write it as
\begin{equation}
  \label{eq:169}  
  \left. \Tr \left( G_{k,cq} \mathcal{F}_1 \right)^2 \right\rvert_{\mathrm{ss}} = - i 2 \int_Q \delta
  \chi(-Q)^T \Sigma(Q) \delta \chi(Q),
\end{equation}
where we set the fields to their stationary values. $\Sigma(Q)$ can be
visualized as consisting of one-loop diagrams with four external legs two of
which are attached to the condensate (cf.\ the second diagram on the RHS of
Eq.~\eqref{eq:27}) and is given by
\begin{equation}
  \label{eq:170}
  \Sigma_{ij}(Q) = \frac{i}{2} \int_P \tr \left( G_k(P) V_i^{(3)}
    G_k(P + Q) V_j^{(3)} \right),
\end{equation}
where $G_k(Q) = P_k(Q)^{-1}$ with the inverse propagator given by
Eqs.~\eqref{eq:39} and~\eqref{eq:40} to which the cutoff $R_k(q^2)$ has to be
added. For $\phi_c = \phi_c^{*} = \phi_0$ and $\phi_q = \phi_q^{*} = 0$ the
matrices $V_i^{(3)}$ have the structure
\begin{equation}
  \label{eq:171}
  V^{(3)}_i =
  \begin{pmatrix}
    v_{3,i}^H & v_{3,i}^A \\
    v_{3,i}^R & 0
  \end{pmatrix}, \quad v_{3,1}^H = v_{3,2}^H = 0, \quad v_{3,3}^{R/A} = v_{3,4}^{R/A} = 0.
\end{equation}
Inserting this expression in Eq.~\eqref{eq:170} above and taking the causality
structure of the propagator into account, we can rewrite the integrand in the
form ($P_{+} = P + Q$)
\begin{multline}
  \label{eq:172}
  \tr \left( G_k(P) V_i^{(3)} G_k(P_{+}) V_j^{(3)} \right) = \tr \left( G_k^K(P)
    v_{3,i}^H G_k^K(P_{+}) v_{3,j}^H \right) \\ + \tr \left( G_k^K(P) v_{3,i}^H
    G_k^R(P_{+}) v_{3,j}^R \right) + \tr \left( G_k^R(P) v_{3,i}^R G_k^K(P_{+})
    v_{3,j}^H \right) \\ + \tr \left( G_k^K(P) v_{3,i}^A G_k^A(P_{+}) v_{3,j}^H
  \right) + \tr \left( G_k^A(P) v_{3,i}^H G_k^K(P_{+}) v_{3,j}^A \right).
\end{multline}
Then the second and third equalities in Eq.~\eqref{eq:171} imply that
$\Sigma(Q)$ has the same causality structure as the inverse propagator. For the
retarded block we find
\begin{multline}
  \label{eq:173}
  \Sigma^R_{ij}(Q) = \frac{i}{2} \int_P \left[ \tr \left( G_k^K(P) v_{3,i + 2}^H
      G_k^R(P_{+}) v_{3,j}^R \right) \right. \\ \left. + \tr \left( G_k^A(P)
      v_{3,i + 2}^H G_k^K(P_{+}) v_{3,j}^A \right) \right],
\end{multline}
where now the indices $i$ and $j$ take the values $1,2$, and the Keldysh
component is given by
\begin{equation}
  \label{eq:174}
  \Sigma^K_{ij}(Q) = \frac{i}{2} \int_P \tr \left( G_k^K(P) v_{3,i + 2}^H
    G_k^K(P_{+}) v_{3,j + 2}^H \right).  
\end{equation}
The frequency integrals appearing in Eqs.~\eqref{eq:173} and~\eqref{eq:174} can
be evaluated by straightforward application of the resiude theorem: $G_k^R(Q)$
has simple poles $\omega_{1,2}^R$ given by Eq.~\eqref{eq:42} with $A q^2$ and $D
q^2$ replaced by $p_A(q^2)$ and $p_D(q^2)$ respectively. While the poles of the
advanced propagator $\omega_{1,2}^A$ are complex conjugate to the poles of the
retarded propagator, $G_k^K(Q)$ has poles at both $\omega_{1,2}^R$ and
$\omega_{1,2}^A$. We omit the lengthy expression for $\Sigma(Q)$ after frequency
integration.

Combining Eqs.~\eqref{eq:73} and~\eqref{eq:152}, the flow equation for
frequency- and momentum-dependent part of the the bare inverse propagator can be
written as
\begin{equation}
  \label{eq:175}
  \partial_t \left( \bar{P}(Q) - \bar{P}(0) \right) = - z^T \left( \dtt
    \Sigma(Q) \right) z,
\end{equation}
with the matrix $z$ defined in Eq.~\eqref{eq:133}. Inserting this expression in the
flow equations for the wave-function renormalization $Z$ and the gradient
coefficient $\bar{K}$, Eqs.~\eqref{eq:76} and~\eqref{eq:75} respectively, we
find after some algebra,
\begin{align}
  \label{eq:176}
  \eta_Z & = - \frac{1}{2} \partial_{\omega} \tr \left[ \left( \id + \sigma_y
    \right) \dtt \Sigma^R(Q) \right] \Bigr\rvert_{Q = 0}, \\ \label{eq:177}
  \partial_t \bar{K}/Z & = \partial_{q^2} \left( \dtt \Sigma^R_{22}(Q) + i \dtt
    \Sigma^R_{12}(Q) \right) \Bigr\rvert_{Q = 0}.
\end{align}
The real and imaginary parts of the anomalous dimension $\eta_Z$, which appear
in the flow equations~\eqref{eq:164} of the real quartic and sextic couplings,
are then given by
\begin{align}
  \label{eq:178}
  \ezr & = \Re \eta_Z = - \frac{1}{2} \partial_{\omega} \tr \left( \sigma_y \dtt \Sigma^R(Q) \right)
  \Bigr\rvert_{Q = 0}, \\ \label{eq:179} \ezi & = \Im \eta_Z = - \frac{i}{2}
  \partial_{\omega} \tr \left( \dtt \Sigma^R(Q) \right) \Bigr\rvert_{Q = 0}.
\end{align}
Here we used the relation $\Sigma^R(Q) = \Sigma^R(-Q)^{*}$ which implies
$\partial_{\omega} \Sigma^R(0) = - \partial_{\omega} \Sigma^R(0)^{*}$. To
further evaluate $\ezr$ and $\ezi$ we switch to \mathematica{}. The derivatives
with respect to the frequency can be carried out without any difficulty and
$\dtt$ can be calculated as in Eq.~\eqref{eq:158} above. Again the integral over
spatial momenta is facilitated by the $\theta$-function contained in $\dtt
R_{k,a}(x)$ and can be carried out analytically.

Finally, for the real and imaginary parts of the renormalized kinetic
coefficient $K = \bar{K}/Z = A + i D$ we have
\begin{align}
  \label{eq:180}
  \partial_t A & = \beta_A = \Re \partial_t K = \ezr A - \ezi D - \bar{\eta}_A A,
  \\ \label{eq:181} \partial_t D & = \beta_D = \Im \partial_t K = \ezr D + \ezi A -
  \bar{\eta}_D D,
\end{align}
where using $\partial_{q^2} \Sigma^R(0) = \partial_{q^2} \Sigma^R(0)^{*}$ (note
that $\Sigma(Q)$ depends only on the norm squared $q^2$ of the spatial momentum)
we may express the quantities $\bar{\eta}_A$ and $\bar{\eta}_D$ defined in
Eq.~\eqref{eq:162} as
\begin{equation}
  \label{eq:182}
  \begin{split}  
    \bar{\eta}_A & = - \frac{1}{A}
    \partial_{q^2} \dtt \Sigma^R_{22}(Q) \bigr\rvert_{Q = 0} = -
    \frac{1}{2A} \partial_q^2 \dtt \Sigma^R_{22}(Q) \bigr\rvert_{Q = 0},
    \\ \bar{\eta}_D & = - \frac{1}{D}
    \partial_{q^2} \dtt \Sigma^R_{12}(Q) \bigr\rvert_{Q = 0} = - \frac{1}{2
      D} \partial_q^2 \dtt \Sigma^R_{12}(Q) \bigr\rvert_{Q = 0}.
  \end{split}
\end{equation}
We will proceed with the evaluation of these expressions in the next section.

\subsection{Computation of gradient coefficient anomalous dimensions}
\label{sec:comp-grad-coeff}

As the cutoff Eq.~\eqref{eq:43} is a non-analytic function of the momentum, the
evaluation of the derivatives in Eq.~\eqref{eq:182} requires some care. In this
section we present two approaches to this problem: The first one was introduced
by Wetterich in Ref.~\onlinecite{wetterich08:_funct} and the second one makes use of
Morris' lemma.\cite{morris94:_the_exact_renor_group_and_approx_solut} Our
starting point is Eq.~\eqref{eq:173} in which we set the external frequency
$\omega$ to zero. Using the shorthand $\int_{\mathbf{p}} = \int \frac{d^d
  \mathbf{p}}{\left( 2 \pi \right)^d}$ we may write
\begin{equation}
  \label{eq:183}
  \Sigma^R(0,\mathbf{q}) = \int_{\mathbf{p}} \sigma^R(p_A, p_D, p_{A+}, p_{D+}).
\end{equation}
Here and in the following for the sake of brevity we will omit the arguments in
$p_a \equiv p_a(x)$ and $p_{a\pm} \equiv p_a(x_{\pm})$ for $a = A,D$, $x = q^2$
and $x_{\pm} = \abs{\mathbf{p} \pm \mathbf{q}}^2$. The integrand in the above
expression is given by the integral over the frequency component of the internal
momentum $P = \left( \nu,\mathbf{p} \right)$,
\begin{multline}
  \label{eq:184}
  \sigma_{ij}^R(p_A, p_D, p_{A+}, p_{D+}) = \frac{i}{2} \int \frac{d \nu}{2
    \pi} \left[ \tr \left( G_k^K(P) v_{3,i + 2}^H G_k^R(P_{+}) v_{3,j}^R \right)
  \right. \\ \left. + \tr \left( G_k^A(P) v_{3,i + 2}^H G_k^K(P_{+}) v_{3,j}^A
    \right) \right].
\end{multline}
Our notation makes explicit that the momentum dependence of the regularized
propagator $G_k(Q)$ is contained in the functions $p_a(q^2)$ introduced in
Eq.~\eqref{eq:45}. Inserting Eq.~\eqref{eq:184} in the expressions for the
anomalous dimensions Eq.~\eqref{eq:182} we find
\begin{equation}
  \label{eq:185}
  \begin{split}
    \bar{\eta}_A & = - \frac{1}{2 A} \dqn \int_{\mathbf{p}} \dtt
    \sigma^R_{22}(p_A, p_D, p_{A+}, p_{D+}), \\ \bar{\eta}_D & = - \frac{1}{2 D}
    \dqn \int_{\mathbf{p}} \dtt \sigma^R_{12}(p_A, p_D, p_{A+}, p_{D+}).
  \end{split}
\end{equation}
In the following we will discuss the evaluation of $\bar{\eta}_A$ while we will
only state the result for $\bar{\eta}_D$. Let us begin by introducing the
abbreviations $\partial_a \equiv \partial_{p_a(x)}$ and $\partial_{a\pm}
\equiv \partial_{p_a(x_{\pm})}$. In the integrand we omit the arguments and
write $\sigma^R_{22+} \equiv \sigma^R_{22}(p_A, p_D, p_{A+}, p_{D+})$ and
$\sigma^R_{22-} \equiv \sigma^R_{22}(p_{A-}, p_{D-}, p_A, p_D)$. We recall that
the derivative $\dtt$ acts only on the cutoff, hence we have
\begin{equation}
  \label{eq:186}
  \bar{\eta}_A = \frac{1}{2 A} \dqn \int_{\mathbf{p}} \sum_a \dtt
  R_{k,a}(x) \partial_a \left( \sigma^R_{22+} + \sigma^R_{22-} \right),
\end{equation}
where we performed a change of integration variables $\mathbf{p} \to \mathbf{p}
- \mathbf{q}$ in the second term.

\subsubsection{Wetterich's method}
\label{sec:wetterichs-method}

Following Ref.~\onlinecite{wetterich08:_funct} we introduce new variables: With $y = x
- k^2$ and $z = \left( x - k^2 \right) \theta(x - k^2) = y \theta(y)$ we have
\begin{equation}
  \label{eq:187}
  p_a(x) = a \left( k^2 + z \right).
\end{equation}
We now use the fact that an expansion of the integrand in Eq.~\eqref{eq:186} in
powers of $z_{\pm}$ is effectively equivalent to an expansion in $q^2$: Below we
will see that due to the $\theta$-functions contained in $z_{\pm}$ and $\dtt
R_{k,a}(x)$ the integration over $\mathbf{p}$ is restricted to a region that is
$O(q)$ for $q \to 0$. In this region $p \approx k$ and the prefactor of the
$\theta$-function in the definition of $z_{\pm}$, therefore, is also
$O(q)$. Hence we may restrict ourselves to the first order in the expansion
\begin{equation}
  \label{eq:188}
  a \partial_a \sigma^R_{22\pm} = a \partial_a \sigma^R_{22\pm}
  \bigr\rvert_{z_{\pm} = 0} +
  A_{\pm} z_{\pm} + O \left( z_{\pm}^2 \right),
\end{equation}
where the coefficient of the linear term is
\begin{equation}
  \label{eq:189}
  A_{\pm} = a \partial_a \sum_b b \partial_{b \pm} \sigma^R_{22\pm} \bigr\rvert_{z_{\pm} = 0}.
\end{equation}
The zeroth order term does not depend on $q$ and can be discarded from the
expression for $\bar{\eta}_A$ which now becomes
\begin{equation}
  \label{eq:190}
  \bar{\eta}_A = \frac{1}{2 A} \dqn \int_{\mathbf{p}} \sum_a
  \frac{1}{a} \dtt R_{k,a}(x) \left(  A_{+} z_{+} + A_{-} z_{-} \right).
\end{equation}
Inserting here the explicit expressions for $z_{\pm} = y_{\pm} \theta(y_{\pm})$
we find
\begin{equation}
  \label{eq:191}
  \bar{\eta}_A = - \frac{1}{2 A} \dqn \left( B_{+} + B_{-} \right),
\end{equation}
where using Eq.~\eqref{eq:161} we have
\begin{equation}
  \label{eq:192}
  B_{\pm} = \sum_a \int_{\mathbf{p}} \left[ \left( 2 - \bar{\eta}_a
    \right) k^2 + \bar{\eta}_a x \right] \theta(k^2 - x) \theta(y_{\pm}) A_{\pm} y_{\pm}.
\end{equation}
Due to the first $\theta$-function only momenta $\mathbf{p}$ within a circle of
radius $k$ centered at the origin contribute to the integral (hence we may set
$p_a(x) = a k^2$ in $A_{\pm}$), while the second $\theta$-function excludes all
$\mathbf{p}$ inside a circle of radius $k$ centered at $\mp \mathbf{q}$. In the
resulting area of integration -- which is itself $O(q)$ as anticipated above --
we have $p \approx k$ for $q \to 0$. Without loss of generality we choose
$\mathbf{q} = \left( q,0,\dotsc \right)$ and decompose the integral as
$\int_{\mathbf{p}} = \int_{\mathbf{p}_t} \int_{-\infty}^{\infty} \frac{d p_1}{2
  \pi}$, where $p_1$ is the component in the direction of $\mathbf{q}$, i.e.,
$\mathbf{p} = \left( p_1,\mathbf{p}_t \right)$, and $\mathbf{p}_t \in \R^{d -
  1}$. The integrand does not depend on the direction of $\mathbf{p}_t$, hence,
using (this relation holds for $d \geq 2$; for $d = 1$ there is no integration
over $\mathbf{p}_t$)
\begin{equation}
  \label{eq:193}
  \int_{\mathbf{p}_t} f(x_t) = 2 v_{d - 1} \int_0^{\infty} d x_t x_t^{(d - 3)/2},
\end{equation}
where the integration variable on the RHS is $x_t = p_t^2$, we have
\begin{equation}
  \label{eq:194}
  B_{\pm} = \int_0^{\infty} d x_t 
  \int_{-\infty}^{\infty} d p_1 \theta(k^2 - x) \theta(y_{\pm}) b_{\pm},
\end{equation}
where
\begin{equation}
  \label{eq:195}
  b_{\pm} = \frac{v_{d - 1}}{\pi} x_t^{(d - 3)/2} \sum_a \left[ \left( 2 - \bar{\eta}_a \right) k^2 +
    \bar{\eta}_a x \right] A_{\pm} y_{\pm}.
\end{equation} 
In Eq.~\eqref{eq:194} the $\theta$-functions restrict the range of integration
to
\begin{equation}
  \label{eq:196}
  k^2 - p_1^2 - x_t > 0, \quad \left( p_1 \pm q \right)^2 + x_t - k^2 > 0.
\end{equation}
The first of these inequalities allows for a solution for $p_1$ only if $0 < x_t
< k^2$. Then it implies
\begin{equation}
  \label{eq:197}
  -\alpha < p_1 < \alpha.
\end{equation}
where $\alpha = \sqrt{k^2 - x_t}$. The second inequality is equivalent to
\begin{equation}
  \label{eq:198}
  p_1 > \alpha \mp q \quad \vee \quad p_1 < -\alpha \mp q.
\end{equation}
For $B_{+}$ we have to consider the upper sign. Then Eq.~\eqref{eq:197} and the
first inequality Eq.~\eqref{eq:198} have the joint solution
\begin{equation}
  \label{eq:199}
  \max \left\{ -\alpha,\alpha - q \right\} < p_1 < \alpha.
\end{equation}
Splitting the integration over $x_t$ into two ranges $0 < x_t < x_{t0}$ where
$x_{t0} = k^2 - q^2/4$ and $x_{t0} < x_t < k^2$ we can specify the maximum
explicitly as
\begin{equation}
  \label{eq:200}
  \max \left\{ -\alpha,\alpha - q \right\} =
  \begin{cases}
    \alpha - q & \text{for } 0 < x_t < x_{t0}, \\
    -\alpha & \text{for }  x_{t0} < x_t < k^2.
  \end{cases}
\end{equation}
The second inequality Eq.~\eqref{eq:198} and Eq.~\eqref{eq:197} do not have a common
region of validity, and we find
\begin{equation}
  \label{eq:201}
  B_{+} = \int_0^{x_{t0}} d x_t \int_{\alpha - q}^{\alpha} d p_1 b_{+} + \int_{x_{t0}}^{k^2} d
  x_t \int_{-\alpha}^{\alpha} d p_1 b_{+}.
\end{equation}

Let us now consider $B_{-}$: Eq.~\eqref{eq:197} and the second
inequality Eq.~\eqref{eq:198} are solved by
\begin{equation}
  \label{eq:202}
  -\alpha < p_1 < \min \left\{ \alpha,-\alpha + q \right\}.
\end{equation}
where in the same ranges of $x_t$ as above the minimum is
\begin{equation}
  \label{eq:203}
  \min \left\{ \alpha,-\alpha + q \right\} =
  \begin{cases}
    -\alpha + q & \text{for } 0 < x_t < x_{t0}, \\
    \alpha & \text{for } x_{t0} < x_t < k^2.
  \end{cases}
\end{equation}
The first inequality Eq.~\eqref{eq:198} and Eq.~\eqref{eq:197} can not be fulfilled at
the same time. Thus we have
\begin{equation}
  \label{eq:204}
  B_{-} = \int_0^{x_{t0}} d x_t \int_{-\alpha}^{-\alpha + q} d p_1 b_{-} +
  \int_{x_{t0}}^{k^2} d x_t \int_{-\alpha}^{\alpha} d p_1 b_{-}.
\end{equation}
Now it is straightforward to carry out the integral over $x_t$ in both $B_{+}$
and $B_{-}$ and we obtain the result
\begin{equation}
  \label{eq:205}
  B_{\pm} = \frac{4 v_d}{d} k^{d + 2} q^2 \sum_a A_{\pm}.
\end{equation}
Inserting this in Eq.~\eqref{eq:191} and using that setting $z_{\pm} = 0$ in
Eq.~\eqref{eq:189} is the same as setting $q = 0$ and $p = k$ we find
\begin{equation}
  \label{eq:206}  
  \bar{\eta}_A = - \frac{4 v_d}{d A} k^{d + 2} \sum_{a,b} a b \partial_a
  \left[ \partial_{b+} \sigma^R_{22+} + \partial_{b-} \sigma^R_{22-}
  \right]_{q = 0,p = k}.
\end{equation}
Both terms on the RHS give the same contribution. Then, carrying out a similar
analysis for $\bar{\eta}_D$ yields
\begin{equation}
  \label{eq:219}
  \begin{split}
    \bar{\eta}_A & = - \frac{8 v_d}{d A} k^{d + 2} \sum_{a,b} a b \partial_a
      \partial_{b+} \sigma^R_{22+} \bigr\rvert_{q = 0,p = k}, \\ \bar{\eta}_D &
    = - \frac{8 v_d}{d D} k^{d + 2} \sum_{a,b} a b \partial_a
    \partial_{b+} \sigma^R_{12+} \bigr\rvert_{q = 0,p = k},
  \end{split}
\end{equation}
The remaining derivatives can straightforwardly be
performed using \mathematica.

\subsubsection{Morris' lemma}
\label{sec:morris-lemma}

The same results can also be obtained by a direct evaluation of the
derivatives in Eq.~\eqref{eq:186},
\begin{multline}
  \label{eq:207}
  \bar{\eta}_A = \frac{1}{2 A} \int_{\mathbf{p}} \sum_{a,b} \dtt
  R_{k,a}(x) \partial_a \biggl\{ \sum_c \partial^2_{b+,c+} \sigma^R_{22+}
  p_{b+}' p_{c+}' \left( \partial_q x_{+} \right)^2 \\ + \partial_{b+}
  \sigma^R_{22+} \left[ p_{b+}'' \left( \partial_q x_{+} \right)^2 +
    p_{b+}' \partial_q^2 x_{+} \right] + \left( + \to - \right) \biggr\}
  \biggr\rvert_{q = 0}
\end{multline}
Upon setting $q = 0$ in the terms in braces, $x_{\pm}$ are replaced by $x$. Then
we may drop all terms that include the product $\dtt R_{k,a}(x) p_b'$ as it
contains $\theta$-functions that do not have a common support: According to
Eq.~\eqref{eq:161} $\dtt R_{k,a}(x)$ is proportional to $\theta(k^2 - x)$, while
$p_b'(x) = b \theta(x - k^2)$. With $\partial_q x_{\pm} \rvert_{q = 0} = \pm 2
\mathbf{p} \cdot \unitvec{q}$ (here $\unitvec{q}$ denotes the vector of unit
length in the direction of $\mathbf{q}$) we find
\begin{equation}
  \label{eq:208}
  \bar{\eta}_A = \frac{2}{d A} \int_{\mathbf{p}} x
  \sum_{a,b} \dtt R_{k,a}(x) p_b'' \partial_a \left[ \partial_{b+}
    \sigma^R_{22+} + \partial_{b-} \sigma^R_{22-} \right]_{q = 0},
\end{equation}
where we used
\begin{equation}
  \label{eq:209}
  \int_{\mathbf{p}} \left( \mathbf{p} \cdot \unitvec{q} \right)^2 f(p) =
  \frac{1}{d} \int_{\mathbf{p}} p^2 f(p)
\end{equation}
The second derivative $p_b''(x) = b \delta(x - k^2)$ contains a
$\delta$-function and, therefore, we set $p = k$ in the terms in brackets. (Note
that $p_a(x)$ is continuous at $x = k^2$.) Then, Using Morris' lemma according
to which we can replace $\delta(x) \theta(x) \to \frac{1}{2} \delta(x)$ when
this combination is multiplied by a function that is continuous at $x = 0$, we
have
\begin{equation}
  \label{eq:210}
  \dtt R_{k,a}(x) p_b''(x) = - \frac{a b k}{2} \delta(p - k).
\end{equation}
Evaluating the integral over $p$ with the aid of the $\delta$-function
reproduces the result Eq.~\eqref{eq:206}.

\bibliography{bibliography}

\end{document}